\documentclass[11pt,a4paper,twoside,notitlepage,authoryear]{elsarticle}
\journal{Journal of the Mechanics and Physics of Solids}

\usepackage{natbib}
\usepackage[colorlinks]{hyperref}
\usepackage[margin=0.75in]{geometry}
\usepackage{amsmath, amssymb, amsfonts, amsthm}
\usepackage{times}
\usepackage{textcomp}
\usepackage{graphicx}
\usepackage{amssymb}
\usepackage{listings}
\usepackage{rotating}
\usepackage{epstopdf}
\usepackage{notoccite}
\usepackage{amsmath}
\usepackage[ruled,vlined]{algorithm2e}
\numberwithin{equation}{section}
\usepackage{svg}
\usepackage[font={small,it}]{caption}
\usepackage{nomencl}
\usepackage{epstopdf}
\newcommand{\avg}[1]{\left\langle{#1}\right\rangle}
\newcommand{\vol}{\mathcal{V}}
\newcommand{\lavg}[1]{\overline{#1}}

\newcommand{\Revision}[1]{{\color{black}#1}}
\newcommand{\boldface}[1]{\boldsymbol{#1}}  

\newcommand{\bfb}{\boldface{b}}

\newcommand{\bfp}{\boldface{p}}
\newcommand{\bfq}{\boldface{q}}

\newcommand{\bfu}{\boldface{u}}

\newcommand{\bfx}{\boldface{x}}

\newcommand{\bfz}{\boldface{z}}
\newcommand{\Rset}{\mathbb{R}}
\def\id{{\bfI}}

\newcommand{\bfA}{\boldface{A}}
\newcommand{\bfB}{\boldface{B}}
\newcommand{\bfC}{\boldface{C}}

\newcommand{\bfF}{\boldface{F}}

\newcommand{\bfI}{\boldface{I}}

\newcommand{\bfK}{\boldface{K}}

\newcommand{\bfN}{\boldface{N}}

\newcommand{\bfS}{\boldface{S}}
\newcommand{\bfT}{\boldface{T}}
\newcommand{\bfU}{\boldface{U}}

\newcommand{\T}{^{\mathrm{T}}} 

\def\dd{\;\!\mathrm{d}}
\newcommand{\bfSigma}{\boldsymbol{\Sigma}}
\DeclareMathOperator{\tr}{tr}
\newcommand{\bfphi}{\boldsymbol{\phi}}
\newcommand{\bfzeta}{\boldsymbol{\zeta}}
\newcommand{\st}[1]{\qquad\text{#1}\qquad}
\newcommand{\be}{\begin{equation*}}
\newcommand{\ee}{\end{equation*}}
\newcommand{\bfsigma}{\boldsymbol{\sigma}}
\newcommand{\bfXi}{\boldsymbol{\Xi}}
\newcommand{\bfkappa}{\boldsymbol{\kappa}}
\newcommand{\calN}{\mathcal{N}}

\graphicspath{{./}}
\begin{document}

\begin{frontmatter}
\title{
	{Nonequilibrium thermomechanics of Gaussian phase packet crystals:\\ application to the quasistatic quasicontinuum method}}
\author[eth]{Prateek Gupta\corref{mycorrespondingauthor}}
\ead{pgupta@ethz.ch}
\author[caltech]{Michael Ortiz}
\ead{ortiz@aero.caltech.edu}
\author[eth]{Dennis M.~Kochmann\corref{mycorrespondingauthor}}
\ead{dmk@ethz.ch}
\address[eth]{Mechanics \& Materials Lab, Department of Mechanical and Process Engineering \\ ETH Z\"urich, 8092 Z\"urich, Switzerland}
\address[caltech]{Graduate Aerospace Laboratories, California Institute of Technology, Pasadena, CA 91125, USA}

\begin{abstract}
The quasicontinuum \Revision{(QC)} method was originally introduced to bridge across length scales \Revision{by coarse-graining an atomistic ensemble to significantly larger continuum scales at zero temperature, thus overcoming the crucial length-scale} limitation of classical atomic-scale simulation techniques while solely relying on atomic-scale input (in the form of interatomic potentials). An associated challenge lies in bridging across time scales to overcome the time-scale limitations of atomistics \Revision{at finite temperature}. To address the biggest challenge, bridging across both length and time scales, only a few techniques exist, and most of those are limited to conditions of constant temperature. Here, we present a new \Revision{general} strategy for the \Revision{\textit{space-time coarsening}} of an atomistic ensemble, which introduces thermomechanical coupling. \Revision{Specifically, we evolve the statistics of an atomistic ensemble in phase space over time by applying the Liouville equation to an approximation of the ensemble's probability distribution (which further admits a variational formulation). To this end, we approximate a crystalline solid as a lattice of lumped correlated Gaussian phase packets occupying atomic lattice sites, and we investigate the resulting quasistatics and dynamics of the system}. By definition, phase packets account for the dynamics of crystalline lattices at finite temperature through the statistical variances of atomic momenta and positions. We show that momentum-space correlation allows for an exchange between potential and kinetic contributions to the crystal's Hamiltonian. Consequently, local adiabatic heating due to atomic site motion is captured. Moreover, \Revision{in the quasistatic limit,} the governing equations reduce to the minimization of thermodynamic potentials \Revision{(similar to maximum-entropy formulation previously introduced for finite-temperature QC), and they yield the local equation of state, which we derive for isothermal, isobaric, and isentropic conditions. Since our formulation without interatomic correlations precludes irreversible heat transport, we demonstrate its combination with thermal transport models to describe realistic atomic-level processes, and we discuss opportunities for capturing atomic-level thermal transport by including interatomic correlations in the Gaussian phase packet formulation. Overall, our Gaussian phase packet approach} offers a promising avenue for finite-temperature non-equilibrium quasicontinuum \Revision{techniques, which may be combined with thermal transport models and extended to other approximations of the probability distribution as well as to exploit the variational structure.}
\end{abstract}

\end{frontmatter}

\section{Introduction}
\label{sec: intro}
Crystalline solids exhibit physical and chemical transport phenomena across wide ranges of length and time scales. This includes the transport of charges~\citep{butcher1986theory,ziman2001electrons}, heat ~\citep{ziman2001electrons}, and mass~\citep{weiner2012statistical}, as well as mechanical failure. Understanding such phenomena is crucial from both a fundamental scientific standpoint as well as to further advance technologies ranging from solid-state batteries~\citep{kim2014atomistic} to thermal management systems~\citep{hicks1993thermoelectric} to failure-resistant metallic structural components~\citep{hirth1980effects} -- all exposed to complex dynamic conditions \Revision{varying over time scales of a few microseconds to several hours and length scales of a few nanometers to a meter}. \Revision{Such a variety of length and time scales underlying the transport phenomena calls for simulation techniques that capture the physical processes across} all length and time scales involved. While continuum mechanics and related finite element (FE) and phase field methods have been successful at modeling physical processes at relatively large length scales (typically micrometers and above)
and time scales (milliseconds and above)~\citep{hirth1980effects,mendez2018diffusive}, molecular statics (MS) and molecular dynamics (MD) have been successful at elucidating the physics of various transport phenomena at atomic-level length scales (angstroms to tens of nanometers) and time scales (femto- to nanoseconds)~\citep{tuckerman2010statistical}. Where a higher level of accuracy is required, methods such as Density Functional Theory (DFT) or the direct computation of Schr\"{o}dinger's equation have aimed \Revision{at capturing the quantum coupling of molecular-level physics.}
All of the aforementioned techniques specialize in the approximate ranges of length and time scales mentioned above. However, each of those techniques poses restrictive assumptions at smaller scales while becoming prohibitively costly at larger scales, hence making scale-bridging techniques attractive~\citep{srivastava2014limit,van2020roadmap}. \Revision{For instance, the state-of-the-art DFT-based multiscale modeling techniques developed by~\citet{motamarri2020dft} make ab-initio accuracy available to larger atomic ensembles, simulating systems consisting of approximately 4000 atoms with a large computational cost met by massively parallel supercomputers.}

\Revision{Several concurrent scale-bridging techniques have been developed over the past few decades, particularly focusing on multiscale thermomechanical modeling of crystalline materials (cf.~\citet{xu2019modeling} for a detailed review of some of the available techniques). The \emph{atomistic-to-continuum} (AtoC) method developed by \cite{wagner2008atomistic} utilizes a coupling between a pre-determined atomistic domain and an overlaid continuum domain, discretized using a suitable FE method. The continuum domain is used for providing a heat bath to the atomistic domain, simultaneously minimizing the difference between the temperature fields in both domains, hence establishing a two-way coupling. In the \emph{bridging-domain method} (BDM), the pre-determined atomistic and continuum subdomains are coupled by imposing a weak displacement compatibility condition at the intersecting nodes~\citep{belytschko2003coupling}. Nodal mechanical forces are computed using the total Hamiltonian of the system constructed using the atomistic and continuum subdomains as well as the weak compatibility conditions. \citet{chen2009reformulation} developed the \emph{concurrent atomistic-continuum} (CAC) method in which microscopic balance equations, derived using the theory of \citet{irving1950statistical}, are solved to determine the thermomechanical deformation of atomic sites within large continuum-scale subdomains. Unlike the AtoC and BDM methods, the CAC method can capture lattice defects, such as dislocations and cracks, in the continuum subdomain. The \emph{coupled atomistic/discrete-dislocation} (CADD) method of \citet{shilkrot2002coupled} is another multiscale method that allows the movement of dislocation defects across the atomistic and continuum subdomains. While CAC requires only the interatomic potential as the constitutive input, CADD requires reduced-order continuum constitutive models to determine long-range elastic stress fields. All of the aforementioned methods require a-priori knowledge of atomistic and continuum subdomains, and they solve the thermomechanical deformation in both subdomains using distinct methodologies. This, unfortunately, inhibits a seamless transition from atomistic- to continuum-scale subdomains. Furthermore, finite-temperature variants of all those techniques typically use time-integration at the scale of atomic vibrations to resolve the available phonon modes in the domain~\citep{chen2017effects}, thus not allowing for temporal coarsening.}

The quasicontinuum (QC) method of~\citet{tadmor1996quasicontinuum} is \Revision{a concurrent scale-bridging} method that aims to solve the problem of spatial \Revision{upscaling} from atomistic to continuum length scales via intermediate mesoscopic scales~\citep{miller1998quasicontinuum}. Starting from a standard atomistic setting, a carefully selected set of representative degrees of freedom (\Revision{associated with} \emph{repatoms}) reduces the computational costs and admits simulating continuum-scale problems by restricting atomistic resolution to where it is in fact needed. Different flavors of QC have been proposed based on the interpolation of forces on repatoms~\citep{knap2001analysis}, or based on approximating the total Hamiltonian of the system using quadrature-like \emph{summation and sampling rules}~\citep{eidel2009variational,dobson2010accuracy, gunzburger2010quadrature, espanol2013gamma}. For example, the nonlocal energy-based formulation of \citet{amelang2015summation} enables a seamless spatial scale-bridging within the QC setup, which does not require a strict separation of (nor apriori knowledge about) atomistic and non-atomistic (coarsened) subdomains within the simulation domain. The capabilities of this fully nonlocal QC technique have been demonstrated, e.g., by large-scale quasistatic total-Lagrangian simulations of dislocation interactions during nano-indentation \citep{amelang2015summation}, nanoscale surface and size effects \citep{AmelangKochmann2015}, and void growth and coalescence \citep{tembhekar2017automatic}. In this work, we adopt the \Revision{theoretical nonlocal, energy-based QC formulation} of \citet{amelang2015summation} in a new, updated Lagrangian \Revision{framework and implementation} for spatial upscaling.

While spatial coarse-graining is thus achieved by approximating an \Revision{atomic ensemble by} a subset of atoms, temporal coarse-graining requires approximate \textit{modeling} due to the unavailability of the trajectories of all the atoms at a given time. Furthermore, \Revision{the} Hamiltonian dynamics of an ensemble of atoms couples length and time scales in the system~\citep{j2007statisticalEvans}, which is why spatial scale-bridging techniques have often been applied to systems at zero temperature or a uniform temperature only \citep{tadmor2013finite}. One way to model uniform temperature is to apply a global ergodic assumption for every atomic site, thus yielding space-averaged trajectories of atoms as phase-averaged trajectories. Suitable global thermal equilibrium distributions (such as NVT ensembles \citep{tadmor2011modeling}) are used for phase averaging of trajectories. \Revision{The motion of atoms on these phase-averaged trajectories is governed by the phase averages of interatomic potentials and kinetic energy. Due to thermal softening of interatomic potentials upon phase-averaging, the accessible time-scales in isothermal dynamic simulations increase, hence enabling time coarsening. \citet{kim2014hyper} further increased the accessible time limits of \citeauthor{tadmor2013finite}'s method using hyperdynamics~\citep{voter1997method} to capture thermally activated rare events, such as atomic-scale mass diffusion. However, most prior work has been restricted to local harmonic approximations of interatomic potentials and isothermal deformation \citep{tadmor2011modeling, tadmor2013finite} at a uniform temperature. 
Using Langevin dynamics is an alternative strategy}, in which a stochastic thermal forcing is added to the dynamics of atoms to account for thermal vibrations~\citep{qu2005finite, marian2009finite}. Unfortunately, such an approach poses a time integration restriction even for systems in thermodynamic equilibrium (uniform temperature), which is why it is computationally costly. 

\citet{kulkarni2008variational} introduced a variational \Revision{mean-field} approach for modeling non-uniform temperature, which approximates the global distribution function of the ensemble as a product of local entropy-maximizing (or \emph{max-ent}) distributions, constraining the local frequency of atomic vibrations \Revision{by} using the local equipartition of energy of every atom. This local ergodic assumption yields time-averaged trajectories as phase-averaged trajectories and thus enables the definition of non-uniform temperature and internal energy.  However, the interatomic independence or statistically uncorrelated local distributions, inherent in that approach, precludes the transport of energy across the atoms. As a remedy, \citet{kulkarni2008variational} modeled thermal transport using the FE setup of the QC formulation and empirical bulk-thermal conductivity values. \citet{venturini2014atomistic} extended the max-ent approach to non-uniform interstitial concentrations of solute atoms in crystalline solids to also include mass diffusion. Specifically, they modeled transport using linear Onsager kinetics, derived from a local dissipation inequality. \Revision{Combining~\citeauthor{venturini2014atomistic}'s max-ent-based approach to include non-uniform interstitial concentrations, \citet{mendez2018diffusive} used an Arrhenius-type master-equation model to achieve diffusive transport in long-term atomistic simulations. In the specific case of isothermal diffusion of impurities in crystals, the max-ent framework of \citet{venturini2014atomistic} resembles the \emph{diffusive molecular dynamics} (DMD) formulation of \citet{li2011diffusive}, who used an isotropic Gaussian atomic density cloud to model uncorrelated vibrations of atoms.} 

\Revision{We here present a similar nonequilibrium finite-temperature formulation which approximates the global distribution function of the ensemble} based on Gaussian Phase Packets (GPP)\Revision{, which is a different ansatz from the max-ent one and exhibits time evolution governed by the Liouville equation.} Previously, \citet{ma1993approximate} studied the dynamic Gaussian Phase Packet (GPP) \Revision{ansatz} as a \emph{trajectory-sampling} technique for uniform-temperature MD simulations. They approximated the distribution function of each atom in an ensemble as a correlated Gaussian distribution with the covariance matrix as the mean-field or phase space parameter. Such an approximation yields the evolution equations of the covariance matrix by either directly substituting \Revision{the GPP ansatz into} the Liouville equation (strong form) or by using the Frenkel-Dirac-McLachlan variational principle~\citep{mclachlan1964variational}. The resulting equations\Revision{ -- combined with appropriate ergodic assumptions --} may be integrated in time to infer the locally averaged physical quantities of the system (such as temperature). However, \citet{ma1993approximate} applied the formulation on a small system of atoms relaxing towards a state of thermodynamic equilibrium with uniform temperature. Their work was inspired by the application of GPPs in quantum mechanics by \citet{heller1975time}, who used it for calculating parametrized solutions of the Schr\"{o}dinger's equation. 

In this work, we \Revision{apply} the GPP \Revision{ansatz} to an ensemble of atoms to elucidate the nonequilibrium and (local-thermal) equilibrium thermomechanical quasistatics and dynamics of the system. We show that an approximation of interatomic correlations is required for modeling atomistic-level transport phenomena. In an ensemble of $N$ atoms, such an approximation increases the degrees of freedom by $\mathcal{O}(N^2)$, which is computationally costly. Therefore, we \Revision{assume} uncorrelated atoms or \emph{interatomic independence}. \Revision{Our formulation may be considered as a dynamic extension of the max-ent methods of \citet{kulkarni2008variational} and \citet{venturini2014atomistic} as well as the DMD approach of \citet{li2011diffusive}, to which our theory reduces in the quasistatic limit. We show that incorporating momentum-displacement correlations in a Gaussian ansatz for the distribution function elucidates the local dynamics of atoms and the energy exchange from kinetic to potential, thus dynamically capturing the thermomechanical deformation. When all atoms are assumed to be uncorrelated (such a solid crystal of independent atoms is also known as an \emph{Einstein's solid}), the GPP approximation fails to capture thermal transport due to non-uniform thermomechanical deformation and hence requires additional phenomenological modeling to that end. Moreover, to achieve temporal coarsening, our GPP approach highlights the importance of vanishing interatomic correlations, approaching the quasistatic approximation.} 

\Revision{To simulate irreversible heat flux, we} combine the GPP framework within the quasistatic approximation with \citeauthor{venturini2014atomistic}'s linear Onsager kinetics to model local thermal transport. \Revision{Although our GPP approach under the quasistatic approximation resembles quasistatic max-ent \citep{kulkarni2008variational,venturini2014atomistic}, our correlated Gaussian ansatz highlights the physical significance of assuming vanishing cross-correlations across all degrees of freedom. Furthermore, it emphasizes the need for additional thermodynamic assumptions required for modeling the thermomechanical deformation of the crystal, due to the loss of knowledge of the temporal evolution of the correlations. Our approach furthermore furnishes the quasistatic isothermal max-ent setup with alternative formulations for isobaric and isentropic conditions.} 

\Revision{The remainder of this study is structured as follows.} In Section~\ref{sec: Modeling} we review the fundamentals of nonequilibrium statistical mechanics based on the Liouville equation and the GPP approximation. \Revision{We derive the weak formulation using the variational structure of the Liouville equation, which yields the equations of motion for the phase space parameters}. We also show that the interatomic correlations are of fundamental importance for modeling interatomic heat flux. Since such correlations increase the degrees of freedom significantly, they are neglected in \Revision{subsequent sections while retaining} the phase space correlation of each atom, which we later identify as the \emph{thermal momentum}. In Section~\ref{sec: dynamicsAndQuasistatics} we discuss the importance of thermal momentum in dynamics and the implications of its vanishing limit in quasistatics. We show that, in the \Revision{latter} limit, the \Revision{governing} equations define \Revision{the} local thermomechanical equilibrium of the system and yield a rate-independent local thermal equation of state. To capture thermal transport, we \Revision{adopt} \citeauthor{venturini2014atomistic}'s linear Onsager kinetics-based model. We also discuss the time scale imposed by the Onsager kinetics-based model and its time-stepping constraints. In Section~\ref{sec: QCApplication} we discuss \Revision{our} QC implementation of the local thermomechanical equilibrium equations combined with thermal transport and demonstrate its use in a \Revision{new, update-Lagrangian, distributed-memory} QC solver for coarse-grained atomistic simulations. Finally, Section~\ref{sec: Conclusions} concludes our analysis and discusses limitations and future prospects.

\section{Nonequilibrium thermodynamics of Gaussian Phase Packets}
\label{sec: Modeling}

\begin{figure}[!t]
  \centering
 {\includegraphics[width = 0.55\textwidth]{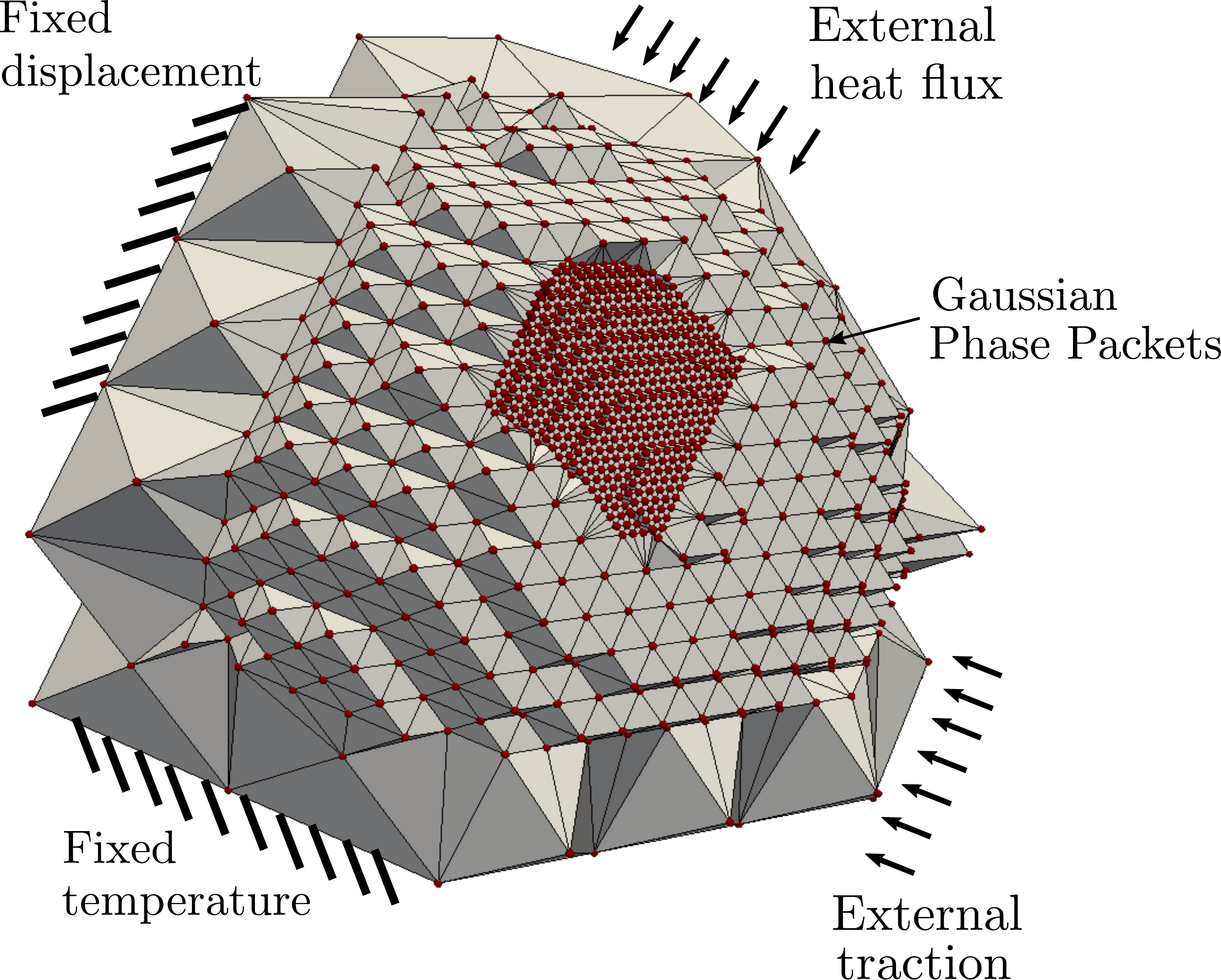}}
 \caption{Schematic illustrating a typical nonequilibrium QC study of a domain with atomistic and coarsened subdomains. All atomic sites are modeled using the Gaussian Phase Packet (GPP) approximation. The transport of energy among the atomic sites is modeled using the linear Onsager kinetics of~\citet{venturini2014atomistic}.}
 \label{fig: SchematicFiniteTemperatureQC}
\end{figure}

\Revision{We begin by} discussing the nonequilibrium modeling of deformation mechanics of crystalline solids, using the GPP approximation in which atoms are treated as Gaussian clouds, centered at the mean phase space positions of the atoms. We briefly review the application of the Liouville equation for analyzing the statistical evolution of a large ensemble of atoms subject to high-frequency vibrations. Such random vibrations are modeled by assuming local phase space coordinates (positions and momenta) drawn from local Gaussian distributions. We discuss the impact of assuming independent Gaussian distributions for each atom (termed \emph{interatomic independence} hereafter) and the corresponding inability of the model to capture nonequilibrium thermal transport. Moreover, the independent Gaussian assumption allows us to formulate the dynamical system of equations governing the atoms and the corresponding mean-field parameters. In subsequent sections, we use the insights gained here to formulate isentropic and non-isentropic quasistatic problems of finite-temperature crystal deformation (see Figure~\ref{fig: SchematicFiniteTemperatureQC} for a schematic description).

\subsection{Hamiltonian dynamics}

The time evolution of an atomic crystal, modeled as an ensemble of classical particles, is fully characterized by the generalized positions $\bfq = \{\bfq_i(t),i=1,\ldots,N\}$ and momenta $\bfp = \{\bfp_i(t),i=1,\ldots,N\}$ of all $N$ particles, which evolve in time according to \Revision{Hamilton's} equations,
\begin{subequations}
 \begin{align}
&\frac{\dd\bfp_i}{\dd t} = -\frac{\partial H}{\partial \bfq_i} = -\nabla_{\bfq_i}V(\bfq) = \bfF_i(\bfq), \\
&\frac{\dd\bfq_i}{\dd t} = \frac{\partial H}{\partial \bfp_i} = \frac{\bfp_i}{m_i},
 \end{align}
 \label{eq: EqOfMotion}
\end{subequations}
where
\begin{equation} 
 H\left(\bfp, \bfq\right) = \sum_{i=1}^N\frac{\bfp_i\cdot\bfp_i}{2m_i} + V(\bfq)
 \label{eq: Hamiltonian}
\end{equation}
is the Hamiltonian of the system, $V({\bfq})$ represents the potential field acting on all atoms in the system, \Revision{$\bfF_i(\bfq)=-\nabla_{\bfq_i}V(\bfq)$ is the net force on the $i^{\mathrm{th}}$ atom}, and $m_i$ is the mass of the $i^{\mathrm{th}}$ atom. \Revision{In classical MD, Equations~\eqref{eq: EqOfMotion} are solved for the trajectories $\left(\bfp_i(t), \bfq_i(t)\right)$ of all atoms over time}. As a consequence, such simulations require femtosecond-level time step \Revision{resolution} \citep{tuckerman2010statistical,tadmor1996quasicontinuum, tadmor1996mixed} and are unable to capture long-time-scale phenomena within computationally feasible times. 

To this end, we consider the statistical treatment of the Hamiltonian dynamics governed by \eqref{eq: EqOfMotion} and \eqref{eq: Hamiltonian}, in which the local vibrations of all atoms about their mean positions are modeled as random fluctuations in the phase space coordinate $\bfz = (\bfp(t), \bfq(t))$. Here and in the following, we use $\bfz\in\Rset^{6N}$ for brevity to represent the momenta and positions of the $N$ particles in three dimensions (3D). It is convenient to introduce the distribution $f(\bfz, t)$ such that the quantity
\begin{equation}
 \dd P = f(\bfp, \bfq, t) \prod_{i=1}^N\dd\bfp_i\prod_{i=1}^N\dd\bfq_i
\end{equation}
is the probability of finding the system of atoms such that the position and momentum of the $i^{\mathrm{th}}$ atom lie within $(\bfq_i, \bfq_i + \dd\bfq_i)$ and $(\bfp_i, \bfp_i + \dd\bfp_i)$, respectively. The \Revision{time evolution of the} probability distribution $f(\bfz,t)$ is governed by the Liouville equation~\citep{j2007statisticalEvans, tadmor2011modeling}:
\begin{equation}
 \frac{\partial f\left(\bfz, t\right)}{\partial t} + i\mathcal{L}f = 0
 \quad\text{with}\quad
  f(\bfz, 0) = f_0(\bfz),\quad \lim_{\bfz\to\infty}f(\bfz, t) = 0,
 \label{eq: Liouville}
\end{equation}
with the Liouville operator $\mathcal{L}$ defined by
\begin{equation}
 i\mathcal{L} = \frac{\partial }{\partial \bfz}\cdot\dot{\bfz} + \dot{\bfz}\cdot\frac{\partial }{\partial \bfz}
 \label{eq: LiouvilleOperator}
\end{equation}
and $\dot{\bfz} = \left(\dot{\bfp},\dot{\bfq}\right)$. Here and in the following, dots denote time derivatives. Given the equations of motion in~\eqref{eq: EqOfMotion} and the Hamiltonian of the system in~\eqref{eq: Hamiltonian}, the Liouville operator in \eqref{eq: LiouvilleOperator} becomes
\Revision{
\begin{equation}
 i\mathcal{L} = \dot{\bfz}\cdot\frac{\partial }{\partial \bfz} + \frac{\partial }{\partial \bfz}\cdot\dot{\bfz} = \dot{\bfp}\cdot\frac{\partial }{\partial \bfp} + \dot{\bfq}\cdot\frac{\partial }{\partial \bfq} + \frac{\partial }{\partial \bfp}\cdot\dot{\bfp} +  \frac{\partial }{\partial \bfq}\cdot\dot{\bfq}= -\frac{\partial H}{\partial \bfq}\cdot\frac{\partial }{\partial \bfp} + \frac{\partial H}{\partial \bfp}\cdot\frac{\partial }{\partial \bfq} = \{\cdot,H\},
 \label{eq: HamiltonianLiouville}
\end{equation}
where $\{\cdot,\cdot\}$ defines the Poisson bracket~\citep{landau2013course}.} We note that \Revision{solving} \eqref{eq: Liouville} combined with \eqref{eq: HamiltonianLiouville} at all times is equivalent to solving the equations of motion~\eqref{eq: EqOfMotion} with \eqref{eq: Hamiltonian}~\citep{j2007statisticalEvans}. In general, \Revision{solving \eqref{eq: Liouville}}  requires a discretization of the $6N$-dimensional phase space $\{\Gamma \subseteq\mathbb{R}^{6N} : \bfz \in \Gamma\}$ and of time for a system of $N$ atoms in 3D. \Revision{To avoid such a general discretization, which is computationally prohibitively expensive, one resorts to physically sensible approximations of the probability distribution $f\left(\bfz, t\right)$. One such approximation is the GPP ansatz discussed in Section~\ref{sec: GPP_generic},} which yields the equations of motion of the mean phase space coordinates and the respective fluctuation auto- and cross-correlations, categorized as phase space or mean-field parameters. \Revision{The derivation of such equations of motion is straightforward if the mean-field parameters can be identified as functions of moments of the random variable $\bfz$. However, a general framework for obtaining such equations of motion can be derived using the variational structure of the Liouville equation, as outlined below.

\subsection{Variational structure and weak formulation of the Liouville equation}
\label{sec: VariationalStructure}

The Liouville equation in \eqref{eq: Liouville} may be identified as the strong form of the equation of motion of $f(\bfz,t)$ in phase space. The weak form of the Liouville equation can be written as 
\begin{equation}
    \int^{T}_{0}\int_{\Gamma}\left(\frac{\partial f}{\partial t} + \{f, H\}\right)\delta \pi\, \dd\bfz \dd t = 0,
\label{eq: weak_form_Liouville}
\end{equation}
where $\delta\pi\in \mathbb{R}$ is a test function in real space (the $\delta$ denoting a variation in the variational-calculus sense), $T$ is some instance of time, $\{\cdot,\cdot\}$ denotes the Poisson brackets, and $\Gamma\subseteq\mathbb{R}^{6N}$ denotes the phase space. Analogous to the Hamiltonian setting of classical mechanics~\citep{lanczos2020variational}, \eqref{eq: weak_form_Liouville} can be identified as the first variation of the functional
\begin{equation}
    \mathcal{A}[\mathcal{Z}]
    =
    \int_0^T\int_{\Gamma}
            \left(\frac{\partial f}{\partial t} + \{f, H\}\right) \pi \,\dd\bfz \dd t
\label{eq: LiouvilleAction}
\end{equation}
with respect to $\pi$, where we abbreviated $\mathcal{Z} = (f,\pi)$. We refer to $\mathcal{A}[\mathcal{Z}]$ as the \emph{Liouville action}. Integrating \eqref{eq: LiouvilleAction} by parts with respect to $\bfz$, we obtain 
\begin{equation}
 \mathcal{A}[\mathcal{Z}]
    =
    \int_0^T\int_{\Gamma}
            \left(\pi\frac{\partial f}{\partial t} - \{f, \pi\} H\right) \dd\bfz \dd t.
\label{eq: LiouvilleHamiltonian}
\end{equation}
The above equation resembles the general definition of the action integral in classical mechanics (cf. Chapter VI in \citet{lanczos2020variational}) and analogously defines the \emph{Liouville Hamiltonian} $\mathcal{H}(\mathcal{Z},t)$ as,
\begin{equation}
\mathcal{H}(\mathcal{Z},t)
    =
   \int_{\Gamma}
        \{f,\pi\}
         H(\bfz)
     \dd\bfz,
    \label{eq: LiouvilleHamiltonianDefinition}
\end{equation}
which identifies $\pi$ as the \emph{Liouville momentum}. Approximation schemes for the Liouville equation can be systematically formulated by restricting the Liouville action in~\eqref{eq: LiouvilleAction} to some parametrized trajectories $\mathcal{Z}(t)$ or \emph{ans\"atze}. Considering the parametrized trajectories as
\begin{subequations}\label{eq:parametrized}
\begin{align}
    & \label{eq: parametrizedF}
    f(\bfz,t) = f\left(\bfz,\xi(t)\right) ,
    \\ & \label{eq: parametrizedPi}
    \pi(\bfz,t)  = v_j\left(\bfz,\xi(t)\right) \eta_j(t),
\end{align}
\end{subequations}
where $\xi = \{\xi_i,i=1,\ldots,P\}$ and $\eta = \{\eta_i,i=1,\ldots,P\}$ denote the arrays of the total of $P$ parameters, while $v_j(\bfz, \xi(t))$ are the test functions, and Einstein's summation convention is used. Differentiating~\eqref{eq: parametrizedF} in time yields
\begin{equation}
 \frac{\partial f}{\partial t} = \frac{\partial f}{\partial \xi_i}\dot{\xi}_i(t) = f u_i\dot{\xi}_i,
 \qquad \text{where}\quad
 u_i = \frac{1}{f}\frac{\partial f}{\partial \xi_i}.
 \label{eq: timeDerivParams}
\end{equation}
Substituting \eqref{eq:parametrized} in \eqref{eq: LiouvilleAction}, considering $v_i = u_i$, and enforcing stationarity with respect to $\eta_i$, we obtain the following set of $P$ equations of motion for parameters $\xi_i(t)$ ($i=1,\ldots,P$):
\begin{equation}
\left(\int_{\Gamma} u_ju_if \dd\bfz\right)\dot{\xi}_i(t) - \int_{\Gamma}\{u_j, H\}f\dd\bfz = 0.
\label{eq: EOM_Parameters_1}
\end{equation}
The phase average of any function $A(\bfz)$ is denoted by $\avg{A(\bfz)}$ and defined as
\begin{equation}
 \avg{A(\bfz)} = \frac{1}{N!\,h^{3N}}\int_{\Gamma}A(\bfz) f(\bfz, t) \dd\bfz,
 \label{eq: phase_avg_def}
\end{equation} 
where $h$ is Planck's constant and the factor $N!\,h^{3N}$ is introduced to normalize by the phase space volume~\citep{landau1980statistical}. Using \eqref{eq: phase_avg_def}, \eqref{eq: EOM_Parameters_1} can be rewritten as
\begin{equation}
\avg{u_ju_i}\dot{\xi}_i(t) - \avg{\{u_j, H\}} = 0.
\label{eq: EOM_Parameters}
\end{equation}
This system of equations yields the equations of motion of parameters $\xi_i(t)$, which parametrize the trajectories of the system governed by \eqref{eq: Liouville} and can be considered as the classical analogue of the Frenkel-Dirac-McLachlan variational principle used in quantum mechanics~\citep{mclachlan1964variational}. For parameters $\xi_i(t)$, which can be directly written as phase-averaged functions of the random variable $\bfz$ (of mathematical form $\avg{A(\bfz)}$ where $A(\cdot)$ is some smooth function), the equations of motion can be derived using the identity \citep{j2007statisticalEvans, zubarev1974nonequilibrium}
\begin{equation}
 \frac{\dd\avg{A}}{\dd t} = \frac{1}{N!\,h^{3N}}\int_{\Gamma} f(\bfz, t) \frac{\dd A}{\dd t} \dd\bfz = \avg{\frac{\dd A}{\dd t}}
 \label{eq: PhaseSpaceDiff}
\end{equation}
(see~\ref{sec: timeEvoPhaseAVG} for a brief discussion). Although not exploited in the following, the variational setting of the Liouville-based approach pursued in this study bears potential with regards to numerical solution schemes, as it allows for a systematic discretization and convergence study of solutions, among others.
}

\subsection{Crystal lattice of Gaussian phase packets}
\label{sec: GPP_generic}

\Revision{The above framework requires an ansatz for the probability distribution. As discussed in Section~\ref{sec: intro}, the mean-field approximation-based temporal-coarsening approaches \citep{kulkarni2008variational, li2011diffusive, venturini2014atomistic} start by constructing an ansatz for the distribution $f(\bfz)$ \textit{directly at steady state}. If the Hamiltonian $H$ is constrained as in~\citet{venturini2014atomistic}, then the distribution function $f(\bfz)$ is only a function of $H$, satisfying $i\mathcal{L}f = 0$. While these restrictions may admit quasistatic solutions, they do not provide insight into the dynamic evolution nor do they consider the nature of the process which led to a quasistatic solution. 
In addition, considering only variance constraints on the distribution function~\citep{kulkarni2008variational, li2011diffusive} enforces a steady state of the mean-field parameters, as shown in the section below. 

As a new approach, we here start with a multivariate Gaussian phase packet (GPP) ansatz to the distribution function with no assumptions about vanishing correlations, and we systematically discuss the consequences of eliminating interatomic and cross-correlations of the degrees of freedom.}
The GPP approximation was first introduced in the context of quantum mechanics by \citet{heller1975time} and used for classical systems by \citet{ma1993approximate}. Both \citet{heller1975time} and \citet{ma1993approximate} substituted Gaussian distributions into the Liouville equation~\eqref{eq: Liouville} (Schr\"odinger's equation for quantum systems) to obtain the dynamical evolution of the phase space parameters. 
Following their idea, we approximate the system-wide probability distribution $f(\bfz, t)$ as a multivariate Gaussian distribution, i.e.,
\begin{equation}
 f(\bfz, t) = \frac{1}{Z(t)}e^{-\frac{1}{2}\left(\bfz - \overline{\bfz}(t)\right)\T\bfSigma^{-1}(t)\left(\bfz - \overline{\bfz}(t)\right)},
 \label{eq: GPP_FULL}
\end{equation}
where $\bfSigma$ is the $6N\times 6N$ covariance matrix composed of the interatomic and momentum-displacement correlations, $\overline\bfz$ represents the vector of all atoms' mean positions and momenta, and the partition function $Z(t)$ is defined by
\begin{equation} 
 Z(t) = \frac{1}{N!h^{3N}}\int_{\mathbb{R}^{6N}} e^{-\frac{1}{2}\left(\bfz - \overline{\bfz}(t)\right)\T\bfSigma^{-1}(t)\left(\bfz - \overline{\bfz}(t)\right)}d\bfz = \frac{(2\pi)^{3N}}{N!\,h^{3N}}\sqrt{\det\bfSigma}.
 \label{eq: PartitionFunctionDef}
\end{equation}
\Revision{Phase averaging of $\bfz$ using \eqref{eq: phase_avg_def} with \eqref{eq: GPP_FULL}} confirms that $\overline\bfz(t) = \langle\bfz(t)\rangle$. We further conclude that $\bfSigma$ can be written as a block-matrix with components
\begin{equation}
 \bfSigma_{ij}=\frac{1}{N!\,h^{3N}}\int_{\mathbb{R}^{6N}}\left(\bfz_i - \overline{\bfz}_i\right)\otimes\left(\bfz_j - \overline{\bfz}_j\right)f(\bfz, t)\dd\bfz = \avg{\left(\bfz_i - \overline{\bfz}_i\right)\otimes\left(\bfz_j - \overline{\bfz}_j\right)},
 \label{eq: covarianceAndPhaseAvg}
\end{equation}
such that each block represents the correlation among displacements and momenta of atoms $i$ and $j$. 

\Revision{Consequently, instead of time-resolving the positions and momenta of all atoms (as in MD), we may equivalently track the time evolution of the atomic ensemble through the phase space parameters $\left(\overline{\bfz}(t), \bfSigma(t)\right)$, which includes the mean momenta and positions of all atoms as well as the covariance matrix.}  \Revision{Since parameters $\overline{\bfz}(t)$ and $\bfSigma(t)$ can be identified as phase functions, application of \eqref{eq: PhaseSpaceDiff} to $\overline{\bfz}(t)$ and $\bfSigma(t)$ yields the dynamical equations that govern their time evolution:}
\begin{align}
    \frac{\dd \overline{\bfz}}{\dd t} = \avg{\dot{\bfz}},
\qquad 
    \frac{\dd \bfSigma_{ij}}{\dd t} = \avg{\left(\dot{\bfz}_i - \dot{\lavg{\bfz}}_i\right) \otimes \left({\bfz}_j - {\lavg{\bfz}_j}\right)} + \avg{\left({\bfz}_i - {\lavg{\bfz}_i}\right) \otimes \left(\dot{\bfz}_j - \dot{\lavg{\bfz}}_j\right)}.
\label{eq: phaseSpaceEq}
\end{align}
\Revision{Application of \eqref{eq: EOM_Parameters} to $\overline{\bfz}(t)$ and $\bfSigma(t)$ yields identical equations, as derived in \ref{sec: variationalFormulationGaussian}}.
Let us further specify the second equation in \eqref{eq: phaseSpaceEq}. Writing the components of covariance matrix $\bfSigma_{ij}$ as
\begin{equation}
 \bfSigma_{ij} = \left(\begin{matrix}
                    \bfSigma^{(\bfp,\bfp)}_{ij} &  \bfSigma^{(\bfp,\bfq)}_{ij} \\
                     \bfSigma^{(\bfq,\bfp)}_{ij} &  \bfSigma^{(\bfq,\bfq)}_{ij}
                       \end{matrix}
\right),
\label{eq: lumped_covariance}
\end{equation}
and assuming that all atoms have the same mass $m$ \Revision{without loss of generality},
identity~\eqref{eq: PhaseSpaceDiff} yields the following time evolution equations for the submatrices identified above:
\begin{subequations}
\begin{equation}
\frac{\dd \bfSigma^{(\bfp,\bfp)}_{ij}}{\dd t} = \avg{\bfF_i(\bfq)\otimes\left(\bfp_j - \overline{\bfp}_j\right)} + \avg{\left(\bfp_i - \overline{\bfp}_i\right)\otimes\bfF_j(\bfq)},
\label{eq: heat_flux_atomistic}
\end{equation}
\begin{equation}
\frac{\dd \bfSigma^{(\bfp,\bfq)}_{ij}}{\dd t}= \avg{\bfF_i(\bfq)\otimes\left(\bfq_j - \overline{\bfq}_j\right)} + \frac{\bfSigma^{(\bfp, \bfp)}_{ij}}{m}, 
\label{eq: thermo_mechanical_2}
\end{equation}
\begin{equation}
\frac{\dd \bfSigma^{(\bfq,\bfp)}_{ij}}{\dd t}= \avg{\left(\bfq_i - \overline{\bfq}_i\right)\otimes{\bfF}_j} + \frac{\bfSigma^{(\bfp, \bfp)}_{ij}}{m},
\label{eq: thermo_mechanical_3} 
\end{equation}
\begin{equation}
\frac{\dd \bfSigma^{(\bfq,\bfq)}_{ij}}{\dd t} = \frac{\bfSigma^{(\bfp, \bfq)}_i + \bfSigma^{(\bfq, \bfp)}_i}{m}.
\label{eq: thermo_mechanical_1}
\end{equation}
\label{eq: lumpedMesodynamics}
\end{subequations}

Equations~\eqref{eq: lumpedMesodynamics} govern the evolution of the pairwise momentum and displacement correlations of atoms $i$ and $j$, and they must be solved to obtain the interatomic correlations at all times. Equations~\eqref{eq: thermo_mechanical_3}-\eqref{eq: thermo_mechanical_1} govern the thermomechanical coupling of the crystal. Note that we may identify the pairwise kinetic tensor $\bfSigma^{(\bfp, \bfp)}_{ij}$ as a measure of temperature, so that \eqref{eq: thermo_mechanical_2} and \eqref{eq: thermo_mechanical_3} describe the evolution of the system-wide distribution as a result of unbalanced pairwise virial tensors and kinetic tensors (see \citet{admal2010unified} for the tensor virial theorem). The virial tensor $\avg{\bfF_i(\bfq)\otimes\left(\bfq_j - \overline{\bfq}_j\right)}$ changes with changing displacement correlations of atoms due to varying extents of atomic vibrations $\bfSigma^{(\bfq, \bfq)}_{ij}$, thus coupling \eqref{eq: thermo_mechanical_1} with \eqref{eq: thermo_mechanical_2} and \eqref{eq: thermo_mechanical_3}. The right-hand side of \eqref{eq: heat_flux_atomistic} resembles the tensor form of the interatomic heat current~\citep{saaskilahti2015frequency, lepri2003thermal} and changes with varying correlation matrices $\bfSigma^{\bfq,\bfp}_{ij}$, thus coupling~\eqref{eq: heat_flux_atomistic} with \eqref{eq: thermo_mechanical_2} and \eqref{eq: thermo_mechanical_3}. Consequently, the imbalance between virial and kinetic tensors in equations \eqref{eq: thermo_mechanical_2} and \eqref{eq: thermo_mechanical_3} drives the phase space motion of the system of particles, resulting in the time evolution of $\bfSigma^{(\bfp, \bfp)}_{ij}$ and $\bfSigma^{(\bfq, \bfq)}_{ij}$.

\Revision{In the following, it will be helpful to} identify the entropy $S$ of the atomic ensemble as
\begin{equation}
 S = -k_B\avg{\ln f} = k_B\left({3N}\left[1 + \ln(2\pi)\right] - \ln\left(N!\right) + {\ln\left(\frac{\sqrt{\det\bfSigma}}{h^{3N}}\right)} \right) = S_0 + k_B\ln\left(\frac{\sqrt{\det\bfSigma}}{h^{3N}}\right),
 \label{eq: entropyDef}
\end{equation}
where $k_B$ is Boltzmann's constant, and $S_0$ is a constant for a given system with a \Revision{fixed} number of atoms. The entropy rate of change follows as
\begin{equation}
 \frac{\dd S}{\dd t} = \frac{k_B}{2\det\bfSigma}\frac{\dd (\det\bfSigma)}{\dd t}.
 \label{eq: entropyRateFull}
\end{equation}

\Revision{Equations}~\eqref{eq: phaseSpaceEq} govern the phase space motion of a system of atoms and contain \Revision{a total of $6N + 36 N (N+1)/2$ evolution} equations, solving which is even more computationally expensive than solving the state-space governing equations of MD. \Revision{Therefore, as} a simplifying assumption \citet{ma1993approximate} and \citet{heller1975time} assumed the statistical independence of atoms (\Revision{or} of states in the quantum analogue), which implies
\begin{equation}
 \bfSigma_{ij} = \mathbf{0} \quad\mathrm{for} \ i\neq j.
 \label{eq: InteratomicIndependence}
\end{equation}
\Revision{Although this assumption severely limits the applicability of \eqref{eq: phaseSpaceEq}, it admits a beneficial analytical treatment of the time evolution problem and allows us to gain new insight into the nature of correlations and, especially, into the quasistatic limit. Therefore, we use \eqref{eq: phaseSpaceEq} in the following. As we will demonstrate below, a consequence of assumption \eqref{eq: phaseSpaceEq} is that heat transport cannot be resolved correctly.} Consequently, we may apply the phase space evolution equations
\eqref{eq: phaseSpaceEq} and \eqref{eq: lumpedMesodynamics}
only for isentropic (reversible) finite temperature simulations of quasistatic and dynamic processes.

\subsection{Independent Gaussian phase packets}
\label{sec: IndependentGPP}

Since solving the full system of evolution equations is prohibitively expensive, as discussed above, \Revision{we apply} \eqref{eq: InteratomicIndependence} and assume non-zero correlations between the position and momentum of each individual atom, but no cross-correlations between different atoms. To this end, we apply the GPP approximation to a single atom $i$, which yields the multivariate Gaussian distribution of the phase space coordinate $\bfz_i$ as
\begin{equation}
 f_i(\bfz_i, t) = \frac{1}{Z_i}e^{-\frac{1}{2}\left(\bfz_i-\overline{\bfz}_i\right)\T\bfSigma^{-1}_i\left(\bfz_i-\overline{\bfz}_i\right)}
 \qquad\text{so that}\qquad
 f(\bfz,t) = \prod_{i=1}^N f_i(\bfz_i, t),
 \label{eq: singleGPP}
\end{equation}
where the phase space parameters $(\overline{\bfz}_i, \bfSigma_i)$ denote the mean phase space coordinate and variance of the $i^\text{th}$ atom, respectively, defined as
\begin{equation}
 \overline{\bfz}_i(t) = \frac{1}{h^3}\int_{\Rset^6}f_i(\bfz_i, t)\bfz_i \dd\bfz_i = \avg{\bfz_i},
 \qquad
 \bfSigma_i = \avg{\bfz_i\otimes\bfz_i}.
\end{equation}
The normalization quantity $Z_i(t)$ may be identified as the single-particle partition function
\begin{equation}
 Z_i(t) =\frac{1}{h^3} \int_{\Rset^6} e^{-\frac{1}{2}\left(\bfz_i-\overline{\bfz}_i\right)\T\bfSigma^{-1}_i\left(\bfz_i-\overline{\bfz}_i\right)} \dd\bfz_i = \left(\frac{2\pi}{h}\right)^3\sqrt{\det\bfSigma_i}.
\end{equation}
$\bfSigma_i$ is the $6\times 6$ covariance matrix of the multivariate Gaussian and accounts for the variance or uncertainty in the momentum $\bfp_i$ and displacement $\bfq_i$ of the $i^\text{th}$ atom.  

\begin{figure}
  \centering
 {\includegraphics[width = 0.75\textwidth]{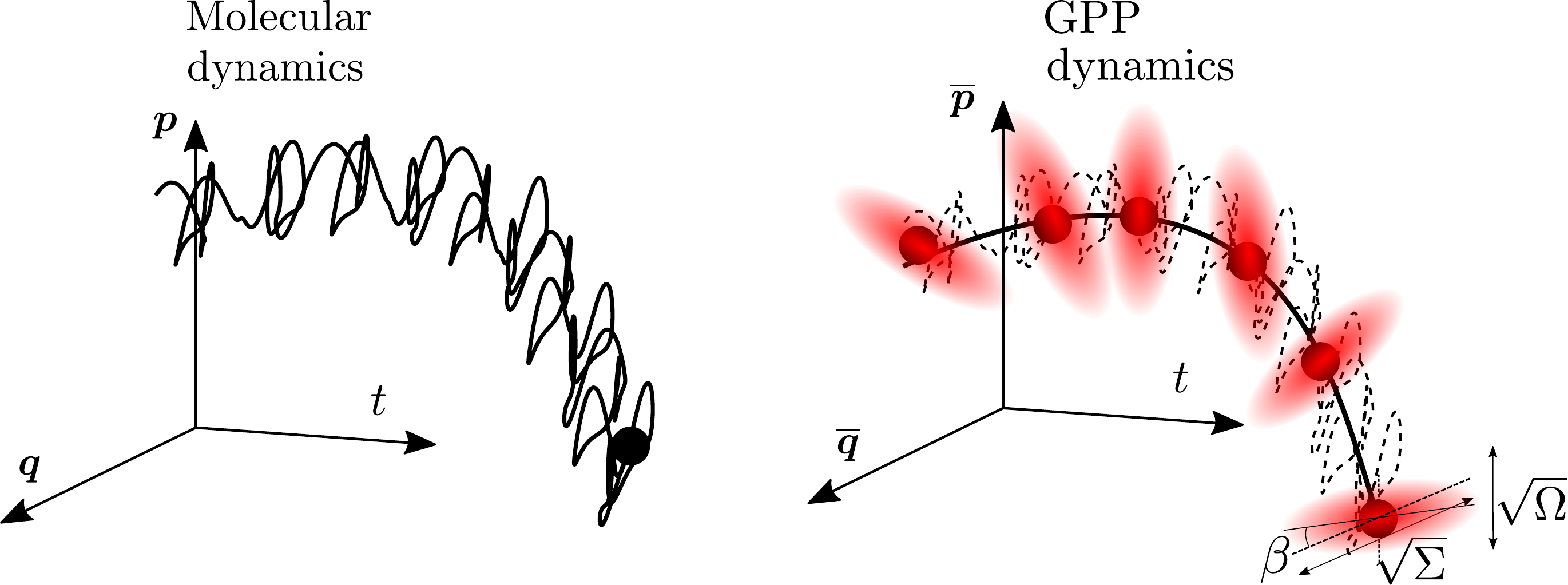}}
 \caption{\Revision{Schematic i}llustration of an atomic trajectory by GPP dynamics (on the right) compared to an MD trajectory (on the left) for a single particle. Small-scale motions of the particle are approximated by parameters $\Omega$, $\Sigma$ and $\beta$ upon averaging over suitable time intervals. As discussed in Section~\ref{sec: QuasiStatics}, in the quasistatic limit, $\Omega$ and $\Sigma$ are proportional to the local temperature \Revision{and $\beta\to 0$}.}
 \label{fig: MD_vs_GPP}
\end{figure}

The assumed interatomic independence eliminates the interatomic correlations as independent variables, thus reducing the total number of equations to $27N + 6N$ for a system of $N$ atoms, which govern the time evolution of the kinetic tensor $\bfSigma^{(\bfp, \bfp)}_i$, displacement-correlation tensor $\bfSigma^{(\bfq, \bfq)}$, and the momentum-displacement-correlation tensor $\bfSigma^{(\bfp, \bfq)}_i = \left(\bfSigma^{(\bfq, \bfp)}_i\right)\T$ via
\begin{subequations}
\begin{align}
&\frac{\dd \bfSigma^{(\bfp, \bfp)}_i}{\dd t} = \avg{\bfF_i(\bfq)\otimes\left(\bfp_i - \overline{\bfp}_i\right)} + \avg{\left(\bfp_i - \overline{\bfp}_i\right)\otimes\bfF_i(\bfq)},\\
&\frac{\dd \bfSigma^{(\bfp, \bfq)}_i}{\dd t} = \avg{\bfF_i(\bfq)\otimes\left(\bfq_i - \overline{\bfq}_i\right)} + \frac{\bfSigma^{(\bfp, \bfp)}_{i}}{m}, \\
&\frac{\dd \bfSigma^{(\bfq, \bfq)}_i}{\dd t} = \frac{\bfSigma^{(\bfp, \bfq)}_i + \bfSigma^{(\bfq, \bfp)}_i}{m},
\end{align} 
\label{eq: GPP_EVO_TENSOR}
\end{subequations}
combined with the phase-averaged equations of motion, 
\begin{equation}
 \frac{\dd \avg{\bfp}_i}{\dd t} = \avg{\bfF}_i,
 \qquad
 \frac{\dd \avg{\bfq}_i}{\dd t} = \frac{\avg{\bfp}_i}{m}.
 \label{eq: avgDynamics}
\end{equation}
For simplicity, we further make the spherical distribution assumption that all cross-correlations between different directions of momenta and displacements vanish, hence approximating the \emph{atomic clouds} in phase space as spherical (thus eliminating any directional preference of the atomic vibrations). While being a strong assumption, this allows us to reduce the above tensorial evolution equations to scalar ones. Specifically, taking the trace $\mathrm{tr}\left(\cdot\right)$ of \eqref{eq: GPP_EVO_TENSOR}, we obtain,
\begin{subequations}
 \begin{align}
&\frac{\dd \Omega_i}{\dd t} = \frac{2\avg{\bfF_i(\bfq)\cdot\left(\bfp_i - \overline{\bfp}_i\right)}}{3},\\
&\frac{\dd \Sigma_i}{\dd t} = \frac{2\beta_i}{m},\\
&\frac{\dd \beta_i}{\dd t} = \frac{\avg{\bfF_i(\bfq)\cdot\left(\bfq_i - \overline{\bfq}_i\right)}}{3} + \frac{\Omega_i}{m},
\end{align}
\label{eq: GPP_EVO_TRACE}
\end{subequations}
where we introduced the three scalar parameters
\begin{equation}
 \mathrm{tr}\left(\bfSigma^{(\bfp, \bfp)}_i\right) = 3\Omega_i,
 \qquad
 \mathrm{tr}\left(\bfSigma^{(\bfq, \bfq)}_i\right) = 3\Sigma_i
 \quad
 \mathrm{and}
 \quad
 \mathrm{tr}\left(\bfSigma^{(\bfp, \bfq)}_i\right) = 3\beta_i.
\end{equation}
Equations~\eqref{eq: GPP_EVO_TRACE} are $3N$ coupled scalar ODEs, which determine the changes in the vibrational widths of atoms in phase space ($\Omega_i$ and $\Sigma_i$) and the correlation $\beta_i$ between the displacement and momentum vibrations of the $i^\text{th}$ atom (see \figurename~\ref{fig: MD_vs_GPP}). We note that~\eqref{eq: GPP_EVO_TRACE} are identical to the equations used by \citet{ma1993approximate}, who used the formulation as an optimization procedure for a system of atoms at a uniform constant temperature. 

The physical role of momentum-displacement correlation $\beta_i$ becomes evident upon applying a time-reversal transformation $t\mapsto -t$ to \eqref{eq: avgDynamics} and \eqref{eq: GPP_EVO_TRACE}, which results in the transformations $\left(\overline{\bfq}_i, \Omega_i, \Sigma_i\right) \mapsto \left(\overline{\bfq}_i, \Omega_i, \Sigma_i\right)$ and $\left(\overline{\bfp}_i, \beta_i\right)\mapsto\left(-\overline{\bfp}_i, -\beta_i\right)$. \Revision{Since the correlation $\beta_i$ signifies the momentum of the $i^\text{th}$ atom in phase space, governing the time evolution of the thermal coordinate $\Sigma_i$, $\beta_i$} will be referred to as the \emph{thermal momentum} hereafter. The dynamics and thermodynamics of crystals modeled via the independent GPP approximation can be summarized via eliminating the thermal and dynamical momenta, yielding for every atom $i=1,\ldots,N$
\begin{subequations}
 \begin{align}
&\frac{\dd ^2\avg{\bfq}_i}{\dd t^2} = \frac{\avg{\bfF}_i}{m},\\
&\frac{\dd ^2\Sigma_i}{\dd t^2} = \frac{2\Omega_i}{m^2} + \frac{2\avg{\bfF_i(\bfq)\cdot\left(\bfq_i - \overline{\bfq}_i\right)}}{3m},\\
&\frac{\dd \Omega_i}{\dd t} = \frac{2\avg{\bfF_i(\bfq)\cdot\left(\bfp_i - \overline{\bfp}_i\right)}}{3}.
\end{align}
\label{eq: GPP_EVO_SUMM}%
\end{subequations}
Time reversibility of \eqref{eq: GPP_EVO_SUMM} highlights that \textit{these evolution equations do not capture nonequilibrium irreversible thermal transport at all scales} due to the simplification of \eqref{eq: lumpedMesodynamics} to \eqref{eq: GPP_EVO_TENSOR} under the interatomic independence assumption~\eqref{eq: InteratomicIndependence}. \Revision{As discussed above, this assumption is necessary to keep the number of unknowns tractable  for large ensembles.} 

The reversible entropy fluctuations can be obtained by substituting the independent GPP assumption into \eqref{eq: entropyRateFull}, which gives
\begin{equation}
 \frac{\dd S}{\dd t} = \frac{k_B}{2}\frac{1}{\det\bfSigma}\frac{\dd \det\bfSigma}{\dd t} = \sum_{i=1}^N\left(\frac{k_B}{2}\frac{1}{\det\bfSigma_i}\frac{\dd \det\bfSigma_i}{\dd t}\right) = \sum_{i=1}^N\frac{\dd S_i}{\dd t},
 \label{eq: entropyRateEnsemble}
\end{equation}
where the local entropy fluctuations of the $i^\text{th}$ atom are given by
\begin{align}
 \frac{\dd S_i}{\dd t} = & \frac{k_B}{2}\frac{1}{\det\bfSigma_i}\frac{\dd \det\bfSigma_i}{\dd t} = \frac{k_B}{2\left(\Omega^3_i\Sigma^3_i - \beta^6_i\right)}\frac{\dd }{\dd t}\left(\Omega^3_i\Sigma^3_i - \beta^6_i\right) \nonumber \\ 
 = & \frac{k_B}{2\left(\Omega^3_i\Sigma^3_i - \beta^6_i\right)}\left(\frac{6\Omega_i\beta_i}{m}\left[(\Omega_i\Sigma_i)^2 - \beta^4_i\right] + 2\left[\Omega^2_i\Sigma^3_i \bfF_i\cdot(\bfp_i-\lavg{\bfp}_i) - \beta^5_i\bfF_i\cdot(\bfq_i - \lavg{\bfq}_i)\right]\right).
 \label{eq: entropyRateGPP}
\end{align}
The above equation shows that the local entropy fluctuation $\dot S_i$ is proportional to the thermal momentum $\beta_i$. We will return to relation \eqref{eq: entropyRateGPP} when discussing specific types of interatomic potentials in subsequent sections.

Since the interatomic independence assumption results in an incorrect calculation of the interatomic heat flux in \eqref{eq: heat_flux_atomistic}, one needs to model irreversible thermal transport, e.g., using the linear kinetic potential framework used by \citet{venturini2014atomistic}, as discussed in the following.

\section{Dynamics and Quasistatics of independent GPPs}
\label{sec: dynamicsAndQuasistatics}

We proceed to analyze the dynamic behavior of the GPP equations~\eqref{eq: GPP_EVO_SUMM} (under the interatomic independence assumption) and subsequently deduce the quasistatic behavior as a limit case. For instructive purposes (and because the (quasi-)harmonic assumption plays a frequent role in atomistic analysis), we apply the equations to both harmonic and anharmonic potentials \Revision{for the} atomic interactions within the crystal. \Revision{Although the GPP-based ansatz for the probability density consists of only quadratic functions in $\bfq$ (resembling harmonic interactions), the force $\bfF_i(\bfq)$ is derived from the potential $V(\bfq)$ and can be anharmonic in general. Such an approximation is classically known as the \emph{quasi-harmonic approximation} because, at equilibrium, the distribution resembles that of a canonical ensemble of harmonic oscillators, even though the interatomic forces are anharmonic. In the following,} we show that, for a quasistatic change in the thermodynamic state 
(i.e., driving the mean dynamical and thermal momenta $\lavg{\bfp}$ and $\beta$, respectively, to zero), the information \Revision{about} the evolution of $\Omega$ is lost (cf.~\eqref{eq: GPP_EVO_SUMM}). Correspondingly, we may assume a specific nature of the thermodynamic process of interest (e.g., isothermal, isentropic, \Revision{or isobaric}) to determine the change in $\Omega$ of the GPP atoms during the quasistatic change in the thermodynamic state. Alternatively, the decay of correlation $\beta(t)$ can be modeled empirically to obtain the change in $\Omega$, if the nature of the thermodynamic process is unknown.

\subsection{Dynamics}

\subsubsection{Harmonic approximation}
\label{sec: dynamics_harmonic}

As a simplified case that admits analytical treatment, we \Revision{first} consider a harmonic approximation of the interaction potential $V(\bfq)$, writing
\begin{equation}
 V(\bfq) = V_0 + \frac{1}{2}\sum_{i=1}^N\sum_{j\in\mathcal{N}(i)}(\bfq_i - \bfq_j)\T\bfK(\bfq_i - \bfq_j),
 \label{eq: HarmonicPotential}
\end{equation}
where $\bfK\in\Rset^{3\times 3}$  is the local harmonic dynamical matrix, $V_0$ is the equilibrium potential between the atoms approximated as independent GPPs, and $\mathcal{N}(i)$ represents the neighborhood of the $i^{\mathrm{th}}$ atom. \Revision{Inserting potential~\eqref{eq: HarmonicPotential} into the evolution equations~\eqref{eq: GPP_EVO_SUMM} leads to}
\begin{equation}
\frac{\dd ^2\avg{\bfq}_{i}}{\dd t^2} = -2\sum_{j\in\mathcal{N}(i)}\bfK\avg{\bfq_i - \bfq_j}
\label{eq: GPP_DYNAMIC_HARM}
\end{equation}
and
\begin{subequations}
 \begin{align}
&\frac{\dd ^2\Sigma_i}{\dd t^2} = \frac{2\Omega_i}{m^2} - \frac{4n_i\tr(\bfK)}{m}\Sigma_i,\\
&\frac{\dd \Omega_i}{\dd t} = -4n_i\tr(\bfK)\beta_i,
\end{align}
\label{eq: GPP_EVO_HARM}%
\end{subequations}
where $n_i\tr(\bfK)$ \Revision{serves} an effective force constant, which depends on the \Revision{number $n_i$} of immediate neighbors of the $i^{\mathrm{th}}$ atom.
Equations~\eqref{eq: GPP_EVO_HARM} show that the mean mechanical displacement $\avg{\bfq}_i$ of \Revision{the $i^{\mathrm{th}}$ atom} is decoupled from its thermodynamic displacements $\Omega_i$ and $\Sigma_i$ for a harmonic potential field between the atoms. The resulting decoupled thermodynamic equations
\begin{equation}
 \frac{\dd \Omega_i}{\dd t} = -4n_i\tr(\bfK)\beta_i,
 \qquad
 \frac{\dd \Sigma_i}{\dd t} = \frac{2\beta_i}{m},
 \quad\text{and}\quad
 \frac{\dd \beta_i}{\dd t} = \frac{\Omega_i}{m} - 2n_i\tr(\bfK)\Sigma_i,
 \label{eq: GPP_THERMO_HARMONIC}
\end{equation}
exhibit the following independent eigenvectors $(\bfphi_0, \bfphi_+, \bfphi_-)$ and corresponding eigenvalues $(\omega_0, \omega_+, \omega_-)$:
\begin{equation}
 \bfphi_0 = \left(\begin{matrix}
                 2mn_i\tr(\bfK)\\
                 1\\
                 0
                \end{matrix}
\right), \quad
\bfphi_{\pm} = \left(\begin{matrix}
                               -2mn_i\tr(\bfK)\\
                               1\\
                               -\frac{im\omega_{\pm}}{2}
                              \end{matrix}\right),
                              \qquad
                              \omega_0 = 0,
                              \quad 
                              \omega_{\pm} = \Revision{\pm2}\sqrt{\frac{2n_i\tr(\bfK)}{m}},
\label{eq: HarmonicPotentialEigs}
\end{equation}
so that \Revision{the} general homogeneous solution
\begin{equation}
 \left(\begin{matrix}
        \Omega_i(t)\\
        \Sigma_i(t)\\
        \beta_i(t)
       \end{matrix}\right) = a_0\bfphi_0 + a_+\bfphi_+e^{i\omega_+ t} + a_-\bfphi_-e^{i\omega_- t},
\end{equation}
is composed of constant and oscillatory components. Coefficients $(a_0, a_+, a_-)$ are determined by the initial thermodynamic state of each atom. The constant component $\bfphi_0$ corresponds to $\beta = 0$ and $\Omega_i/\Sigma_i = 2mn_i\tr(\bfK)$. 
To interpret the three terms within this solution, let us formulate the excess internal energy, which for the harmonic approximation \eqref{eq: HarmonicPotential} becomes
\begin{equation}
 E = \avg{H} = \avg{V(\bfq)} + \sum_{i=1}^N\frac{\avg{|\bfp_i|^2}}{2m}  
 = \sum_{i=1}^N\left(\frac{\Omega_i}{2m} + n_i\tr(\bfK)\Sigma_i\right).
 \label{eq: InternalEnergy_Harmonic}
\end{equation}
By insertion into \eqref{eq: InternalEnergy_Harmonic}, it becomes apparent that the constant component $\bfphi_0$ with $\Omega_i = 2mn_i\tr(\bfK)\Sigma_i$ has equal average kinetic and potential energies. This equipartition of energy implies that $\bfphi_0$ corresponds to the thermodynamic equilibrium. Consequently, the components $\bfphi_{\pm}$ correspond to oscillations about the equilibrium state $\bfphi_0$ with frequencies $\omega_{\pm} = 2\sqrt{2n_i\tr(\bfK)/m}$. \Revision{Therefore, numerical time integration of the GPP equations unfortunately incurs a \textit{similar computational cost as a standard MD simulation} of the same system (although an equilibrium solution can be identified, oscillations about the equilibrium occur at frequencies on the same order as atomic vibrations). Thus, even though mean motion and statistical information have been separated, the interatomic independence assumption within the GPP framework prevents significant temporal upscaling. Further note that, due} to the decoupling of the thermodynamic equations~\eqref{eq: GPP_THERMO_HARMONIC} from the dynamic equation of motion~\eqref{eq: GPP_DYNAMIC_HARM}, \textit{a harmonic GPP lattice exhibits no thermomechanical coupling} and hence displays no \Revision{heating nor cooling} under external stress (\Revision{i.e., it does not expand due to local heating and vice-versa}). 

Finally, by substituting the harmonic potential~\eqref{eq: HarmonicPotential} into~\eqref{eq: entropyRateGPP}, we obtain the reversible fluctuations in entropy of atom $i$ for a system of atoms in a harmonic field:
\begin{equation}
 \frac{\dd S_i}{\dd t} = \frac{3k_B n_i\tr(\bfK)\beta_i\dot{\beta_i}\left(\Omega^2_i\Sigma^2_i - \beta^4_i\right)}{\Omega^3_i\Sigma^3_i - \beta^6_i}.
 \label{eq: entropyFluctHarmonic}
\end{equation}

\subsubsection{Anharmonic thermomechanical effects}
\label{sec: dynamics_anharmonic}

As \Revision{the above} harmonic approximation of the interatomic potential renders the GPP crystal thermomechanically decoupled, we next study the effects of anharmonicity in the potential. As the simplest possible extension of the harmonic potential \Revision{\eqref{eq: HarmonicPotential}}, we consider
\begin{equation}
 V(\bfq) = V_0+ \frac{1}{2}\sum_{i=1}^N\sum_{j\in\mathcal{N}(i)}(\bfq_i - \bfq_j)\T\bfK(\bfq_i - \bfq_j) + \frac{1}{6}\sum_{i=1}^N\sum_{j\in\mathcal{N}(i)}(\bfq_i - \bfq_j)\T\bfzeta\left[(\bfq_i - \bfq_j)\otimes(\bfq_i - \bfq_j)\right],
 \label{eq: CubicPotential}%
\end{equation}
where $\bfK\in\Rset^{3\times 3}$ is the local dynamical matrix, and $\bfzeta$ denotes a constant anharmonic third-order tensor. With this potential, the evolution equations \eqref{eq: GPP_EVO_SUMM} become
\begin{align}
 \frac{\dd ^2\avg{\bfq}_{i}}{\dd t^2} & = -2\sum_{j\in\mathcal{N}(i)}\bfK\avg{\bfq_i - \bfq_j} - 2\sum_{j\in\mathcal{N}(i)}\bfzeta\avg{(\bfq_i - \bfq_j)\otimes\left(\bfq_i - \bfq_j\right)},\nonumber\\
 &=-2\sum_{j\in\mathcal{N}(i)}\bfK\avg{\bfq_i - \bfq_j} - 2\sum_{j\in\mathcal{N}(i)}\bfzeta\left[\left(\Sigma_i + \Sigma_j\right)\id +  \avg{\bfq_i - \bfq_j}\otimes\avg{\bfq_i - \bfq_j}\right]
 \label{eq: ANHARM_DYN}
\end{align}
and
\begin{subequations}
 \begin{align}
  \frac{\dd ^2\Sigma_i}{\dd t^2} 
  & = \frac{2\Omega_i}{m^2} -\frac{4}{m}\left(n_i\tr(\bfK)\Sigma_i + \sum_{j\in\mathcal{N}(i)}\zeta_{lmn}\avg{(\bfq_i - \bfq_j)_l(\bfq_i - \bfq_j)_m(\bfq_i - \avg{\bfq_i})_n}\right),\nonumber\\
  &
  =\frac{2\Omega_i}{m^2}-\frac{4\Sigma_i}{m}\left(n_i\tr(\bfK) - \sum_{j\in\mathcal{N}(i)}\zeta_{lmn}\left(\delta_{ml}\avg{\bfq_{j}}_n + \delta_{nl}\avg{\bfq_{j}}_m\right)\right),\\
  \frac{\dd \Omega_i}{\dd t} 
  & = -4n_i\tr(\bfK)\beta_i - 4\sum_{j\in\mathcal{N}(i)}\zeta_{lmn}\avg{(\bfq_i - \bfq_j)_l(\bfq_i - \bfq_j)_m(\bfp_i - \avg{\bfp_i})_n}\nonumber,\\
 &=-4\beta_i\left(n_i\tr(\bfK) - \sum_{j\in\mathcal{N}(i)}\zeta_{lmn}\left(\delta_{ml}\avg{\bfq_{j}}_n + \delta_{nl}\avg{\bfq_{j}}_m\right)\right),
 \end{align}
\label{eq: GPP_EVO_ANHARM}%
\end{subequations}
where $\zeta_{lmn}$ are the components of $\bfzeta$, $(\cdot)_l$ denotes the $l^\text{th}$ component of vector $(\cdot)$, and $\delta$ represents Kronecker's delta (and we use Einstein's summation convention, implying summation over $l,m,n$).
Note that the second term in \eqref{eq: ANHARM_DYN} couples the mechanical perturbations \Revision{from} the thermodynamic perturbations of atom $i$ and its neighbors. Moreover, since equations~\eqref{eq: GPP_EVO_ANHARM} contain products of the thermodynamic variables $\Sigma$ and $\beta$ with the mean mechanical displacements $\avg{\bfq}$, the anharmonic potential leads to thermomechanical coupling in the GPP evolution equations. 

Due to the apparent harmonic nature of most standard interatomic potentials, the time scale of the GPP equations~\eqref{eq: ANHARM_DYN} and \eqref{eq: GPP_EVO_ANHARM}, when being applied to common potentials, is comparable to the time scale of atomic vibrations, since the system exhibits eigenfrequencies of $2\sqrt{2n_i\tr(\bfK)/m}$ for a pure harmonic potential (cf.~\eqref{eq: HarmonicPotentialEigs}). \Revision{Consequently, numerical time integration of the GPP equations -- in the anharmonic as in the harmonic potential case -- incurs computational costs comparable to MD, so that (as discussed previously for the harmonic case) the interatomic independence assumption prevents temporal upscaling.}

\subsection{Quasistatics and thermal equation of state}
\label{sec: QuasiStatics}

Using the insight gained from the time evolution equations~\eqref{eq: GPP_EVO_HARM} and~\eqref{eq: GPP_EVO_ANHARM}, we proceed to study the GPP equations within the quasistatic approximation. The latter yields a system of coupled nonlinear equations, whose solution yields the thermodynamic \Revision{equilibrium} state of the crystal \Revision{with atoms modeled using the GPP ansatz.} In the quasistatic approximation, the GPP equations \eqref{eq: GPP_EVO_SUMM} with mean mechanical momentum $\lavg{\bfp}_i = 0$ and thermal momentum $\beta_i=0$ reduce to the following steady-state equations:
\begin{subequations}
\begin{equation}
 \avg{\bfF_i(\bfq)} = 0,
 \label{eq: GPP_mechanical}
\end{equation} 
\begin{equation}
 \frac{\Omega_i}{m} + \frac{\avg{\bfF_i(\bfq)\cdot(\bfq_i -\lavg{\bfq}_i)}}{3}= 0,
 \label{eq: GPP_Thermal}
\end{equation}
 \label{eq: GPP_QS}%
\end{subequations}
which are to be solved for the equilibrium parameters $(\lavg{\bfq}_i, \Sigma_i, \Omega_i)$ for each atom \Revision{$i=1,\ldots,N$. 
Substitution of the quasistatic limits $\lavg{\bfp}_i \rightarrow 0, \beta_i\rightarrow 0$ in \eqref{eq: GPP_EVO_SUMM} yields that, at thermomechanical equilibrium, the solution of equations~\eqref{eq: GPP_QS} corresponds to the equilibrium mean displacements $\lavg{\bfq}_i$ and the equilibrium displacement-variance $\Sigma_i$ of all atoms. Therefore, \eqref{eq: GPP_QS} provides only two equations for the three unknowns $(\lavg{\bfq}_i, \Sigma_i, \Omega_i)$ per atom.
Analogous to the loss of information about the evolution of the mean mechanical momentum $\lavg{\bfp}_i(t)$, the quasistatic limit only states that, at final thermodynamic equilibrium, the thermal momentum $\beta_i(t)$ has decayed to~$0$, and $\Omega_i$ and $\Sigma_i$ are related by equation~\eqref{eq: GPP_Thermal}.  (Note that substitution of $\beta_i = 0$ in \eqref{eq: GPP_EVO_SUMM} yields trivially $\dd \Omega_i/\dd t = 0$ at quasistatic thermomechanical equilibrium.) To determine the evolution of $\Omega_i(t)$ during the thermomechanical relaxation of the system towards equilibrium, a model $\beta_i(t)$ is required to complement \eqref{eq: GPP_QS}. Hence, the quasistatic approximation results in the loss of information about the thermodynamics of the process through which the system is brought to the thermomechanical equilibrium. In the following, we consider different such thermodynamic processes and derive the respective process constraint, which -- together with \eqref{eq: GPP_QS} -- yields the result of a thermomechanical relaxation.}

\Revision{To physically approximate the nature of a thermodynamic process}, we assume that the ergodic hypothesis holds for quasistatic processes, i.e.,
\begin{equation}
 \avg{A(\bfz)} = \frac{1}{\tau}\int^{\tau}_{0}A(\bfz) \dd t,
 \label{eq: ergodic}
\end{equation}
where $\tau$ is a sufficiently large time interval, over which the evolution of the system is assumed quasistatic. In the ergodic limit, the momentum variance becomes
\begin{equation} 
 \Omega_i = mk_B T_i,
 \label{eq: OmegaTemperature}
\end{equation}
where $T_i$ is the local temperature of the $i^\text{th}$ atom.
\Revision{As a consequence, to model an \emph{isothermal} relaxation, we must keep $T_i$ and -- by \eqref{eq: OmegaTemperature} -- $\Omega_i$ constant for each atom, so that $\Omega_i$ is known for all atoms and the quasistatic equations~\eqref{eq: GPP_QS} can then be solved for $(\lavg{\bfq}_i, \Sigma_i)$}.

By contrast, \textit{isentropic} equilibrium parameters $(\lavg{\bfq}_i, \Sigma_i, \Omega_i)$ can be obtained by keeping $S_i$ fixed for each atom \Revision{during the relaxation}. From \eqref{eq: entropyDef} we know that
\begin{equation}
 S_i = S_{0,i} + 3k_B\ln\left(\frac{\sqrt{\Omega_i\Sigma_i}}{h}\right) = \tilde{S}_0 + {3k_B}S_{\Sigma,i} + {3k_B}S_{\Omega,i} = \text{const}.,
 \label{eq: entropyGPP_QS}
\end{equation}
where $\tilde{S}_0=S_{0,i}-3k_B\ln h$. Upon using suitable dimensional constants of unit values, parameters \Revision{
\be
    S_{\Omega,i}= \frac{1}{2}\ln\Omega_i
    \st{and} 
    S_{\Sigma,i}= \frac{1}{2}\ln\Sigma_i
\ee}%
may be interpreted as dimensionless momentum-variance and displacement-variance entropies, respectively. In the following, it will be convenient to use $S_{\Omega,i}$ and $S_{\Sigma,i}$ as the mean free parameters instead of $\Omega_i$ and $\Sigma_i$. 

\Revision{Analogously, \textit{isobaric}} conditions can be derived from the system-averaged Cauchy stress tensor \citep{admal2010unified}
\begin{equation}
 \bfsigma = -\frac{1}{\vol}\sum_i\avg{\frac{\bfp_i\otimes\bfp_i}{m} + \bfF_i(\bfq)\otimes(\bfq_i - \lavg{\bfq}_i)},
 \label{eq: VirialStress}
\end{equation}
where $\vol$ is the volume of the system. The average hydrostatic pressure $p$ of the system is
\begin{equation}
 p = -\frac{\mathrm{tr}(\bfsigma)}{3} =  \frac{1}{\vol}\sum_i\left(k_B T_i + \frac{\avg{\bfF_i(\bfq)\cdot(\bfq-\lavg{\bfq}_i)}}{3}\right).
 \label{eq: VirialPressure}
\end{equation}
Hence, setting $p=\text{const}.$ in \eqref{eq: VirialPressure} is the \textit{isobaric} constraint, subject to which equations~\eqref{eq: GPP_QS} become
\begin{equation}
\avg{\bfF_i(\bfq)} = 0,
\qquad
\left(k_B T_i - \frac{p\vol}{N}\right) + \frac{\avg{\bfF_i(\bfq)\cdot(\bfq-\lavg{\bfq}_i)}}{3} 
= 0.
\label{eq: Isobaric_GPP_QS}
\end{equation}
Solving these equations, in which $\Omega_i=m\left(k_B T_i - \frac{p\vol}{N}\right)$ serves as the momentum variance corrected for pressure $p$, yields the equilibrium parameters $(\lavg{\bfq}_i, S_{\Sigma, i}, S_{\Omega, i})$ for a given externally applied pressure $p$. Note that for a system of non-interacting atoms, the quasistatic equations subjected to an external pressure $p$ reduce to the ideal gas equation $p\vol = N k_B T$. Consequently, \eqref{eq: Isobaric_GPP_QS} shows that the quasistatic approximation yields the thermal equation of state of the system accounting for the interatomic potential, which enables the thermomechanical coupling within the crystal. The three constraints on $\Omega_i$ for isentropic, isothermal, and isobaric processes -- based on the above discussion -- are summarized in Table~\ref{tab: thermodynamicProcessAssumption}. 

\begin{table}[!h]
\setlength{\tabcolsep}{1em}
\def\arraystretch{1.5}
\centering
\begin{tabular}{c|c|c|c}
{Process}& {Isentropic}  &  {Isothermal} &  {Isobaric} \\
\hline
{Constraint}	& $\Omega_i\Sigma_i = \mathrm{const.}$  &  \Revision{$\Omega_i = k_Bm T_i=\text{const}.$} & \Revision{$N\left(\Omega_i - k_BmT_i\right)/\vol = -pm =\text{const.}$}
\end{tabular}\quad
 \caption{Summary of the thermodynamic constraints on $\Omega_i$ for the different assumptions about the thermodynamic process under which the system is brought to equilibrium.} 
 \label{tab: thermodynamicProcessAssumption}
\end{table}

\subsection{Helmholtz free energy minimization}
\label{sec: free_energy_minimization}

The solution $(\lavg{\bfq}, S_\Sigma, S_\Omega)$ of the local equilibrium relations~\eqref{eq: GPP_QS} may be re-interpreted as a minimizer of the Helmholtz free energy $\mathcal{F}$ (note that $(\lavg{\bfq}, S_\Sigma, S_\Omega)$ denotes the whole set of parameters of all $N$ atoms constituting the system). The Helmholtz free energy $\mathcal{F}$ is defined as
\begin{equation}
 \mathcal{F}(\lavg{\bfq}, S_\Sigma, S_\Omega) = \inf_{S}\left\{E(\lavg{\bfq}, S_\Sigma, S) - \sum_i\frac{\Omega_iS_i}{k_B m}\right\},
 \label{eq: FreeEnergy}
\end{equation}
with the internal energy of the system being
\begin{equation}
 E(\lavg{\bfq}, S_\Sigma, S) = \avg{H} = \sum_i\left(\frac{3\Omega_i}{2m} + \avg{V_i(\bfq)}\right).
 \label{eq: InternalEnergy}
\end{equation}
The definition~\eqref{eq: FreeEnergy} implies the local thermodynamic equilibrium definition
\begin{equation}
 \frac{\Omega_i}{k_B m} = \frac{\partial E}{\partial S_i}, 
\end{equation}
which can be verified using \eqref{eq: entropyGPP_QS} and \eqref{eq: InternalEnergy}. In addition, minimization of $\mathcal{F}$ with respect to the parameter sets $\lavg{\bfq}$ and $S_{\Sigma}$, subject to any of the thermodynamic constraints in Table~\ref{tab: thermodynamicProcessAssumption} for updating $S_{\Omega}$, yields equations~\eqref{eq: GPP_QS}, i.e., 
\begin{subequations}
\begin{align}
 &-\frac{\partial \mathcal{F}}{\partial \lavg{\bfq}_i} = 0 \implies \avg{F_i(\bfq)} = 0,
\end{align}
and
\begin{align}
 &-\frac{\partial \mathcal{F}}{\partial S_{\Sigma,i}} = 0 \implies \frac{3\Omega_i}{m} +  \avg{F_i(\bfq)\cdot(\bfq_i - \lavg{\bfq}_i)} = 0.
 \label{eq: FreeEnergyEOS}
\end{align}  
\label{eq: FreeEnergyMinimization}
\end{subequations}
\Revision{A detailed derivation of \eqref{eq: FreeEnergyEOS} is provided in \ref{sec: GPP_QC_FreeEnergy}. }

\begin{figure}[!b]
 \centering
 {\includegraphics[width = \textwidth]{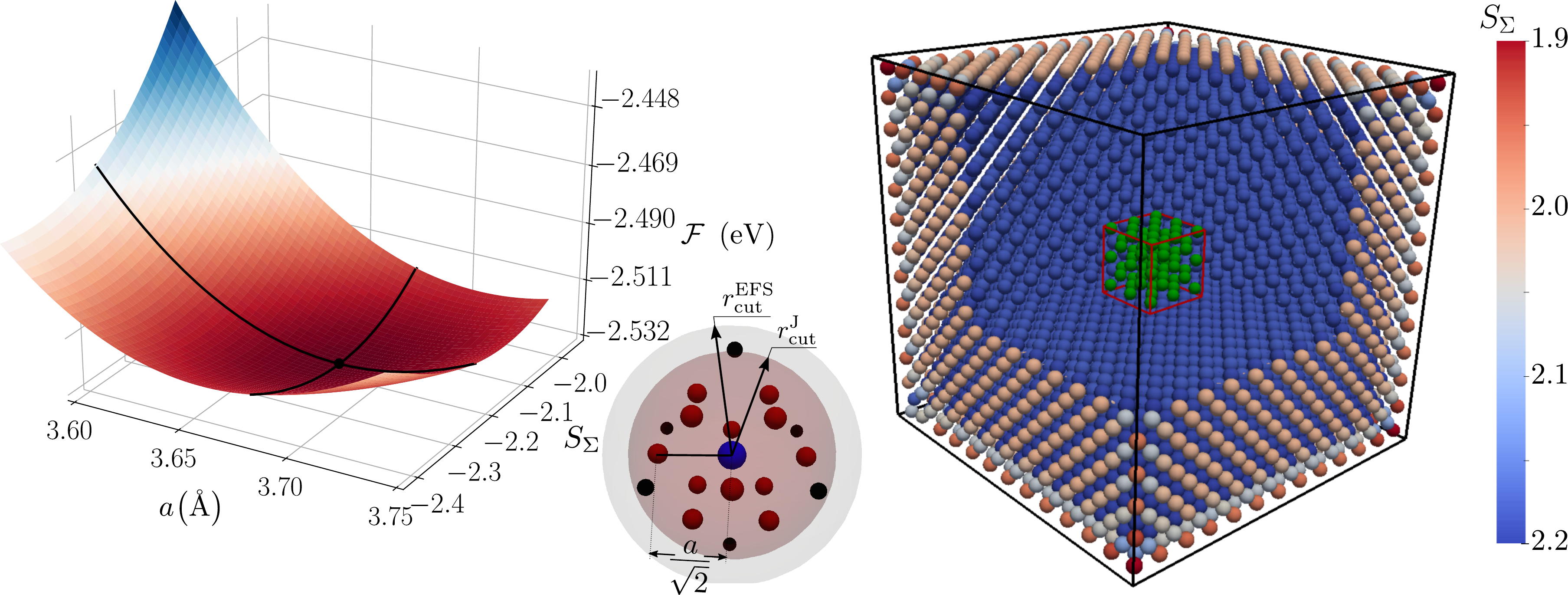}}
  \put(-480,180){$(a)$}
 \put(-230,180){$(b)$}
  \caption{\Revision{Infinite-crystal and free-standing nano-cube simulation domains used for calculating the thermal expansion of a Cu single-crystal. $(a)$ Surface plot of the Helmholtz free energy $\mathcal{F}$ vs.\ lattice parameter $a$ and displacement entropy $S_{\Sigma}$ for \citeauthor{johnson1988analytic}'s (\citeyear{johnson1988analytic}) EAM potential for FCC copper at $T = 1000~\mathrm{K}$. Also shown is the setup for computing $\mathcal{F}$ for a single atom in an infinite crystal with uniform lattice spacings: the free energy of the central atom (blue) is computed using the nearest neighbors (red) for the potential of \citet{johnson1988analytic} ($r^{\mathrm{J}}_{\mathrm{cut}} = 3.5~\mathrm{\AA}$) and up to second-nearest neighbors for the EAM potential of \citet{dai2006extended} ($r^{\mathrm{EFS}}_{\mathrm{cut}}=4.32~\mathrm{\AA}$). 
  $(b)$ Free-standing nano-cube composed of $12\times 12 \times 12$ FCC unit cells of pure single-crystalline copper, illustrating the spatial variation of $S_{\Sigma}$ due to a varying number of neighbors for atoms in the bulk vs.\ near the surface, at $T=1000~\mathrm{K}$ modeled using the potential of \citet{johnson1988analytic}. Thermal expansion is measured using the volume of the inner domain of the crystal (outlined in red). Atoms inside are marked in green. }}
  \label{fig: thermal_expansion_setup}
\end{figure}

\begin{figure}[!t]
 \centering
 {\includegraphics[width = \textwidth]{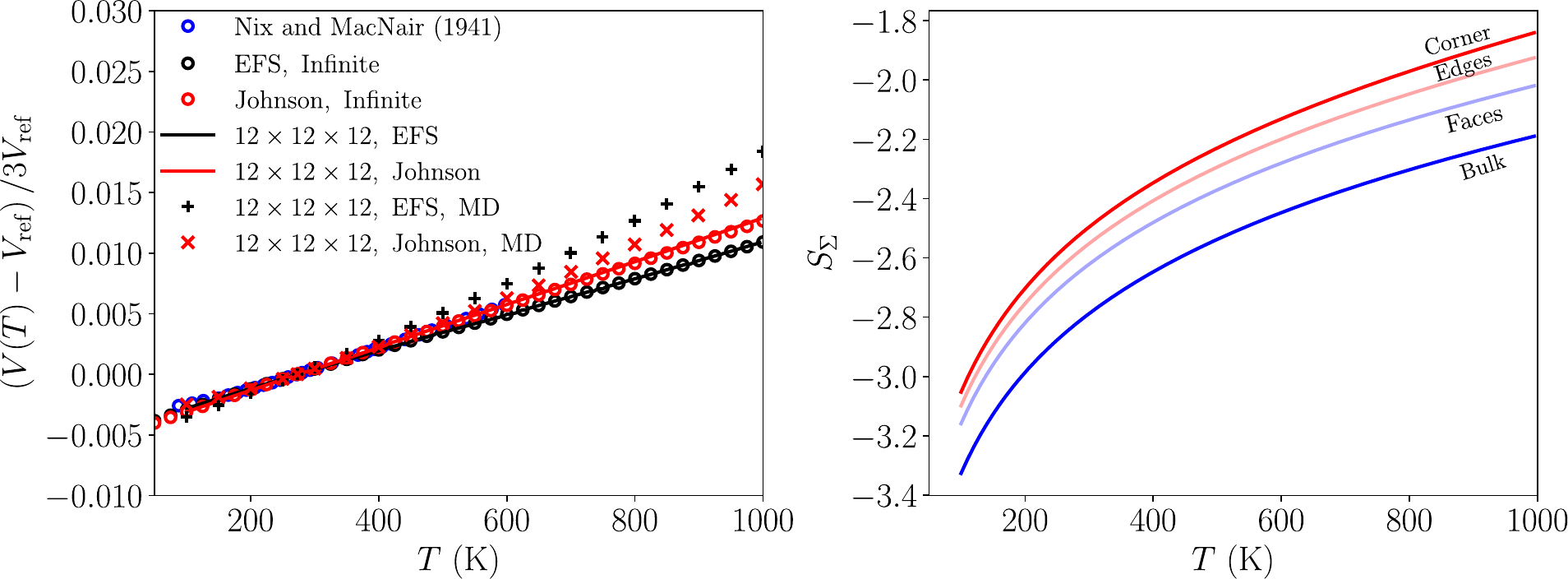}}
  \put(-480,180){$(a)$}
 \put(-230,180){$(b)$}
  \caption{\Revision{Thermal expansion of a free-standing nano-cube ($12\times 12 \times 12$ atomic unit cells) and of an infinite crystal of pure single-crystalline Cu, modeled using the Extended Finnis-Sinclair potential of \citet{dai2006extended} and the exponentially decaying potential of \citet{johnson1988analytic}. $(a)$ Comparison of computed volumetric changes with experimental data from \citet{nix1941thermal} and MD results ($V_{\mathrm{ref}} = V(T=273~\mathrm{K})$), where $V(T)$ is the volume of the inner subdomain of the nano-cube in Figure~\ref{fig: thermal_expansion_setup}$(b)$ for the finite-cube calculation and $a^3(T)$ for the infinite-crystal calculation, at temperature $T$. $(b)$ Variation of the displacement-variance entropy $S_{\Sigma}$ with temperature for the nano-cube calculation. As the temperature increases, the vibrational kinetic energy increases, resulting in an increase of $\Sigma$ at thermal equilibrium due to the equipartition of energy. Due to different numbers of interacting atomic neighbors, atoms on the corners, edges, faces, and in the bulk exhibit different values of $S_{\Sigma}$ (see \figurename~\ref{fig: thermal_expansion_setup}). Note that the value of $S_{\Sigma}$ depends on the interatomic potential (shown results are for potential of  \citet{johnson1988analytic}).}}
  \label{fig: thermal_expansion}
\end{figure}

\Revision{Equations~\eqref{eq: GPP_QS} (subject to a suitable thermodynamic constraint) are identical to the max-ent-based formulations of \citet{kulkarni2008variational} and \citet{venturini2014atomistic}. \citet{kulkarni2008variational} developed the max-ent formulation by enforcing constraints on variances of momenta and displacements of the atoms to obtain an ansatz for $f(\bfz)$, which is a special case of the GPP ansatz with no correlations (see Section~\ref{sec: GPP_generic}). \citet{venturini2014atomistic} developed the max-ent formulation by generalizing the grand-canonical ensemble to nonequilibrium situations and allowing non-uniform thermodynamic properties among atoms. However, for computational feasibility, \citet{venturini2014atomistic} invoked a trial-Hamiltonian procedure to justify the Gaussian formulation of the distribution function (for single-species cases), thus rendering their ansatz also a special case of our GPP ansatz with no correlations. In the above sections, we have shown that \emph{such special case arises only as a result of the quasistatic approximation, which enforces vanishing mechanical and thermal momenta}. Consequently, within the quasistatic assumption, the GPP based local-equilibrium equations~\eqref{eq: GPP_QS} are identical to those of \citet{kulkarni2008variational} and \citet{venturini2014atomistic}. Moreover, subject to the isothermal constraints, GPP quasistatic equations~\eqref{eq: GPP_QS} are identical to those used by~\citet{li2011diffusive} as well.

\subsection{Isothermal validation: thermal expansion of a Cu single-crystal}

Within the scope of the present study, we aim to validate the computational implementation of the local-equilibrium equations~\eqref{eq: GPP_QS} within a finite-temperature, updated-Lagrangian 3D QC framework, which will be detailed and used in Section~\ref{sec: QCApplication} to perform coarse-grained nonequilibrium thermomechanical simulations. For numerical validation of the isothermal quasistatic setting, we here compute the thermal expansion coefficient of a single-crystal of pure copper (Cu). We use an in-house developed updated-Lagrangian QC library, which is based on a simplicial mesh of tetrahedral elements in 3D and which adopts the seamless coarsening strategy of \citet{amelang2015summation}, so that all regions in a simulation (including the fully-resolved domain) are meshed. For computing phase-space averages, we use third-order multivariable Gaussian quadrature~\citep{stroud1971approximate, kulkarni2007coarse}.}

\Revision{Figure~\ref{fig: thermal_expansion} illustrates solutions of the isothermal local-equilibrium relations for varying fixed uniform temperature $T_i = T$ for all atoms within a perfect Cu single-crystal, modeled using the EAM potentials of \citet{dai2006extended} and \citet{johnson1988analytic}. To simulate the expansion of an infinite crystal, we determine the lattice parameter $a$ and displacement-variance entropy $S_{\Sigma}$ which minimize the Helmholtz free energy $\mathcal{F}$ of an atom under the influence of its full centrosymmetric neighborhood in a periodic lattice (cf. \figurename~\ref{fig: thermal_expansion_setup}$(a)$). \citet{kulkarni2007coarse} used the same infinite-crystal formulation, in which the mechanical forces are evaluated using the lattice spacing as the independent variable (see also the appendix of \citet{tembhekar2018fully}). To demonstrate the effect of free surfaces, we repeat the same calculation for a free-standing Cu nano-cube containing $12\times 12 \times 12$ FCC unit cells, subject to the local-equilibrium equations~\eqref{eq: GPP_QS}, isothermal constraint, and free boundaries at various temperature values in our QC solver at full resolution (cf. \figurename~\ref{fig: thermal_expansion_setup}$(b)$). Results are also included in Figure~\ref{fig: thermal_expansion}(a). As temperature $T$ increases, the lattice spacing $a$ increases. In turn, as the spacing between atoms increases, the displacement-variance entropy $S_{\Sigma}$ also increases (cf.~\figurename~\ref{fig: thermal_expansion}$(b)$). Further evident from that graph, $S_{\Sigma}$ is not uniform within the crystal, owing to varying numbers $n_i$ of interacting neighboring atoms on the corners, edges, and faces as well as within the bulk of the crystal. Sufficiently far away from the free boundaries, each atom is embedded in an approximately perfect infinite crystal and hence characterized by uniform values of the displacement entropy. To minimize the effects of free boundaries, we compute the deviation in volume of a bounding box enveloping a $2\times 2 \times 2$ volume at the center of the crystal  with temperature (outlined in red in figure~\ref{fig: thermal_expansion}$(c)$ at the center of the nano-cube) to compute the thermal expansion coefficient in the free-standing cube case. To validate our results, we further report comparable thermal expansion data obtained from the MD code LAMMPS~\citep{plimpton1995fast}, using an NPT ensemble with periodic boundaries (labeled `MD' in ~\figurename~\ref{fig: thermal_expansion}$(a)$).

The comaprison shows that the GPP-based formulation (which in this isothermal setting is equivalent to the max-ent formulation) yields reliable data for the chosen potentials with marginal errors up to half the melting temperature. At higher temperature, the GPP results under-predict the thermal expansion -- as can be expected -- with errors being significantly smaller for the nearest-neighbor-based potential of \citet{johnson1988analytic}. We note that the GPP-based results depend on the chosen third-order multivariable Gaussian quadrature for computing phase-space averages. Even though higher-order quadrature has been shown to yield improved accuracy \citep{tembhekar2018fully} (at considerably increased computational cost), a detailed case study of the impact of the order of quadrature is beyond the present scope.
}


\begin{figure}[!b]
 \centering
 {\includegraphics[width = 0.9\textwidth]{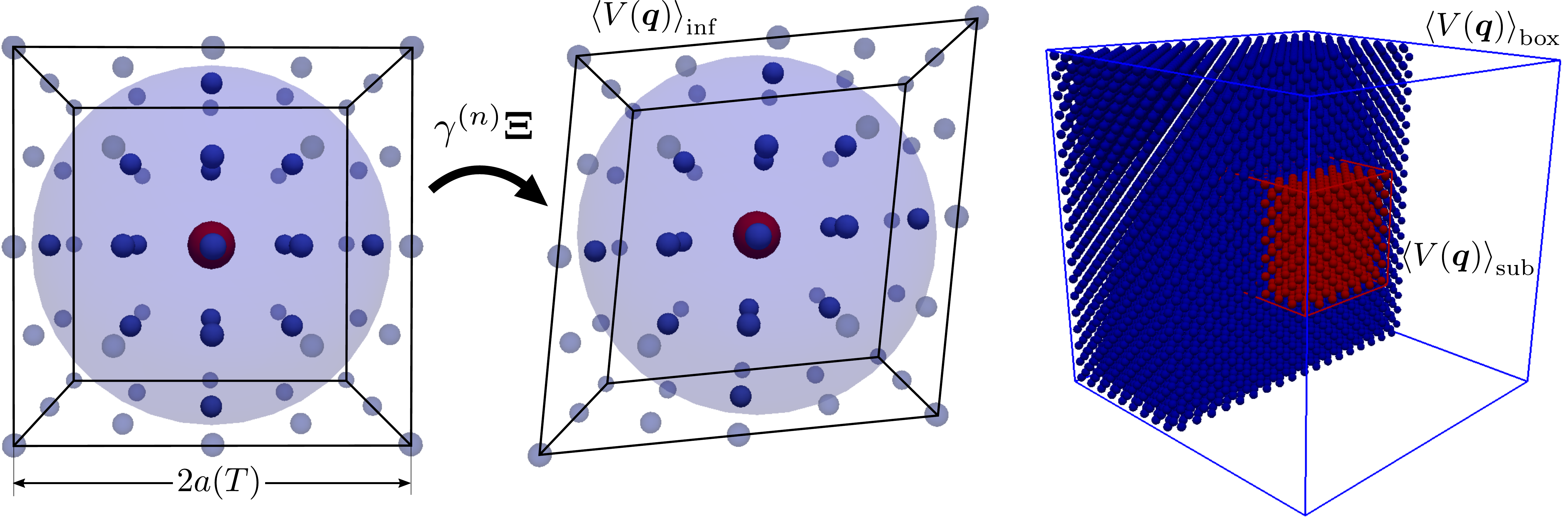}}
 \put(-450, 140){$(a)$}
  \put(-140, 140){$(b)$}
  \caption{\Revision{Domains used for calculating the elastic constants $C_{11}, C_{44}$ and the bulk modulus $\kappa$. $(a)$ Domain of $2\times 2\times 2$ FCC unit cells of Cu initialized with equilibrium lattice spacing $a(T)$ and displacement variance entropy $S_{\Sigma}$ at temperature $T$ undergoing the deformation measured by the strain tensor $\gamma^{(n)}\Xi$. The central atom (red) interacts with the full neighborhood and is relaxed isentropically, keeping the rest of the atoms mechanically fixed. Its phase-averaged potential $\avg{V(\bfq)}_{\mathrm{inf}}$ models the average phase-averaged potential of an infinite crystal. $(b)$ Domain of $18\times 18 \times 18$ FCC unit cells of Cu. The domain is initially relaxed isothermally at the temperature at which the elastic constants are to be obtained. Afterwards, all the atoms are displaced according to the affine deformation~\eqref{eq: DisplacementElasticicty}. Blue atoms in the outer-layer are held mechanically fixed. Red atoms, initially bounded within the red-outlined box of size equivalent to $6\times6\times 6$ FCC unit-cells, are relaxed mechanically. All the atoms are kept thermally fixed since the elastic constants correspond to the curvature of the free energy against mechanical deformation only~\citep{venturini2011topics}.
   $\avg{V(\bfq)}_{\mathrm{sub}}$ is the phase-averaged potential of the red atoms and $\avg{V(\bfq)}_{\mathrm{box}}$ is the phase-averaged potential of all atoms in the domain, both averaged over the respective number of atoms. As $\gamma^{(n)}$ increases, both phase-averaged potentials vary. Curvatures of both $\avg{V(\bfq)}_{\mathrm{sub}}$ and $\avg{V(\bfq)}_{\mathrm{box}}$ with respect to $\gamma^{(n)}$ at the respective energy minima are used for computing the elastic constants.}}
\label{fig: elastic_constants_setup}
\end{figure}

\subsection{\Revision{Isentropic validation: elastic constants of a Cu single-crystal}}

\Revision{Going beyond uniform thermal expansion, we proceed to benchmark the GPP-based formulation with isentropic process conditions by calculating the uniaxial and shear components of the linear elastic stiffness tensor ($C_{11}$ and $C_{44}$, respectively, in Voigt-Kelvin notation, see~\citet{reddy2007introduction}), and the bulk modulus $\kappa$ -- which we collectively refer to as \textit{elastic constants} in the following. A a sample system we again choose single-crystalline pure Cu.

Calculation of the aforementioned elastic constants ($C_{11}$, $C_{44},$ and $\kappa$) was previously used as a benchmark by \citet{venturini2011topics} for a crystal with impurities, and by \citet{tembhekar2018fully}, who all used the max-ent formulation (without specifying the isentropic process condition).}
\Revision{To compute the elastic constants, we follow a procedure similar to the one by~\citet{amelang2015summation}. We consider a domain of $18\times 18 \times 18$ FCC unit-cells of Cu, subject to the local equilibrium equations~\eqref{eq: GPP_QS} and the isentropic process constraint (see \figurename~\ref{fig: elastic_constants_setup}$(b)$), so as to model an isentropic deformation process around the undeformed ground state at a given (initial) temperature~$T$. We use the isentropic constraint for relaxing the deformed crystal, because the elastic constants are measured in an adiabatic setting~\citep{overton1955temperature}. Initially, the domain is relaxed isothermally with free boundaries. After the initial relaxation, atoms are displaced according to, 
\begin{equation}
 \bfq^{(n)}_i = \bfq^{(0)}_i + \gamma^{(n)}\bfXi\cdot \bfq^{(0)}_i,
 \label{eq: DisplacementElasticicty}
\end{equation}
where $\gamma^{(n)} = n \,\Delta \gamma_c$ is the engineering strain measure at the $n^{\mathrm{th}}$ deformation step, $\Delta \gamma$ is the strain increment, and $\bfXi$ is the base deformation matrix. For $C_{11}, C_{44}, $and $\kappa$, it is defined as, respectively,
\begin{equation}
 \bfXi_{11} = \left[\begin{matrix}
                     1 & 0 & 0 \\
                     0 & 0 & 0 \\
                     0 & 0 & 0 
                    \end{matrix}\right],
                    \qquad
                    \bfXi_{44} = \left[\begin{matrix}
                     0 & 1 & 0 \\
                     1 & 0 & 0 \\
                     0 & 0 & 0 
                    \end{matrix}\right],
                    \qquad
                    \bfXi_{\kappa} = \left[\begin{matrix}
                     1 & 0 & 0 \\
                     0 & 1 & 0 \\
                     0 & 0 & 1
                    \end{matrix}\right].
\label{eq: base_deformation_definition}
\end{equation}

\begin{figure}[!t]
 \centering
 {\includegraphics[width = \textwidth]{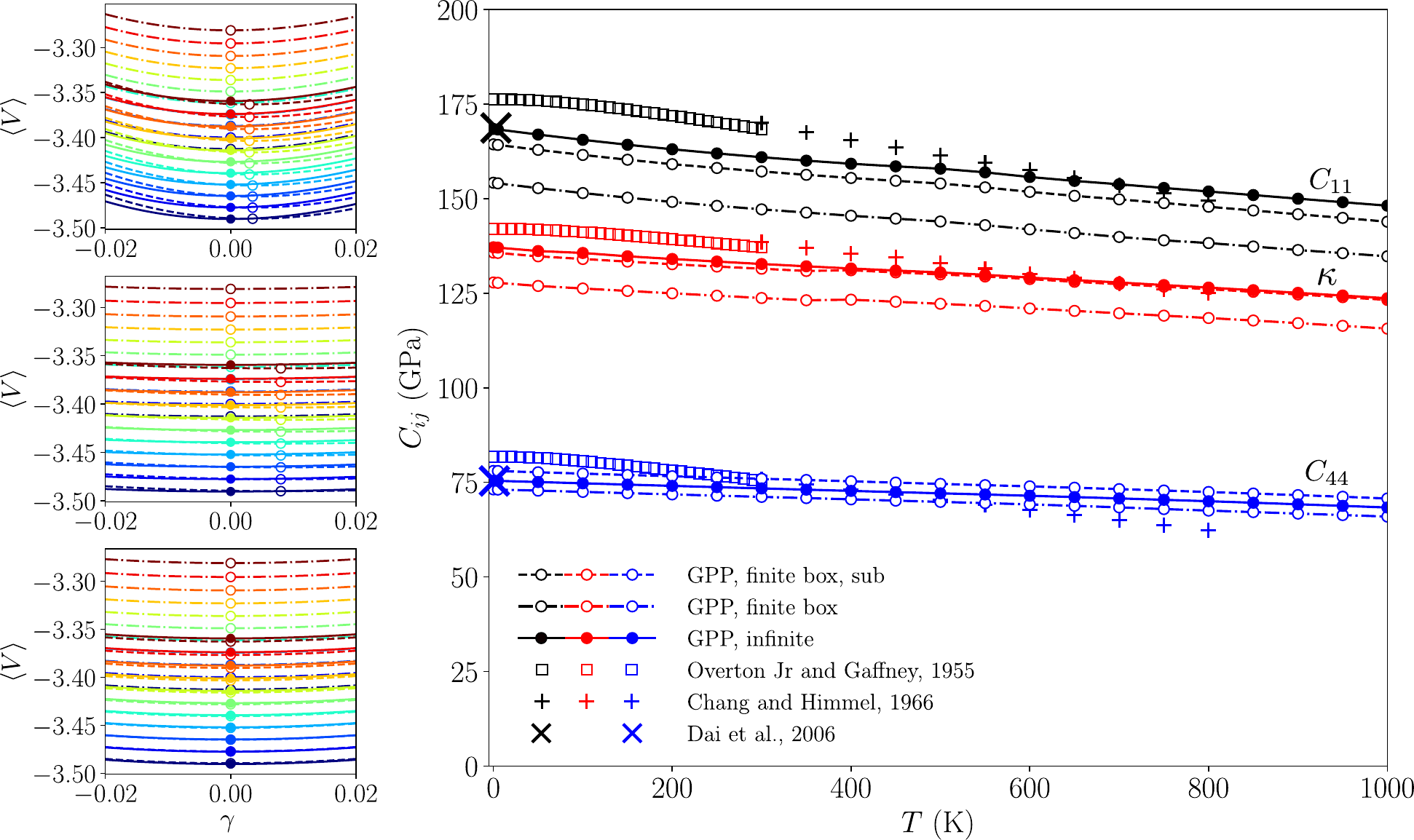}}
   \put(-490,290){$(a)$}
   \put(-490,200){$(b)$}
   \put(-490,110){$(c)$}
   \put(-355,290){$(d)$}
   \caption{\Revision{Variation of $\avg{V(\bfq)}_{\mathrm{box}}$ (chained line),  $\avg{V(\bfq)}_{\mathrm{sub}}$ (dashed line), and $\avg{V(\bfq)}_{\mathrm{inf}}$ (solid line) vs.\ the engineering strain measure $\gamma$ for the deformation matrix $(a)~\bfXi_{11}$, $(b)~\bfXi_{\kappa}$, $(c)~\bfXi_{44}$ for temperatures from $0$ K to $1000$ K. Solid circles mark where $\avg{V(\bfq)}_{\mathrm{inf}}$ is minimal ($\gamma = \gamma_c$), and open circles mark the analogous for $\avg{V(\bfq)}_{\mathrm{box}}$ and $\avg{V(\bfq)}_{\mathrm{sub}}$. $(d)$  Elastic constants of a Cu single-crystal (black: $C_{11}$, red: $\kappa$, blue: $C_{44}$) for temperatures from $0$~K to $1000$~K, obtained using the curvature of $\avg{V(\bfq)}_{\mathrm{box}}$ (chained line, open circles), $\avg{V(\bfq)}_{\mathrm{sub}}$ (dashed line, open circles), and $\avg{V(\bfq)}_{\mathrm{inf}}$ (solid line, solid circles) at their respective minima, using the EAM potential of \citet{dai2006extended}, compared against the experimental data from \citet{chang1966temperature} and \citet{overton1955temperature}. }}
\label{fig: elastic_constants}
\end{figure}

To model an infinite crystal without size effects imposed by free surfaces, we also consider a domain of $2\times 2 \times 2$ FCC unit cells of Cu, initialized using the lattice parameter $a$ and displacement-variance entropy $S_{\Sigma}$ which minimize the Helmholtz free energy at the given temperature (see Figure~\ref{fig: thermal_expansion_setup}$(a)$). The domain is deformed using \eqref{eq: DisplacementElasticicty}, isentropically relaxing the central atom while keeping its neighborhood mechanically fixed by the external deformation (Figure~\ref{fig: elastic_constants_setup}$(a)$). As the atoms are displaced, the free energy of each atom changes due to changing potential energy. For the infinite-crystal model, the atom at the center interacts with its whole neighborhood, which deforms as per~\eqref{eq: DisplacementElasticicty} and exhibits a change in its phase-averaged potential $\avg{V(\bfq)}_{\mathrm{inf}}$. The affine deformation mimics the straining of a perfect infinite crystal. The phase-averaged potential of the central atom, $\avg{V(\bfq)}_{\mathrm{inf}}$, is measured for all increments $n$ to compute the elastic constants. When modeling a finite-sized, free-standing crystal, relaxation is performed while holding the atoms in a layer touching the free boundaries of the domain mechanically fixed (blue atoms in \figurename~\ref{fig: elastic_constants_setup}$(b)$). The phase-averaged potential of the whole domain, $\avg{V(\bfq)}_{\mathrm{box}}$, and that of a sub-domain within the bulk of the domain, $\avg{V(\bfq)}_{\mathrm{sub}}$ (also averaged over the number of respective type of atoms), are measured for all $n$.  We use a $4^{\mathrm{th}}$ degree polynomial fit through the phase-averaged potentials to obtain continuous functions $\avg{V(\bfq)}_{\mathrm{box}}(\gamma)$, $\avg{V(\bfq)}_{\mathrm{sub}}(\gamma)$, and $\avg{V(\bfq)}_{\mathrm{inf}}(\gamma)$, and to compute the respective curvatures at $\gamma^{(n)} = \gamma_c$, for which $\avg{V(\bfq)}_{\mathrm{sub}}$,  $\avg{V(\bfq)}_{\mathrm{box}}$, and $\avg{V(\bfq)}_{\mathrm{inf}}$ are minimal. The elastic constants at temperature $T$ are obtained from the curvatures via \citep{venturini2011topics}
\begin{equation}
 C_{11}(T) = \frac{1}{\vol(\gamma_c,T)}\frac{\partial^2 \avg{V(\bfq)}}{\partial \gamma^2}\Big|_{\gamma = \gamma_c}, ~C_{44}(T) = \frac{1}{4\vol(\gamma_c,T)}\frac{\partial^2 \avg{V(\bfq)}}{\partial \gamma^2}\Big|_{\gamma = \gamma_c}, ~ \kappa(T) = \frac{1}{9\vol(\gamma_c,T)}\frac{\partial^2 \avg{V(\bfq)}}{\partial \gamma^2}\Big|_{\gamma = \gamma_c},
 \label{eq: elasticConstantsCurvature}
\end{equation}
where $\vol(\gamma_c,T)$ is the atomic volume at $\gamma_c$ and temperature $T$. We note that, by definition, the change in free energy $\mathcal{F}(\lavg{\bfq}, S_\Sigma)$ with mechanical strain measure $\gamma$ is used for computing the elastic constants~\citep{venturini2011topics,phillips2001crystals}. Since the change in $\mathcal{F}(\lavg{\bfq}, S_\Sigma)$ is identical to the change in the phase-averaged potential $\avg{V(\bfq)}$ for only mechanical deformations, we have used $\avg{V(\bfq)}$ in \eqref{eq: elasticConstantsCurvature}.

Figure~\ref{fig: elastic_constants} shows the change of $\avg{V(\bfq)}_{\mathrm{box}}$, $\avg{V(\bfq)}_{\mathrm{sub}}$, and $\avg{V(\bfq)}_{\mathrm{inf}}$ with $\gamma$ for various temperatures and the elastic constants $C_{11}, C_{44},$ and $\kappa$, obtained using the GPP-based local-equilibrium equations~\eqref{eq: GPP_QS}, and experimental data~\citep{overton1955temperature, chang1966temperature}. Results were computed for decreasing values of $\Delta \gamma$ to ensure convergence. The reported results correspond to $\Delta \gamma = 0.0005$. The computed elastic constants display similar thermal softening as observed in experiments, yet they exhibit an offset at all temperatures since the EAM potential of \citet{dai2006extended} was calibrated using the elastic constants at room temperature, which are treated as those at 0 K in the present formulation (see~\tablename~4 in \citet{dai2006extended} and \tablename{ 3.1} in \citet{kittel1976introduction}). At 0 K, the values of $C_{11}$ and $C_{44}$ obtained from the infinite-crystal model ($C_{11}=168.41$ GPa, $C_{44}$ = 75.41 GPa) match exactly the values reported by~\citet{dai2006extended}. Those obtained from the finite-crystal simulation setup ($C_{11}=164.34$ GPa, $C_{44}$ = 78.01 GPa using $\avg{V(\bfq)}_{\mathrm{sub}}$ and $C_{11}=154.20$ GPa, $C_{44}$ = 73.12 GPa using $\avg{V(\bfq)}_{\mathrm{box}}$) deviate from those reported by~\citet{dai2006extended} due to the size-effects posed by the free surfaces, as expected. The overall behavior captures the thermal softening of pure Cu well.
}

\subsection{Linear Onsager kinetics}
\label{sec: Dissipation}

\Revision{
We reiterate that equations~\eqref{eq: FreeEnergyMinimization} are identical to the max-ent framework, as derived and utilized previously~\citep{kulkarni2008variational, ariza2012hotqc, venturini2014atomistic, ponga2015finite, tembhekar2018fully}. However, the max-ent framework is based on a different motivation and does not rely on the physical insight gained  in Sections~\ref{sec: GPP_generic},~\ref{sec: dynamics_harmonic}, and~\ref{sec: dynamics_anharmonic} about the dynamics of atoms at long and short time intervals. Moreover, the GPP framework clearly highlights the information loss as a result of the quasistatic approximation (otherwise hidden in the max-ent framework), which also enables the modeling of thermomechanical deformation of crystals under various thermodynamical conditions (e.g., isentropic, isobaric, or isothermal processes). Furthermore, recall that in Section~\ref{sec: GPP_generic} we showed that the interatomic independence assumption precludes the framework from capturing any changes in local temperature due to unequal temperatures of neighboring atoms. Such an assumption is also made in the max-ent framework. 
In reality, a non-uniform temperature distribution arises readily in non-uniformly deformed crystals (differences in local atomic spacings lead to different local temperatures).  As both max-ent and the present GPP formulation cannot model heat flux as a consequence of such local temperature diffferences, we require explicit thermal transport models to capture heat transport. To this end, we here adopt the linear Onsager kinetics model introduced by \citet{venturini2014atomistic}, based on a quadratic dissipation potential -- as discussed in the following.}

The total rate of change of entropy of an atom can be decomposed into a reversible and an irreversible change, i.e.,
\begin{equation}
 \frac{\dd S_i}{\dd t} = \frac{q_{i,\mathrm{rev}}}{T_i} + \frac{q_{i,\mathrm{irrev}}}{T_i},
 \label{eq: entropy_Decomposition}
\end{equation}
where $q_{i,\mathrm{rev}}$ is the reversible heat addition and $q_{i,\mathrm{irrev}}$ is the irreversible change signifying the net influx of heat into the atom due to a non-uniform temperature distribution. In a dynamic system (see Sections~\ref{sec: dynamics_harmonic} and~\ref{sec: dynamics_anharmonic}), the reversible change is composed of fluctuations about the equilibrium state due to the local harmonic nature of the interatomic potential, and it is proportional to the thermal momentum $\beta$. Within the quasistatic approximation, the information about the evolution of $\beta(t)$ as the system relaxes towards the equilibrium is lost since $\beta\rightarrow0$ is imposed, thus rendering the reversible changes in entropy unknown. Therefore, such a reversible change is imposed implicitly by the thermodynamic constraints (see Table~\ref{tab: thermodynamicProcessAssumption}). 

For an \textit{isolated} system of atoms, the reversible heat \Revision{exchange} vanishes ($q_{\mathrm{rev}} = 0$), and the system relaxes to equilibrium adiabatically, which we here term \emph{free relaxation}. Note that this free relaxation is not isentropic, since the temperature can be non-uniform as a result of an imposed non-uniform deformation of the crystal, resulting in irreversible thermal transport that increases the entropy. We model such irreversible change $q_{i,\mathrm{irrev}}$ by adopting the kinetic formulation introduced by \citet{venturini2014atomistic}:
\begin{equation}
 q_{i,\mathrm{irrev}} = \sum_{j\neq i}R_{ij},
 \qquad
 R_{ij} = \frac{\partial \Psi}{\partial P_{ij}},
 \label{eq: PotentialAssumption}
\end{equation}
where $R_{ij}$ is the local, pairwise heat flux, driven by a local pairwise discrete temperature gradient $P_{ij}$ through the kinetic potential $\Psi$. Using the dissipation inequality, \citet{venturini2014atomistic} formulated the discrete temperature gradient as
\begin{equation}
 P_{ij} = \frac{1}{T_i} - \frac{1}{T_j}.
 \label{eq: temperature_gradient}
\end{equation}
Within the linear assumption, the kinetic potential $\Psi$ is modeled as, 
\begin{equation}
 \Psi = \frac{1}{2}\sum_{i=1}^N\sum_{j\neq i} A_{ij}T^2_{ij}P^2_{ij}
 \quad\text{with}\quad
 T_{ij} = \frac{1}{2}\left(T_i + T_j\right),
 \label{eq: LinearPotential}
\end{equation}
where $A_{ij}$ denotes a pairwise heat transport coefficient (which is treated as an empirical constant in this work), and $T_{ij}$ represents the pairwise average temperature. Equations~\eqref{eq: LinearPotential} and~\eqref{eq: PotentialAssumption} yield the  entropy rate kinetic equation for a freely relaxing system of atoms as
\begin{equation}
 \frac{\dd S_i}{\dd t} = \frac{1}{T_i}\sum_{j\neq i} A_{ij}T^2_{ij}P_{ij} = \frac{1}{T_i}\sum_{j\neq i} A_{ij}T^2_{ij}P_{ij}.
\label{eq: entropy_kinetic_linear_onsager}
\end{equation}
\cite{venturini2014atomistic} validated this thermal transport model based on linear Onsager kinetics by demonstrating the capturing of size effects of the macroscopic thermal conductivity of Silicon nanowires. \Revision{\citet{ponga2018unified} used a similar temperature difference-based diffusive transport model to analyse large thermomechanical deformation of carbon nanotubes. They formulated the Arrhenius equation type master-equation model, identical to the one used for mass-diffusion~\citep{zhang2008atomistically} problems and validated their setup against Fourier-based heat conduction problems.}

\Revision{To use the model, one needs to calibrate the heat transport coefficients $A_{ij}$, for which experimentally measured values of $\bfkappa_i$ for a given arrangement of atoms can be used. To this end, \citet{venturini2014atomistic} derived the following discrete-to-continuum relation between $A_{ij}$ and the thermal conductivity tensor $\bfkappa_i$ at the location of atom $i$:
\begin{equation}
 \bfkappa_i = \frac{1}{2V_i}\sum_{j=1}^NA_{ij}\left(\lavg{\bfq}_i - \lavg{\bfq}_j\right)\otimes\left(\lavg{\bfq}_i - \lavg{\bfq}_j\right),
 \label{eq: kinetic_coeffs_conductivity}
\end{equation}
where $V_i$ is the volume of the atomic unit cell in the crystal. The obtained values of $A_{ij}$ may be interpreted to capture the interatomic heat current and regarded as the intrinsic property of the material. In the model of \citet{ponga2018unified}, an empirical parameter (exchange rate) was fitted against theoretically or experimentally characterized thermal conductivity values.}

\begin{figure}[!b]
 \centering
 {\includegraphics[width = \textwidth]{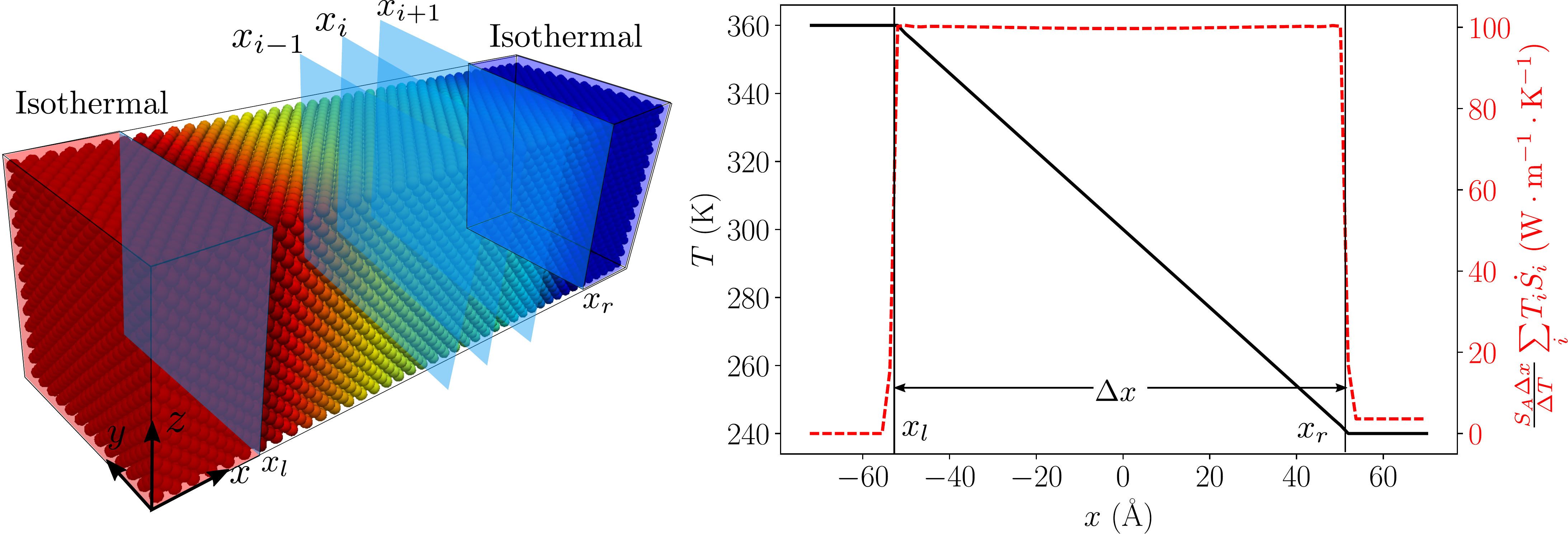}}
 \put(-500,170){$(a)$}
 \put(-260,170){$(b)$}
  \caption{Thermal conduction in a Cu nanowire with square cross-section and edges aligned with the $\langle 100\rangle$-directions. $(a)$~Schematic showing the cross-sectional planes used for computing the macroscopic thermal flux. $(b)$ Spatial variation of temperature (black solid line) and macroscopic heat flux (red dashed line) along the axis of the nanowire.}
  \label{fig: thermal_conduction}
\end{figure}

\Revision{It is important to point out that  equation~\eqref{eq: kinetic_coeffs_conductivity} was derived from first-order approximations of temperature differences between interacting atoms, using Taylor expansions \citet{venturini2014atomistic}. Depending on the thermal boundary conditions and the size of the domain, temperature differences between interacting atoms may not be negligible. For significant temperature differences, \eqref{eq: kinetic_coeffs_conductivity} is inaccurate, hence, we here identify the values of $A_{ij}$ by simulating a differentially heated Cu nanowire of square cross-section  up to steady state.} 
\Revision{Specifically, we fit the kinetic coefficient $A_0$ for Cu, using the thermal conductivity measurements for Cu nanowires obtained by~\citet{mehta2015enhanced}. To this end, we consider a Cu nanowire of square cross-section of size 145$\times$43$\times$43\AA$^3$, with the central axis of the nanowire oriented along the $x$-direction and all edges aligned with the $\langle100\rangle$-directions (see Figure~\ref{fig: thermal_conduction}$a$). Atoms in the region $x<x_l$ are thermostated at a temperature of 360~K, while the atoms in the region $x>x_r$ are thermostated at a temperature of 240~K. All the boundaries are considered as free surfaces and the system is relaxed isentropically, followed by the diffusive step for thermal transport (as explained in Algorithm~\ref{alg: thermomechanical_algo} below).} As the system evolves according to \eqref{eq: entropy_kinetic_linear_onsager} (using full atomistic resolution everywhere in the nanowire), a uniform macroscopic heat flux is established between $x_l<x<x_r$ (see Figure~\ref{fig: thermal_conduction}$b$). Dividing the nanowire into atomic planes marked by $x_i$, the macroscopic flux across the plane at $x=x_i$ is given by
\begin{equation}
T_i\frac{\dd S_i}{\dd t} = J_{x,i} - J_{x,{i-1}},~
J_{x,i} = \sum^{i}_{j=1}T_j\frac{\dd S_j}{\dd t},
\label{eq: macroscopicFlux}
\end{equation}
where $T_j$ and $dS_j/dt$ are temperature and entropy generation at plane $x_j$. From \eqref{eq: macroscopicFlux} the approximate thermal conductivity $\kappa$ is obtained as
\begin{equation}
 \kappa = \frac{J_{x,i}}{S_A}\frac{\Delta x}{\Delta T},
\label{eq: conductivity_flux}
\end{equation}
where $S_A$ is the cross-sectional area, and $\Delta x $ is the length across which the temperature difference $\Delta T$ is maintained. \Revision{To find $A_0$, we do not use equation~\eqref{eq: kinetic_coeffs_conductivity} since it is valid only for small temperature differences. Instead, we start from a reference guess of $A_0 = 0.1~$nW/K, compute the macroscopic flux $J_{x,i}$ and from it the thermal conductivity $\kappa$ via \eqref{eq: conductivity_flux} and compare it to the experimental value. To obtain an approximate value of $A_0$ for our simulations, we considered $\kappa = 100~\text{W}/(\text{m}\cdot\text{K})$ as determined experimentally by \citet{mehta2015enhanced} (cf.~Figure 3 in \citet{mehta2015enhanced}). The initial guess of $A_0$ is updated using the secant method, until the conductivity value of $\kappa = 100~\text{W}/(\text{m}\cdot\text{K})$ is reached. The obtained value of $A_0 \approx 15.92~\text{eV}/(\text{ns}\cdot\text{K})=2.55~$nW/K is used in all subsequent simulations. We note that the numerical values, however, do not affect the physical modeling framework and are only representative values to be used in subsequent simulations. 

In reality, large deformations cause defects which modify the phonon and electron scattering properties of a crystal, due to which $A_0$ may vary with deformation (the same applies to considerably affine straining of a crystal). However, such effects are beyond the scope of the current work. As shown in Figure~\ref{fig: thermal_conduction}$b$, the temperature profile is linear and the macroscopic thermal flux defined by \eqref{eq: conductivity_flux} is constant at steady state, thus highlighting that the discrete model yields a behavior similar to Fourier's heat conduction law, where material between two isothermal boundaries exhibits a linear temperature distribution and constant heat flux, given that the conductivity is uniform. The model can be extended to deformation-dependent values of $A_0$ -- yet this requires a detailed (and presently unavailable) understand of scattering phenomena in strained crystals. Therefore, we here assume $A_0=\text{const}.$ as a first-order approximation.}

\begin{figure}[!b]
 \centering
 {\includegraphics[width = 0.33\textwidth]{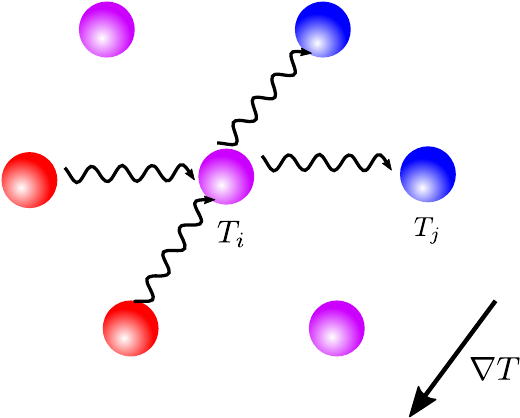}}
  \caption{Schematic illustration of the local entropy dissipation via equation~\eqref{eq: entropy_kinetic_linear_onsager}: atom $i$ interacts with its atomic neighbors, whereby differences in local temperature are responsible for heat flux.}
\label{fig: entropy_dissipation}
\end{figure}

Equations~\eqref{eq: GPP_QS} combined with \eqref{eq: entropy_kinetic_linear_onsager} complete the nonequilibrium thermomechanical atomistic model. Every atom is assumed to be in thermal equilibrium at some local temperature $T_i$, and mechanical and thermal interactions of the atoms with different spacings and different temperatures are modeled using the interatomic potential $V(\bfq)$, the local equation of state (e.g.~\eqref{eq: Isobaric_GPP_QS}), and the entropy kinetic equation~\eqref{eq: entropy_kinetic_linear_onsager}. \Revision{For a general system, the assumed thermodynamic process constraints yield the reversible heat exchange (see \tablename{~\ref{tab: thermodynamicProcessAssumption}}). For instance, in an isothermal constraint, $S_{\Omega,i}$ remains constant and $S_{\Sigma,i}$ changes with mechanical deformation to satisfy the local equation of state (e.g.~\eqref{eq: Isobaric_GPP_QS}) at equilibrium, thus changing the net entropy (see \eqref{eq: entropyGPP_QS}). }

While the nature of the assumed thermodynamic process via which the system relaxes depends on the macroscopic and microscopic boundary conditions, the process is generally assumed quasistatic with respect to the fine-scale vibrations of each atom. 
However, the thermal transport equation~\eqref{eq: entropy_kinetic_linear_onsager} introduces a time scale to the problem governing the irreversible evolution of entropy. If we consider a two-atom toy model, consisting of only two atoms with temperatures $T_i$ and $T_j$ (Figure~\ref{fig: entropy_dissipation}), equations~\eqref{eq: entropy_kinetic_linear_onsager} reduce to
\begin{equation}
\frac{\dd }{\dd t}\left(\begin{matrix}
S_i\\
S_j
\end{matrix}
\right) = \left(\begin{matrix}
A_0T^2_{ij}P_{ij}/T_i\\
A_0T^2_{ji}P_{ji}/T_j
\end{matrix}\right) ,
\label{eq: heat_transport_model_time_stability}
\end{equation}
where we have assumed equal coefficients $A_{ij} = A_0$ for both atoms. Let us further assume the thermomechanical relaxation of $(\lavg{\bfq}, S_{\Sigma})$ takes place through an isentropic process following the quasistatic equations~\eqref{eq: GPP_QS}. When substituting equation~\eqref{eq: entropyGPP_QS}, equation~\eqref{eq: heat_transport_model_time_stability} becomes
\begin{equation}
\frac{\dd }{\dd t}\left(\begin{matrix}
T_i\\
T_j
\end{matrix}
\right) = \frac{2A_0}{3k_B}\left(\begin{matrix}
T^2_{ij}P_{ij}\\
T^2_{ji}P_{ji}
\end{matrix}\right).
\label{eq: temperature_transport}
\end{equation}
The stationary state of equation~\eqref{eq: temperature_transport} is a state with uniform temperature, $T_i = T_j$. When assuming a leading-order perturbation about the stationary state, equation~\eqref{eq: temperature_transport} yields
\begin{equation}
\frac{\dd }{\dd t}\left(\begin{matrix}
T_i\\
T_j
\end{matrix}
\right)\approx\frac{2A_0}{3k_B}\left(\begin{matrix}
T_j - T_i\\
T_i - T_j
\end{matrix}\right),
\label{eq: temperature_transport_linear}
\end{equation}
which shows that, when using a simple first-order explicit-Euler finite-difference scheme for time integration, the time step $\Delta t$ is restricted by
\begin{equation}
    \Delta t \leq \frac{3k_B}{2A_0}
    \label{eq: time_step_constraint}
\end{equation}
for numerical stability. \citet{venturini2014atomistic} used $A_0 = 0.09~$nW/K for the bulk region of a silicon nanowire, which yields a maximum allowed time step of $0.23~$ps. With higher-order explicit schemes, a larger maximum time step may be obtained, but the \Revision{maximum allowed time step} remains at a few picoseconds. Such restriction arises because the bulk thermal conductivity $\bfkappa$ is used to fit the atomistic parameter $A_{ij}$, which reduces the length scale as well as the time scale. Such a restriction also highlights the coupling of length and time scales~\citep{j2007statisticalEvans}. For larger temperature differences and larger systems, the nonlinear equation~\eqref{eq: temperature_transport} yields the following stability limit on time step $\Delta t$ (see~\ref{sec: Entropy_transport_stability} for the derivation)
\begin{equation}
    \Delta t\left(\frac{2A_0}{3k_B}\right)\sum_j\frac{T^2_{ij}}{T_iT_j}\leq 1,
\label{eq: time_step_constraint_nonlinear}
\end{equation}
\Revision{which is stricter than the linear stability limit in \eqref{eq: time_step_constraint}. In summary, even though we simulate quasistatic behavior at, in principle, considerably larger time scales by the GPP-based formulation, the use of the linear thermal transport model severely restricts the usable time step size in simulations.}


\subsection{\Revision{Algorithmic implementation}}

Before applying QC coarse-graining \Revision{to the GPP-based formulation}, let us summarize the proposed thermomechanical transport model for simulating the quasistatic deformation of crystals composed of GPP atoms (see Algorithm~\ref{alg: thermomechanical_algo} for details). Our numerical scheme decomposes each load step as follows:
\begin{itemize}
    \item \textbf{Step 1}: Given the equilibrium parameters $\left(\lavg{\bf{q}}^{(n)}_i, S^{(n)}_{\Sigma, i}, S^{(n)}_{\Omega, i}, S^{(n)}_i\right)$ \Revision{for each atom $i$} from the (previous) $n^{\mathrm{th}}$ load step, an external stress/strain is applied to the system at load step $n+1$.
    \item \textbf{Step 2}: Quasistatic relaxation, solving \eqref{eq: GPP_QS} subject to one of the constraints in Table~\eqref{tab: thermodynamicProcessAssumption} to obtain the intermediate state $\left(\lavg{\bf{q}}^{(*)}_i, S^{(*)}_{\Sigma, i}, S^{(*)}_{\Omega, i}, S^{(*)}_i\right)$ \Revision{for each atom $i$}.
    \item \textbf{Step 3}: Staggered time stepping that alternates between irreversible updates of the total entropy and quasistatic reversible relaxation steps of all variables, until convergence is achieved. Specifically, the total entropy is updated \Revision{irreversibly} from $S^{(*)}_i$ over a time interval $\delta t$, using equation~\eqref{eq: entropy_kinetic_linear_onsager} and explicit forward-Euler updates, \Revision{and reversibly using the assumed thermodynamic process constraints during the subsequent thermomechanical relaxation (see the two contributions in \eqref{eq: entropy_Decomposition}). Time step} $\delta t$ depends on the external stress/strain rate applied. By definition, the thermomechanical relaxation is assumed quasistatic, hence only slow rates with respect to atomistic vibrations can be modeled. However, the irreversible transport imposes a time-scale restriction via \eqref{eq: time_step_constraint_nonlinear}. Consequently, time integration via suitable time steps $\Delta t_k$ must be continued for $K$ steps, such that $\sum^K_{k=1}\Delta t_k = \delta t$. \Revision{The irreversible update of the entropy from $S^{(*),k}_i \rightarrow S^{(**),k+1}_i$ results in a new approximate thermal distribution $S^{(**),k+1}_{\Omega, i}$ via equation~\eqref{eq: entropyGPP_QS} as $S^{(**),k+1}_{\Omega, i} \rightarrow S^{(**),k+1}_i/3k_B - \tilde{S}_0/3k_B - S^{(*),k}_{\Sigma, i}$, generating thermal forces on atoms. Using the irreversibly updated entropy and the approximate thermal distribution, quasistatic relaxation of state $\left(\lavg{\bf{q}}^{(*),k}_i, S^{(*),k}_{\Sigma, i}, S^{(**),k+1}_{\Omega, i}, S^{(**),k+1}_i\right) \rightarrow \left(\lavg{\bf{q}}^{(*),k+1}_i, S^{(*),k+1}_{\Sigma, i}, S^{(*),k+1}_{\Omega, i}, S^{(*),k+1}_i\right)$ follows by solving \eqref{eq: GPP_QS} with a constraint from Table~\ref{tab: thermodynamicProcessAssumption}. This completes a single staggered time step. Note that the update $S^{(**),k+1}_i \rightarrow S^{(*),k+1}_i$ corresponds to the reversible entropy update to satisfy~\eqref{eq: entropyGPP_QS} during the thermomechanical relaxation and depends on the assumed thermodynamic process constraint.} In a full quasistatic setting, the transport equation is driven towards a steady state with $\dot{S}^{(*), K}_i = 0$, which defines the convergence criterion and hence determines the total number of time steps. 
    \item \textbf{Step 4}: Assignment of the final state as  $\left(\lavg{\bf{q}}^{(n+1)}_i, S^{(n+1)}_{\Sigma, i}, S^{(n+1)}_{\Omega, i}, S^{(n+1)}_i\right) =  \left(\lavg{\bf{q}}^{(*),K}_i, S^{(*),K}_{\Sigma, i}, S^{(*),K}_{\Omega, i}, S^{(*),K}_i\right)$, followed by Step 1 until the final load step.
\end{itemize}

\Revision{To complete Step 2, we use a combination of the Fast Inertial Relaxation Engine (FIRE) of \citet{bitzek2006structural} and nonlinear generalized minimal residual (NGMRES) using the Portable, Extensible Toolkit for Scientific Computation (PETSc)~\citep{PETSc}. We employ a forward-Euler finite-difference scheme to update the entropy due to irreversible transport in Step 3. The steps are explained in detail as pseudocode in Algorithm~\ref{alg: thermomechanical_algo}.}

\begin{algorithm}[!h]
\SetAlgoLined
\KwResult{$\left(\lavg{\bf{q}}^{(n+1)}_i, S^{(n+1)}_{\Sigma, i}, S^{(n+1)}_{\Omega, i}, S^{(n+1)}_i\right)$}
\textbf{Input: }$\left(\lavg{\bf{q}}^{(n)}_i, S^{(n)}_{\Sigma, i}, S^{(n)}_{\Omega, i}, S^{(n)}_i\right)$, $\delta t^{(n)}$, ${tol}$\;
$k \gets 0$\;
quasistatic reversible relaxation of $\left(\lavg{\bf{q}}^{(n)}_i, S^{(n)}_{\Sigma, i}, S^{(n)}_{\Omega, i}, S^{(n)}_i\right)$ to $\left(\lavg{\bf{q}}^{(*),k}_i, S^{(*),k}_{\Sigma, i}, S^{(*),k}_{\Omega, i}, S^{(*),k}\right)$ by solving \eqref{eq: GPP_QS} with a constraint from Table~\ref{tab: thermodynamicProcessAssumption} using \Revision{FIRE~\citep{bitzek2006structural} and/or NGMRES~\citep{PETSc}}\;
$t \gets 0$\;
compute $\dot{S}^{(*),k}_i$ using \eqref{eq: entropy_kinetic_linear_onsager}\;
$\dot{S}_i \gets \dot{S}^{(*),k}_i$\;
\While{$t<\delta t^{(n)}$ and $\sqrt{\frac{1}{N}\sum_{i}\dot{S}^{2}_i}> \text{tol}$}{
compute $\Delta t^k$ satisfying the constraint \eqref{eq: time_step_constraint_nonlinear}\;
\Revision{irreversible update of $S^{(*),k}_i$ to $S^{(**),{k+1}}_i$ using \eqref{eq: entropy_kinetic_linear_onsager} and a forward-Euler finite-difference scheme\;
approximate thermal distribution update using \eqref{eq: entropyGPP_QS} such that $S^{(**),k+1}_{\Omega, i} \gets S^{(**),k+1}_i/3k_B - \tilde{S}_0/3k_B - S^{(*),k}_{\Sigma, i}$\;
quasistatic reversible relaxation of $\left(\lavg{\bf{q}}^{(*),k}_i, S^{(*),k}_{\Sigma, i}, S^{(**),k+1}_{\Omega, i}, S^{(**),k+1}\right)$ to $\left(\lavg{\bf{q}}^{(*),{k+1}}_i, S^{(*),{k+1}}_{\Sigma, i}, S^{(*),{k+1}}_{\Omega, i}, S^{(*),{k+1}}\right)$ by solving ~\eqref{eq: GPP_QS} with a constraint from Table~\ref{tab: thermodynamicProcessAssumption}\;}
$k \gets k+1$\;
$\dot{S}_i \gets \dot{S}^{(*),{k}}_i$\;
$t \gets t+\Delta t^{k}$\;
}
$\left(\lavg{\bf{q}}^{(n+1)}_i, S^{(n+1)}_{\Sigma, i}, S^{(n+1)}_{\Omega, i}, S^{(n+1)}_i\right) \gets \left(\lavg{\bf{q}}^{(*),k}_i, S^{(*),k}_{\Sigma, i}, S^{(*),k}_{\Omega, i}, S^{(*),k}\right)$
\caption{Single load step from $n$ to $n+1$ of the quasistatic thermomechanical transport model for irreversible deformation of crystals composed of GPP atoms. For a fully quasistatic transport simulation $\delta t^{(n)} \to \infty$ \Revision{and the thermal gradients are diffused till the RMS entropy generation rate is higher than some tolerance $tol$}.}
\label{alg: thermomechanical_algo}
\end{algorithm}

\section{Finite-temperature updated-Lagrangian quasicontinuum framework based on GPP atoms}
\label{sec: QCApplication}

Having established the GPP framework, we proceed to discuss the application of the thermomechanical and coupled thermal transport model (equations~\eqref{eq: GPP_QS} and \eqref{eq: entropy_kinetic_linear_onsager}, respectively) to an updated-Lagrangian QC formulation for coarse-grained simulations. Previous zero- and finite-temperature QC implementations have usually been based on a total-Lagrangian setting~\citep{ariza2012hotqc, tadmor2013finite, knap2001analysis, amelang2015summation, ponga2015finite, kulkarni2008variational}, in which interpolations are defined and hence atomic neighborhoods computed in an initial reference configuration. Unfortunately, such total-Lagrangian calculations incur large computational costs and render especially nonlocal QC formulations impractical \citep{tembhekar2017automatic}, since atomistic neighborhoods change significantly during the course of a simulation, so that the initial mesh used for all QC interpolations increasingly loses its meaning and atoms that form local neighborhoods in the current configuration may have been considerably farther apart in the reference configuration. Therefore, we here introduce an updated-Lagrangian QC framework to enable efficient simulations involving severe deformation and atomic rearrangement. Moreover, we adopt the fully-nonlocal energy-based QC formulation of \citet{amelang2015summation}, since an energy-based summation rule allows for a consistent definition of the coarse-grained internal energy and all thermodynamic potentials of the system. 

\begin{figure}[!b]
  \centering
 {\includegraphics[width = 0.9\textwidth]{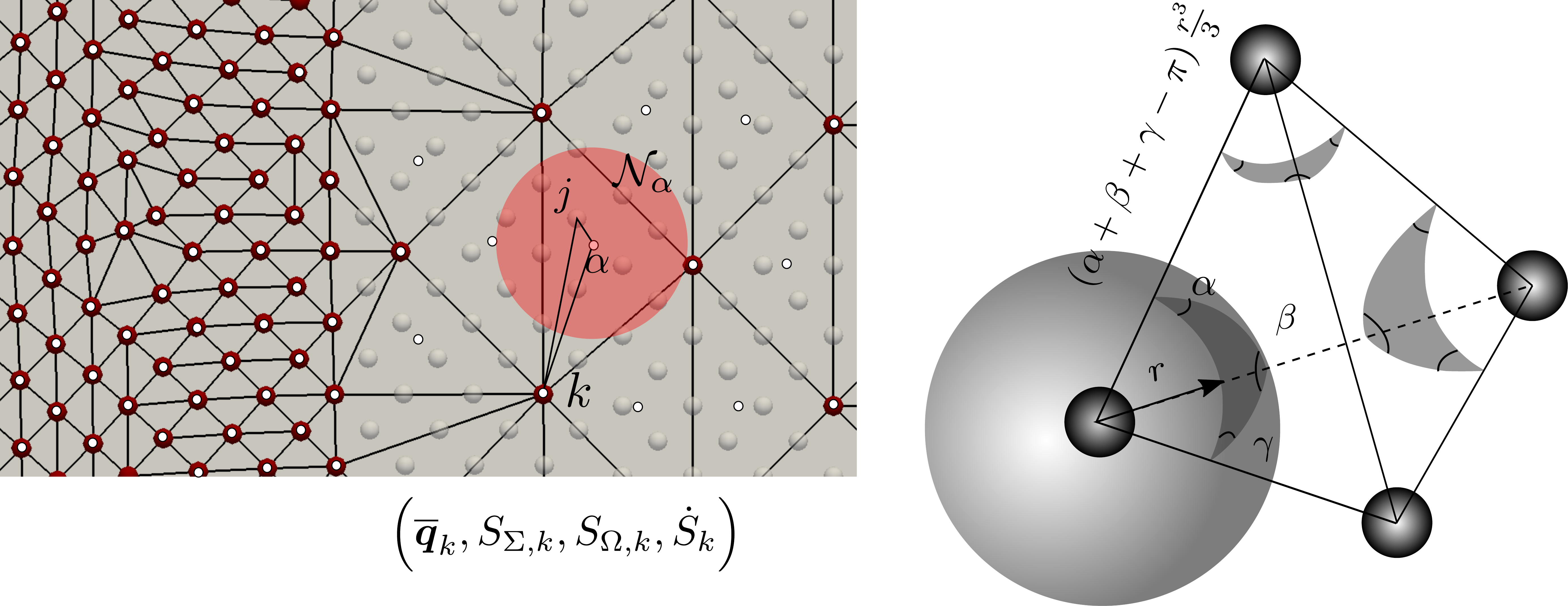}}
 \put(-460,170){$(a)$}
 \put(-160,170){$(b)$}
  \caption{Illustration of the the quasicontinuum (QC) framework based on GPP quasistatics (Eqs.~\eqref{eq: GPP_QS}) combined with the linear Onsager kinetics for thermal transport (Eq.~\eqref{eq: entropy_kinetic_linear_onsager}). $(a)$ The thermomechanical transport parameters $(\overline{\bfq}_k, S_{\Sigma, k}, S_{\Omega,k}, \dot{S}_k)$ are the degrees of freedom of repatom $k$. Repatoms are shown as red circles, sampling atoms as small white circles \Revision{(we use the first-order sampling rule $(0,1^*)$ of~\citet{amelang2015summation})} . All thermodynamic potentials and hence the quasistatic forces and thermal fluxes are computed from a weighted average over a set of sampling atoms (e.g., forces and fluxes for repatom $k$ are governed, among others, by sampling atom $\alpha$ and its atomic neighbors $j$). The fully-nonlocal formulation bridges seamlessly and adaptively from full atomistics to coarse-grained regions. $(b)$ Computation of sampling atom weights for the updated Lagrangian implementation using tetrahedral solid angles.}
 \label{fig: QC_mesh}
\end{figure}

Within the QC approximation, we replace the full atomic ensemble of $N$ GPP atoms (as described in Section~\ref{sec: GPP_generic}) by a total of $N_h\ll N$ \Revision{representative GPP atoms} (\emph{repatoms} for short), each having the thermomechanical transport parameters $(\lavg{\bfq}_k, S_{\Sigma, k}, S_{\Omega,k}, \dot{S}_k)$ as their degrees of freedom (see Figure~\ref{fig: QC_mesh}). The position of each and every atom in the coarse-grained crystal lattice is obtained by interpolation.
For an atom at location $\lavg{\bfq}^h_i$ in the reference configuration, the thermomechanical transport parameters in the current configuration are obtained by interpolation from the respective parameters of the repatoms:
\begin{equation}
 \left(\begin{matrix}
        \lavg{\bfq}^h_i \\
        S^h_{\Sigma, i} \\
        S^h_{\Omega, i} \\
        \dot{S}^h_i\\
       \end{matrix}\right) = \sum^{N_h}_{k=1}\left(\begin{matrix}
        \lavg{\bfq}_k \\
        S_{\Sigma, k} \\
        S_{\Omega, k} \\
        \dot{S}_k\\
       \end{matrix}\right)N_k(\lavg{\bfq}_i),
\label{eq: QC_interpolation}
\end{equation}
where the \Revision{superscript $h$ indicates that a parameter is interpolated} from the $N_h$ repatoms, and $N_k(\lavg{\bfq}^h_i)$ is the shape function/interpolant of repatom $k$ evaluated at $\lavg{\bfq}^h_i$. \Revision{In our updated-Lagrangian setting, the respective previous load step serves as the reference configuration used to define the mesh and the above interpolation.} In the following we use linear interpolation (i.e., constant-strain tetrahedra \Revision{in 3D}), while the method is sufficiently general to extend to other types of interpolants. Based on the interpolated parameters from \eqref{eq: QC_interpolation}, the free energy $\mathcal{F}(\lavg{q},S_{\Sigma},S_{\Omega})$ of the crystal is replaced by the approximate free energy $\mathcal{F}^h$ of the QC crystal with
\begin{equation}
 \mathcal{F}^h(\lavg{q}^h,S^h_{\Sigma},S^h_{\Omega}) = \sum^{N}_{i=1}\left(\frac{\Omega^h_i}{2m_i} - \frac{\Omega^h_i S^h_i}{k_Bm_i}\right) + \avg{V(\bfq)} = \sum^{N}_{i=1}\left(\frac{\Omega^h_i}{2m_i} - \frac{\Omega^h_i S^h_i}{k_Bm_i}+ \avg{V_i(\bfq)}\right),
 \label{eq: repatom_free_energy}
\end{equation}
where we assumed that the decomposition $V(\bfq) = \sum_{i}V_i({\bfq})$ of the interatomic potential holds. Furthermore, we allow the masses of all atoms to be different, denoting by $m_i$ the mass of atom $i$. Equation~\eqref{eq: repatom_free_energy} defines the free energy of the system accounting for \Revision{all $N$ atoms} with their thermomechanical parameters evaluated using the $N_h$ repatoms in a slave-master fashion.

To reduce the computational cost stemming from the summation over all $N$ atoms in \eqref{eq: repatom_free_energy}, sampling rules are introduced, which approximate the full sum by a weighted sum over $N_s\ll N$ carefully selected sampling atoms
\citep{ eidel2009variational, iyer2011field,amelang2015summation,tembhekar2017automatic}: 
\begin{equation}
 \mathcal{F}^h(\lavg{q}^h,S^h_{\Sigma},S^h_{\Omega})\approx \sum^{N_s}_{\alpha=1}w_{\alpha}\left(\frac{\Omega^h_\alpha}{2m_\alpha} - \frac{\Omega_\alpha S_\alpha}{k_Bm_\alpha} + \avg{V_{\alpha}(\bfq)}\right),
 \label{eq: approximateFreeEnergyApp}
\end{equation}
where $w_{\alpha}$ is the sampling weight of the $\alpha^{\mathrm{th}}$ sampling atom.  \Revision{Specifically, we adopt the \textit{first-order optimal summation rule}} of \citet{amelang2015summation}, in which all nodes and the centroid of each simplex (tetrahedron in 3D) are assigned as sampling atoms. \citet{amelang2015summation} computed the sampling atom weights using the geometrical division of the simplices by planes at a distance $r$ from the nodes (Figure~\ref{fig: QC_mesh}$(b)$) and adding the corresponding nodal volume to the respective sampling atom weight, while the rest of the simplex volume was assigned to the Cauchy-Born-type sampling atom at the centroid. Here, we point out that a simpler weight calculation is possible by considering the spherical triangle generated by the intersection of simplex $e$ with a ball of radius $r$ centered at one of the nodes. Considering the arcs of the spherical triangle subtend angles $\alpha$, $\beta$, and $\gamma$ at the opposite points, the area of the triangle is given by $(\alpha+\beta+\gamma - \pi)r^2$. Hence, the approximate volume of the enclosed region is
\begin{equation}
v^e_{\alpha}  \approx \frac{r^3}{3}\left(\alpha+\beta+\gamma - \pi\right),
\label{fig: volume_sphericalTetrahedron}
\end{equation}
and $w_{\alpha} = \rho_e\sum_{e}v^e_{\alpha}$ are the sampling atom weights at nodes, where $\rho_e$ is the density of simplex $e$ (expressed as the number of atoms per unit volume). For the centroid sampling atoms, the remaining volume times $\rho_e$ is assigned as its sampling weight $w_{\alpha}$. Since the deformation is affine within each element $e$ \Revision{owing to the chosen linear interpolation}, sampling atom weights in coarse-grained regions change negligibly in a typical simulation and are therefore kept constant throughout our simulations. In the following we will also need a separate set of repatom weights $\widehat{w}_k$, which we calculate by lumping the sampling atom weights $w_k$ to the repatoms: each repatom receives the weight of itself (each repatom is a sampling atom) plus one quarter of the Cauchy-Born-type centroidal sampling atoms within all adjacent elements $e$:
\begin{equation}
\widehat{w}_k = w_k + \sum_{e}\frac{w_e}{4}.
\label{eq: repatomLump}
\end{equation}

Given the sampling atom weights $w_{\alpha}$, minimization of the approximate \Revision{free energy~\eqref{eq: approximateFreeEnergyApp}} with respect to degrees of freedom $(\lavg{\bfq}_k, S_{\Sigma,k})$ of the $k^{\mathrm{th}}$ repatom yields the local mechanical equilibrium conditions
\begin{equation}
-\frac{\partial \mathcal{F}^h}{\partial \lavg{\bfq}_k} \approx -\sum^{N_s}_{\alpha=1}w_{\alpha}\frac{\partial \avg{V_{\alpha}(\bfq)}}{\partial \lavg{\bfq}_k} = -\sum^{N_s}_{\alpha=1}w_{\alpha}\avg{\frac{\partial V_{\alpha}(\bfq)}{\partial {\bfq}_k}} = 0
\label{eq: mechanical_QC}
\end{equation}
and the corresponding thermal equilibrium conditions
\begin{equation}
-\frac{\partial \mathcal{F}^h}{\partial S_{\Sigma,k}} \approx \sum^{N_s}_{\alpha=1}w_{\alpha}\left(\frac{3\Omega_{\alpha}}{m_{\alpha}}\frac{\partial S_{\Sigma, \alpha}}{\partial S_{\Sigma, k}} - \frac{\partial \avg{V_{\alpha}(\bfq)}}{\partial S_{\Sigma,k}}\right) = \sum^{N_s}_{\alpha=1}w_{\alpha}\left(\frac{3\Omega_{\alpha}}{m_{\alpha}}\frac{\partial S_{\Sigma, \alpha}}{\partial S_{\Sigma, k}} - \avg{\frac{\partial V_{\alpha}(\bfq)}{\partial \bfq}\cdot\frac{\partial \bfq}{\partial S_{\Sigma, k}}}\right)= 0.
\label{eq: thermal_QC}
\end{equation}
Substituting the interpolation from \eqref{eq: QC_interpolation} into \eqref{eq: mechanical_QC} and~\eqref{eq: thermal_QC} yields (for repatoms $k=1,\ldots,N_h$)
\begin{subequations}
\begin{equation}
-\sum^{N_s}_{\alpha=1}w_{\alpha}\left(\sum_{j\in\mathcal{N}(\alpha)}\avg{\frac{\partial V_{\alpha}(\bfq)}{\partial \bfq_j}}N_k(\lavg{\bfq}_j) + \avg{\frac{\partial V_{\alpha}(\bfq)}{\partial \bfq_{\alpha}}}N_{k}(\lavg{\bfq}_{\alpha})\right) = 0
\label{eq: mechanical_QC_interpolation}
\end{equation}
and
\begin{align}
&\sum^{N_s}_{\alpha=1}w_{\alpha}\left[\frac{3\Omega_{\alpha}}{m_{\alpha}}N_k(\lavg{\bfq}_{\alpha}) 
-
\left(\sum_{j\in\mathcal{N}(\alpha)}\avg{\frac{\partial V_{\alpha}(\bfq)}{\partial \bfq_j}\cdot\left(\bfq_j - \overline{\bfq}_j\right)}N_k(\lavg{\bfq}_j) + \avg{\frac{\partial V_{\alpha}(\bfq)}{\partial \bfq_{\alpha}}\cdot\left(\bfq_{\alpha}-\lavg{\bfq}_{\alpha}\right)}N_{k}(\lavg{\bfq}_{\alpha})\right)\right]= 0.
\label{eq: thermal_QC_interpolation}
\end{align}
\end{subequations}
Note that these are similar to the equilibrium conditions derived by \citet{tembhekar2018fully} following \citeauthor{kulkarni2008variational}'s max-ent formulation, although the max-ent formulation bypasses the thermodynamic relevance of the parameters in the dynamic setting. As noted previously in Section~\ref{sec: QuasiStatics}, the local thermal equilibrium equation~\eqref{eq: thermal_QC_interpolation} corresponds to the local equation of state of the system, here providing the equation of state of the coarse-grained quasicontinuum.
Solving equations~\eqref{eq: mechanical_QC} and~\eqref{eq: thermal_QC} subject to one of the constraints from Table~\ref{tab: thermodynamicProcessAssumption} depending upon the assumption of the thermodynamic process yields variables $(\lavg{\bfq}_k, S_{\Sigma, k}, S_{\Omega, k})$ for all repatoms, thus yielding the thermodynamically reversible solution for the deformation of the system. 

To introduce \Revision{seamless coarse-graining of the} linear Onsager kinetic model for irreversible thermal conduction \Revision{governed by equation~\eqref{eq: entropy_kinetic_linear_onsager}}, we solve for the entropy rates $\dot{S}_k$ of all repatoms and evolve the entropy in time for each repatom. To this end, we notice that the first term in equation~\eqref{eq: thermal_QC_interpolation} represents a thermal force due to the thermal kinetic energy of the system. Since that thermal force in our energy-based setting follows a force-based summation rule  \citep{knap2001analysis}, the entropy rate calculation of the repatom $k$ simplifies to
\begin{equation}
\widehat{w}_k\frac{\dd S_k}{\dd t} = \sum^{N_s}_{\alpha=1}w_{\alpha}\frac{\dd S_{\alpha}}{\dd t}N_k(\lavg{\bfq}_{\alpha}) = \sum^{N_s}_{\alpha=1}\frac{w_{\alpha}}{T_{\alpha}}\sum_{j\in\calN(\alpha)}A_{\alpha j}T^2_{\alpha j}P_{\alpha j}N_k(\lavg{\bfq}_{\alpha}),
\label{eq: coarse_grained_heat_flux}
\end{equation}
where $\widehat{w}_k$ is the repatom weight. Note that the kinetic potential in~\eqref{eq: LinearPotential} can, in principle, also be coarse-grained analogously to the free energy in~\eqref{eq: approximateFreeEnergyApp}. However, the resulting calculation of the entropy rates is computationally costly since it involves summation over all repatoms for each sampling atom calculation, which is why this approach is not pursued here. Equations~\eqref{eq: mechanical_QC_interpolation} and \eqref{eq: thermal_QC_interpolation} combined with the thermodynamic constraints in Table~\ref{tab: thermodynamicProcessAssumption} and the coarse-grained thermal transport model in~\eqref{eq: coarse_grained_heat_flux} yield the solution of a generic thermomechanical deformation problem subject to a non-uniform temperature distribution and loading and boundary conditions, as illustrated in Figure~\ref{fig: SchematicFiniteTemperatureQC}.
\Revision{Convergence of the force-based summation rule was analysed by~\citet{knap2001analysis} for repatom forces. Hence, equation~\eqref{eq: coarse_grained_heat_flux} converges to \eqref{eq: entropy_kinetic_linear_onsager} as the coarse-grained mesh is refined down to atomistic resolution (weights $\hat{w}_k$ and $w_{\alpha}$ approach unity, and the dependence on all sampling atoms excluding the repatom vanishes). Consequently, even in the coarse-grained regions, equation~\eqref{eq: coarse_grained_heat_flux} approximates the atomistic thermal transport model governed by \eqref{eq: entropy_kinetic_linear_onsager}. As shown in Section~\ref{sec: Dissipation}, the linear Onsager kinetic model approximates Fourier's law-type thermal transport for a relaxed crystal, since the temperature reaches a linear distribution at steady state and the macroscopic flux reaches a constant value for differentially heated boundaries (neglecting any deformation dependence of the thermal conductivity). Since the deformation in large coarse-grained elements in a QC simulation is expected to be small, the coarse-grained equation~\eqref{eq: coarse_grained_heat_flux} is expected to approximate Fourier's law-type thermal transport sufficiently well.}

\subsection{Updated-Lagrangian QC implementation}
\begin{figure}[!t]
  \centering
 {\includegraphics[width = \textwidth]{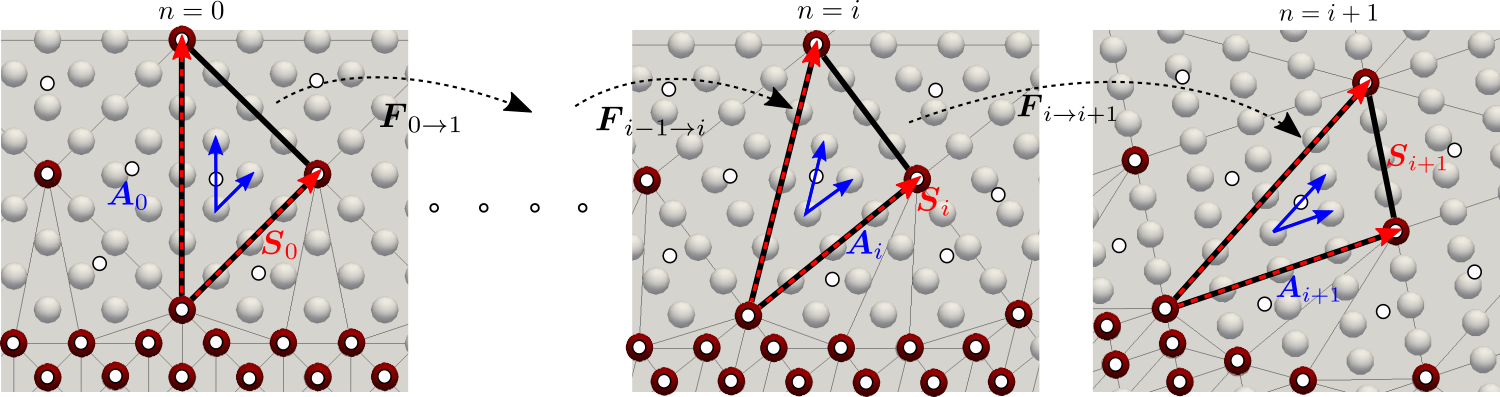}}
  \caption{Illustration of the updated-Lagrangian QC implementation at different external load/strain steps denoted by $n$. The local Bravais basis $\bfA_n$ of the highlighted element is shown in blue, with the edge vectors $\bfS_n$. Repatoms are shown as red circles, sampling atoms as small white circles. Deformation gradient $\bfF_{i\to i+1}$ deforms the edge vectors $\bfS_i$ to $\bfS_{i+1}$ and the local Bravais basis $\bfA_i$ to $\bfA_{i+1}$.}
  \label{fig: updated_lagrangian}
\end{figure}

We implement the thermomechanical local equilibrium relations \eqref{eq: mechanical_QC_interpolation} and \eqref{eq: thermal_QC_interpolation} combined with a thermodynamic constraint from Table~\ref{tab: thermodynamicProcessAssumption} and the coarse-grained thermal transport equation~\eqref{eq: coarse_grained_heat_flux} in an updated-Lagrangian QC setting. The latter is chosen since atoms in regions undergoing large deformation tend to have significant neighborhood changes, for which the initial reference configuration loses its meaning in the fully-nonlocal QC formulation, as illustrated by~\citet{amelang2016fully} and \citet{tembhekar2017automatic}. Consequently, tracking the interatomic potential neighborhoods in the undeformed configuration incurs high computational costs. Alternatively, one could strictly separate between atomistic and coarse-grained regions (as in the local-nonlocal QC method of \citet{tadmor1996quasicontinuum}), yet even this approach suffers from severe mesh distortion in the coarse-grained regions in case of large deformation, and it furthermore does not easily allow for the automatic tracking of, e.g., lattice defects with full resolution \citep{tembhekar2017automatic}. It also requires a-priori knowledge about where full resolution will be required during a simulation. As a remedy, we here deform the mesh with the moving repatoms and we take the deformed configuration from the previous load step as the reference configuration for each new load step, thus discarding the initial configuration and continuously updating the reference configuration.

For every element $e$, we store the three initial edge vectors (i.e., three node-to-node vectors forming a right-handed system) in a matrix $\bfS^e_0$, and the three Bravais lattice vectors indicating the initial atomic arrangement within the element in a matrix $\bfA^e_0$. As the system is relaxed quasistatically under applied loads, all repatoms move to the deformed configuration (e.g., from load step $n=i$ to $n=i+1$), thus deforming the edge vectors of element $e$ from $\bfS^e_{i}$ to $\bfS^e_{i+1}$ (and likewise the Bravais basis from $\bfA^e_i$ to $\bfA^e_{i+1}$\Revision{, see Figure~\ref{fig: updated_lagrangian}}). The incremental deformation gradient of element $e$, from step $i$ to $i+1$, can hence be computed from the kinematic relation
\begin{equation}
\bfF^e_{i\rightarrow{i+1}} = \bfS^e_{i+1} \left(\bfS^{e}_{i}\right)^{-1},
\label{eq: element_kinetmatics}
\end{equation}
which assumes an affine deformation within the element due to the chosen linear interpolation (see Figure~\ref{fig: updated_lagrangian}). As the element deforms, the lattice vectors also deform in an affine manner:
\begin{equation}
\bfA^e_{i+1} = \bfF^e_{i\rightarrow{i+1}} \bfA^e_{i}.
\label{eq: lattice_kinetmatics}
\end{equation}
Consequently, the integer matrix $\bfN$, which contains the numbers of lattice vector hops along the element edges, evaluated as
\begin{equation}
\bfN^e_{i} = \bfS^{e}_{i}\left(\bfA^{e,}_{i}\right)^{-1} = \mathrm{const.},
\label{eq: constant_number_matrix}
\end{equation}
remains constant throughout deformation of a given element $e$. Moreover, each element edge has a constant number vector, denoted by the rows of $\bfN^e_i$ (see Figure~\ref{fig: updated_lagrangian}). That is, in the updated-Lagrangian setting, the number matrix $\bfN^e_{i}$ remains constant during deformation. Such conservation of lattice vector hops along the element edges/faces is particularly useful for adaptive remeshing scenarios, where existing elements may need to be removed and new elements need to be reconnected, with or without changes to the number of lattice sites used for re-connections. The conservation of lattice vector hops can then be used for computing the Bravais lattice vectors local to new elements. The Bravais lattice vectors are used for calculating the neighborhoods of the nodal and centroid sampling atoms belonging to the large elements in the fully nonlocal QC formulation. The local lattice is generated within a threshold radius distance from the sampling atom using those lattice vectors. We \Revision{adopt} the adaptive neighborhood calculation strategy of~\citet{amelang2016fully}, which requires larger threshold radii compared to the interatomic potential cut-off chosen as 
\begin{equation}
r_{th} = r_\text{cut} + r_\text{buff},
\label{eq: thresholdRadius}
\end{equation}
where $r_\text{buff}$ is a buffer distance used for triggering re-computations of neighborhoods, and $r_\text{cut}$ is the interatomic potential cut-off. If the maximum relative displacement among the neighbors with respect to a sampling atom exceeds $r_\text{buff}$, then neighborhoods of the sampling atom are re-computed (see~\citet{amelang2016fully} for details).

Within the region with atomistic resolution, only nodal sampling atoms have finite weights (close to unity) and hence only their neighborhoods are computed. For such neighborhood calculations Bravais lattice vectors are not required. Instead, the unique nodes of all elements within the threshold radius of the sampling atom are added as the neighbors. Consequently, even severely deforming meshes do not require element reconnection/remeshing as long as the deformation \Revision{remains} restricted within the atomistic region, since only nodes of the elements are required. Hence, we use meshes with large atomistic regions in the benchmark cases presented below, to restrict the analysis towards thermodynamics of the deformations. Such simulations do not require adaptive remeshing, the analysis of which is left for future studies.

\subsection{Thermal effects on the shear activation of dislocations}

As a benchmark example, we use the updated-Lagrangian QC method discussed above to analyze the effects of temperature on dislocations and specifically on edge dislocation nucleation under an applied shear stress. We present the analysis for both cases of isothermal and adiabatic \Revision{constraints} (the latter combined with the irreversible entropy transport based on linear Onsager kinetics). \Revision{The adiabatic constraint here signifies that the simulation domain is thermally isolated from the surroundings and there is no heat exchange between the domain boundaries and the surroundings.} For both cases, we generate a pair of dislocations (i.e., a dislocation dipole) using the isotropic linear elastic displacement field solutions of edge dislocations with opposite Burgers' vectors \citep{nabarro1967theory}, \Revision{given by
\begin{subequations}
\begin{align}
 u_1 &= \frac{b}{4\pi\left(1-\nu\right)}\frac{x'_1x'_2}{x'^2_1 + x'^2_2} - \frac{b}{2\pi}\tan^{-1}\left(\frac{x'_1}{x'_2}\right), \\
 u_2 &= -\frac{(1 - 2\nu)b}{8\pi\left(1-\nu\right)}\ln\left(\frac{x'^2_1 + x'^2_2}{b}\right) + \frac{b}{4\pi\left(1-\nu\right)}\frac{x'^2_2}{x'^2_1 + x'^2_2},
 \label{eq: elastic_solutions}
 \end{align}
\end{subequations}
superposed linearly {onto a $32\times25\times1.8$ lab of pure single-crystalline Cu, consisting of 125,632 lattice sites, whose edges are oriented along the slip crystallographic directions. In \eqref{eq: elastic_solutions}, $x'_1$ and $x'_2$ denote the coordinates along the $[110]$- and $[\overline{1}11]$-axes, respectively, with respect to the dislocation centers; $u_1$ and $u_2$ are the displacements along these axes, and $b = \pm\frac{a}{\sqrt{2}}$ is the magnitude of the Burgers' vector}. } 

\begin{figure}[!b]
 \centering
 {\includegraphics[width=\textwidth]{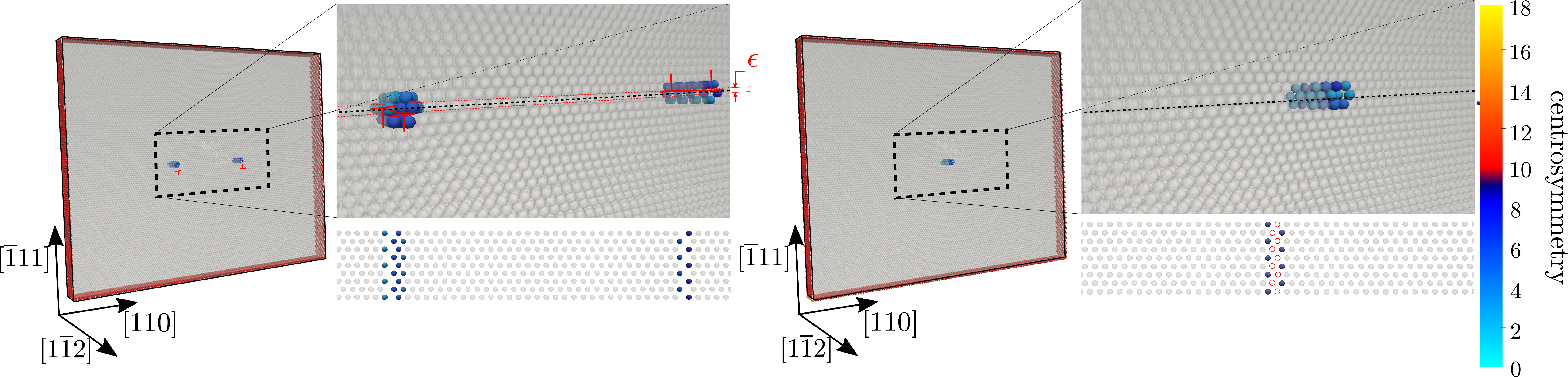}}
\put(-490,120){$(a)$}
 \put(-245,120){$(b)$}
 \caption{Initial conditions of the dislocation dipole setup. $(a)$ Initial condition after displacing the atoms according to the isotropic linear elastic displacement field solution.
 \Revision{Due to the separation $\epsilon>0$ between the slip planes of two dislocation, a line of atoms at the leftmost end remains unaffected in the initial condition and is removed from the domain, thus initiating a void.} $(b)$ Isothermally relaxed state consisting of a void (vacancy column in 3D) created by the annihilation of the dislocations. Shown are 3D views of the full simulation domain with a magnified view of the fully-resolved central region and a top view. Atoms are colored by the centrosymmetry parameter in arbitrary units and shown between threshold values of $2$ to $10$.}
 \label{fig: DislocationDipoleX_Initial}
\end{figure}

\begin{figure}[!b]
 \centering
 {\includegraphics[width=0.95\textwidth]{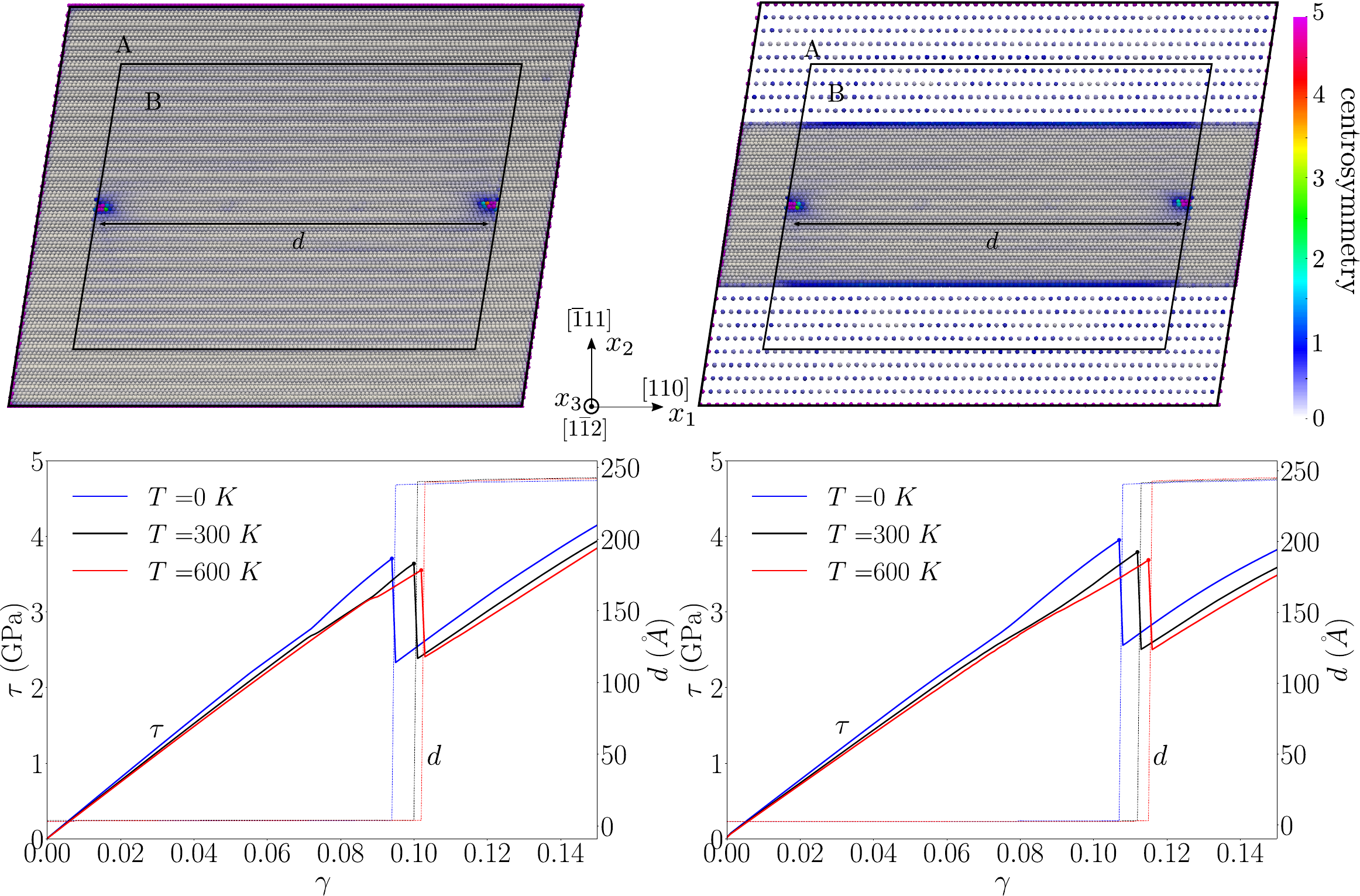}}
  \put(-460,300){$(a)$}
   \put(-235,300){$(b)$}
 \put(-470,140){$(c)$}
  \put(-245,140){$(d)$}
 \caption{Comparison of the isothermal nucleation of dislocation dipoles from a single-atom void, as obtained from fully \Revision{resolved} (left) and QC (right) simulations at varying temperatures in a Cu single-crystal modeled using the EFS potential~\citep{dai2006extended}. Shown is the final sheared state of $(a)$ a snapshot of the \Revision{fully-resolved} simulation and $(b)$ that of the QC simulation. Atoms are colored by the centrosymmetry parameter in arbitrary units. While the atoms in region A are kept fixed, atoms in region B are allowed to relax. (c,d) The shear stress $\tau$ vs.\ the engineering strain $\gamma$ is plotted for $(c)$ the \Revision{fully-resolved} simulation and $(d)$ the QC simulation. The shear stress is evaluated as the net force in the $[110]$-direction on the atoms in region~A per cross-sectional area in the $(\overline{1}11)$ plane. Faces $(1\overline{1}2)$ are periodic.}
 \label{fig: DislocationDipoleX_Stress}
\end{figure}

\begin{figure}[!t]
 \centering
 {\includegraphics[width=0.55\textwidth]{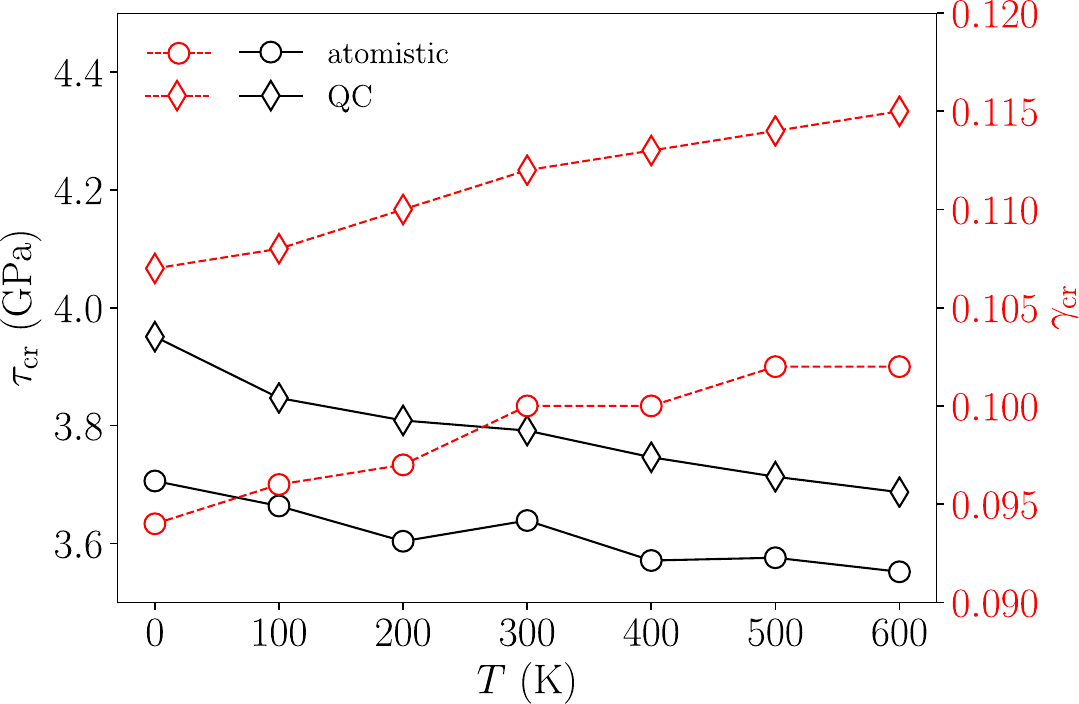}}
 \caption{Comparison of the critical shear stress $\tau_\text{cr}$ and strain $\gamma_{\mathrm{cr}}$ required to nucleate a dislocation dipole from the void as obtained from QC and from \Revision{fully-resolved} simulations at various temperatures. The critical strain $\gamma_{\mathrm{cr}}$ is the external shear strain $\gamma$ at which $\tau_{\mathrm{cr}}$ is achieved.}
 \label{fig: DislocationDipoleXCriticalStressComparison}
\end{figure}

\begin{figure}[!b]
 \centering
 {\includegraphics[width=\textwidth]{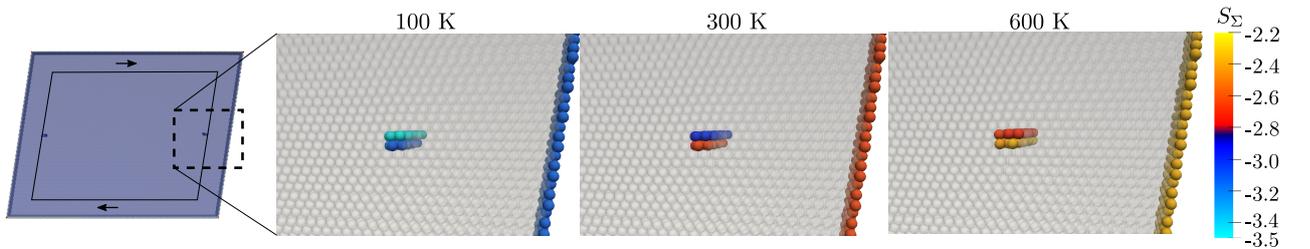}}
 \caption{Variation of the position entropy $S_{\Sigma}$ with temperature inside one of the dislocations nucleated from the void. The highlighted atoms are identified using the centrosymmetry parameter (values $>2$ are shown). As discussed in Section~\ref{sec: free_energy_minimization}, the number of neighbors (and their positions) affects the local interatomic potential of an atom, thus modifying the local variation of positions. Atoms within the dislocations that are closer than the equilibrium interatomic spacing have smaller position variance than those that are further apart. \Revision{Atoms at the boundaries have higher $\Sigma$-values due to smaller numbers of interacting neighbors at the free surface of the domain.}}
 \label{fig: SigmaInDislocations}
\end{figure}

\begin{figure}[!t]
 \centering
 {\includegraphics[width = 0.6\textwidth]{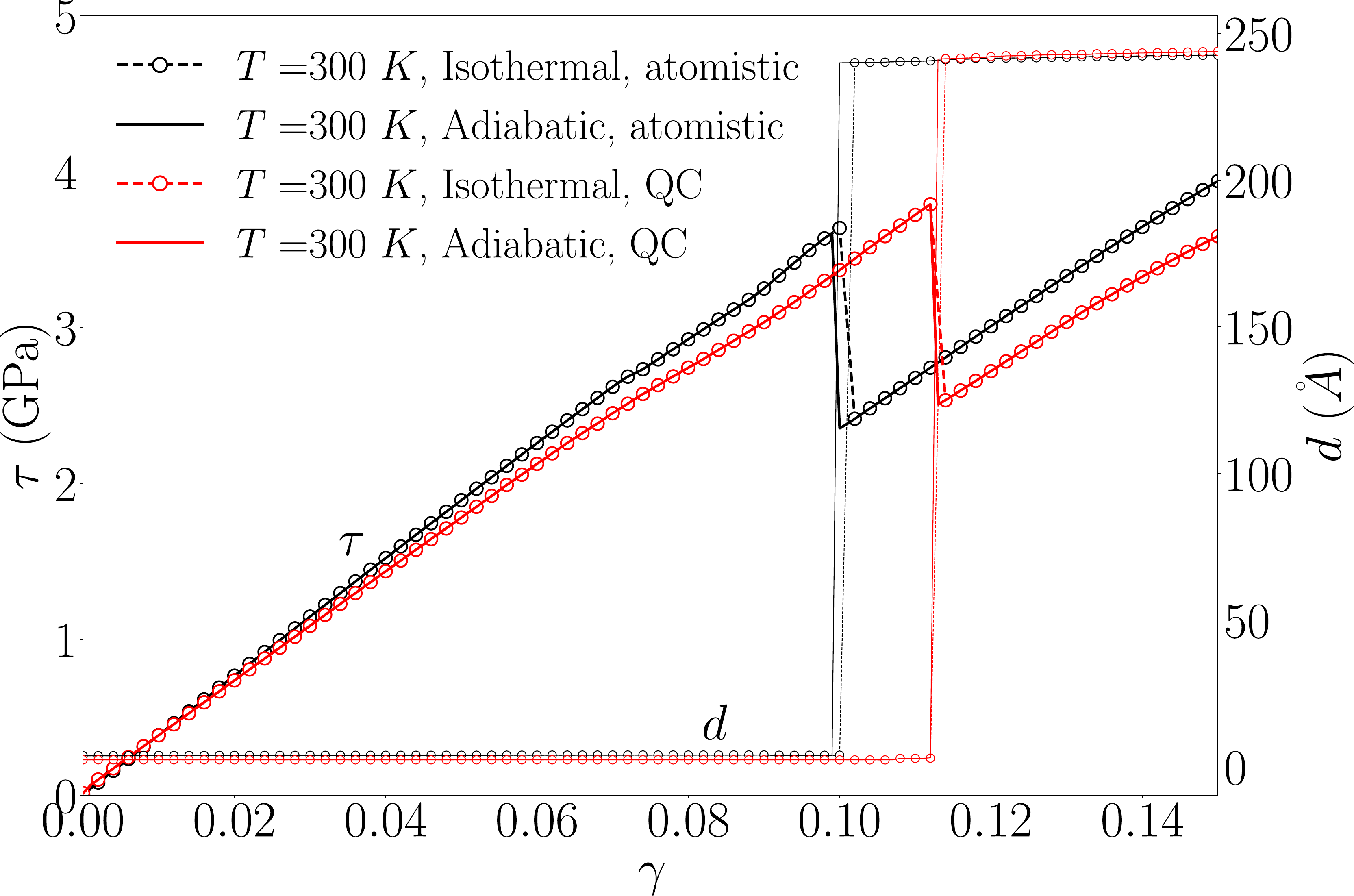}}
 \caption{Shear stress $\tau$ on the $(\overline{1}11)$-plane and dislocation separation distance $d$ vs.\ applied shear strain $\gamma$ for isothermal and adiabatic deformations, obtained from both atomistic and QC simulations.
 Note that the differences between isothermal vs.\ adiabatic data are small, because the temperature increase is not significant \Revision{in this example}. Critical shear stress value deviations are within 6\% (e.g., isothermal QC: 3.7914~GPa, isothermal atomistics: 3.6389~GPa) and critical strain values are within 12\% (isothermal QC: 0.112, isothermal atomistics: 0.100).}
 \label{fig: DislocationShearIsothermalVsAdiabatic}
\end{figure}

Figure~\ref{fig: DislocationDipoleX_Initial}$(a)$ shows the \Revision{thus-generated initial configuration of} the edge dislocation dipole with Burgers' vectors $\bfb = \pm\frac{a}{2}[110]$. separated by a \Revision{horizontal} distance of $80$~\AA. \Revision{The centers of the dislocations are separated by a distance of $80$~\AA~in the (horizontal) $[110]$-direction and a small distance of $1\times10^{-9}$~\AA in the (vertical) $[\overline{1}11]$-direction. Imposing such displacement fields causes the slip planes of the two dislocations to separate in $x_2$ and leaves a line of atoms on the $x_2 = 0$-plane at the left end of the domain with zero displacements. These atoms are removed from the simulation domain, thus creating void at the edge of the domain.} Displacements are restricted to the $(1\overline{1}2)$ plane, while the simulation domain is set up in 3D with periodic boundary conditions on opposite out-of-plane faces. After initial relaxation, the dislocations annihilate each other due to their interacting long-range elastic field. The result is a line defect in the form of a through-thickness (non-straight) vacancy column (in the following for simplicity referred to as a \textit{void}), as shown in Figure~\ref{fig: DislocationDipoleX_Initial}($b$). \Revision{This void was created in the initial configuration, when the non-displaced atoms were removed from the simulation domain, and it diffuses to the center of the domain during the initial relaxation. We note that this is a direct consequence of separating the slip planes of the two dislocations in our simulation. If the slip planes are identical, then the dislocations annihilate and form a perfect single-crystal. We deliberately create the void in this fashion, since it ensures that only two dislocations of opposite orientation are nucleated  when the domain is externally sheared. After relaxation, we load} the simulation domain in simple shear (moving the top and bottom faces relative to each other), while computing the effective applied shear stress from the atomic forces. \Revision{Periodic boundary conditions are imposed on $\left(1\overline{1}2\right)$-surfaces, while the rest of the boundaries are included within region~A, which is mechanically fixed during relaxation. At sufficient applied shear, the void} will nucleate and emit a dislocation dipole, whose activation energy and behavior depend on temperature. For an assessment of the accuracy of the QC framework, we carry out both fully \Revision{resolved} (125,632 atoms) and QC simulations (52,246 repatoms) in isothermal and adiabatic settings. \Revision{QC simulations are performed on a mesh generated by coarse-graining in the $x_2$-direction. All three lattice vectors are expanded by a factor of $4$ in the coarse-grained region. The atomistic region extends fully in the $x_1$- and $x_3$-directions and up to $\pm 51$~\AA in the $x_2$-direction. Coarse-graining is performed only in the $x_2$-direction to prevent the dislocations colliding with the atomistic and coarse-grained subdomain interface.} We acknowledge that the QC setup is relatively simple and there is not yet a significant reduction in the total number of degrees of freedom nor does it involve automatic mesh refinement. Yet, this study presents a simple and instructive example highlighting the efficacy and accuracy of the GPP-based QC formulation introduced in previous sections.

\subsubsection{Isothermal}

In face-centered cubic (FCC) crystals, edge dislocations preferably glide on the close-packed crystallographic $\{\overline{1}11\}$-planes \citep{hull2001introduction}. As the \Revision{initial void} is strained under shear deformation, dislocations nucleate from the void at a sufficient level of applied shear, propagating in opposite directions, as shown in Figure~\ref{fig: DislocationDipoleX_Stress}. We apply a shear deformation to all repatoms in the slab such that, at the $n^\mathrm{{th}}$ load step, 
\begin{equation}
\lavg{q}^{(n)}_{k,1} = \lavg{q}^{(n-1)}_{k,1} + \Delta\gamma\,\lavg{q}^{(n-1)}_{k,2},
\end{equation}
where indices $1$ and $2$ refer to the respective components of the mean position vector in the chosen coordinate system, and $\Delta\gamma$ is the applied shear strain increment. As the strain is applied, repatoms in the inner region~B (Figure~\ref{fig: DislocationDipoleX_Stress}) are relaxed, while keeping those in the outer region~A mechanically fixed to impose the average shear strain. Note that, due to small deformation in the atomic neighborhoods, the displacement-variance entropy $S_{\Sigma}$ of repatoms close to the interface between regions A and B changes and, hence, all repatoms in the domain are thermally relaxed assuming an \textit{isothermal} relaxation (cf.~Table~\ref{tab: thermodynamicProcessAssumption}). While the shear strain is increased, the horizontal component of the force on all repatoms in region~A is computed. The effective shear stress $\tau$ on the $(\overline{1}11)$-plane is computed by normalizing the net horizontal force by the cross-sectional area of the slab. Results are shown in Figure~\ref{fig: DislocationDipoleX_Stress}$(c)$ and $(d)$. Once the stress reaches a critical value, the stress drops as dislocations nucleate from the void and move to the ends of region~B. We observe that the critical stress value decreases slowly with temperature (see Figure~\ref{fig: DislocationDipoleXCriticalStressComparison}). Moreover, the value of the critical stress obtained from a fully atomistic simulation and the quasicontinuum simulation are within about 6\% of each other (see Figure~\ref{fig: DislocationDipoleXCriticalStressComparison}), both capturing the apparent thermal plastic softening \Revision{typical for Cu \citep{ShimEtAl2016}}.

\begin{figure}[!b]
 \centering
 {\includegraphics[width = \textwidth]{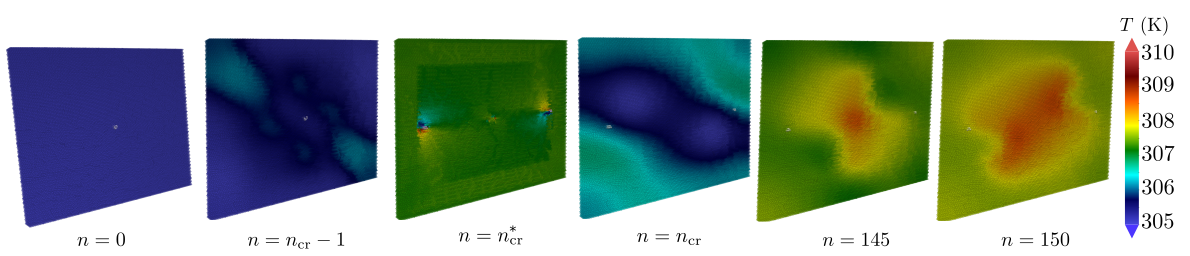}}
 \caption{Local variation of temperature as the slab with void is deformed adiabatically. In the intial stages, the temperature rises slowly due to the external deformation. At the critical shear strain $\gamma_{\mathrm{cr}} = n_{\mathrm{cr}}\Delta \gamma$, dislocations nucleate from the void and move along the $[110]$-direction. The temperature rise of those atoms within the dislocations causes a rapid increase in the temperature of the slab, as may be expected from the heat generated by work hardening. The temperature of a few atoms within the dislocations at the intermediate stage $n=n^*_{\text{cr}}$ before the irreversible transport exceeds the color bar range.}
\label{fig: DislocationActivation_Atomistic_Adiabatic}
\end{figure}

\citet{miller2009unified} studied a similar 2D scenario with a different crystal orientation, in which the dislocation dipole is stable and the dislocations do not annihilate to form the void. In such a case, the (theoretical) critical shear stress corresponds to the shear stress required to cause dislocation movement in the crystallographic plane. \Revision{By contrast}, in our analysis (which is based on a \Revision{more} realistic crystallographic setup \Revision{since the slip planes of the dislocations are close-packed planes of the $\{111\}$-family), the critical shear stress is} defined as the shear stress required to nucleate dislocations from the void-like defect.

\subsubsection{Adiabatic}

To simulate the \Revision{quasistatic} adiabatic \Revision{nucleation} of dislocations under shear, we repeat the above simulations, now with all repatoms in the domain being relaxed \textit{isentropically} (cf.~Table~\ref{tab: thermodynamicProcessAssumption}), followed by the thermal transport model according to the steps discussed in Section~\ref{sec: Dissipation} and Algorithm~\ref{alg: thermomechanical_algo}. \Revision{As noted above, the term \emph{adiabatic} refers to a thermally insulated scenario, in which the domain boundaries do not allow for heat exchange with the surroundings.} We further assume that the strain rate is significantly lower than the time scale imposed by the molecular thermal transport, thus imposing quasistatic conditions for the transport ($\delta t^{(n)}\to \infty$ in Algorithm~\ref{alg: thermomechanical_algo}, see Step~3 at the end of Section~\ref{sec: Dissipation}). 
The initial condition (again, prepared using the isotropic elastic displacement field solutions) is relaxed isothermally at $300$~K, before the adiabatic deformation begins. We compare the adiabatic deformation with the isothermal deformation of the slab at $300$~K. Figure~\ref{fig: DislocationShearIsothermalVsAdiabatic} shows the variation of the shear stress on the $(\overline{1}11)$-plane with external shear strain $\gamma$. Due to the mechanical work done by the external shear deformation, the temperature of the slab increases slightly, causing apparent softening compared to the isothermal deformation. Figure~\ref{fig: DislocationActivation_Atomistic_Adiabatic} shows the spatial variation of temperature as the slab is deformed adiabatically. Before the critical strain, heating caused by local deformation is negligible. As the dislocations are nucleated at the critical strain $\gamma_{\mathrm{cr}}$, the temperature of those atoms around the dislocations changes significantly, as shown by the intermediate stage $(\lavg{\bfq}^*, T^*)$ in Figure~\ref{fig: DislocationActivation_Atomistic_Adiabatic}. \Revision{Due to quasistatic thermomechanical deformation, the temperature field is homogenized to within 1 K, even after dislocation nucleation from the void. Such close-to-isothermal deformation at slow strain rates was previously observed by~\citep{ponga2016dynamic} (who studied strain-rate effects on nano-void growth in magnesium) and by \citep{ponga2018unified} (who studied thermomechanical deformation of carbon nanotubes.} Further plastic deformation causes increased heating of the slab, particularly due to the restricted dislocation motion beyond the interfaces between regions A and B (Figure~\ref{fig: DislocationDipoleX_Stress}). 

As noted above, the critical stress values obtained from QC simulations are within $6$\% of those obtained from the fully \Revision{resolved} simulations. Furthermore, the critical strain values are within $12$\%. Repatoms on the $(110)$- and $(\overline{1}11)$-surfaces, which includes the repatoms on vertices of large elements and the repatoms in the transition region between regions A and B, exhibit both mechanical and thermal spurious forces in 3D~\citep{amelang2015summation,AmelangKochmann2015, tembhekar2017automatic}. These artifacts are expected to be the primary sources of error in the coarse-graining strategy adopted here. Mechanical spurious forces within the energy-based fully nonlocal QC setup were discussed in detail in \citet{amelang2015summation}, and thermal spurious forces are expected to show an analogous behavior. Therefore, we do not study spurious forces here in detail
to maintain the focus on the thermomechanics of the GPP formulation.

\begin{figure}[!t]
 \centering
 {\includegraphics[width = \textwidth]{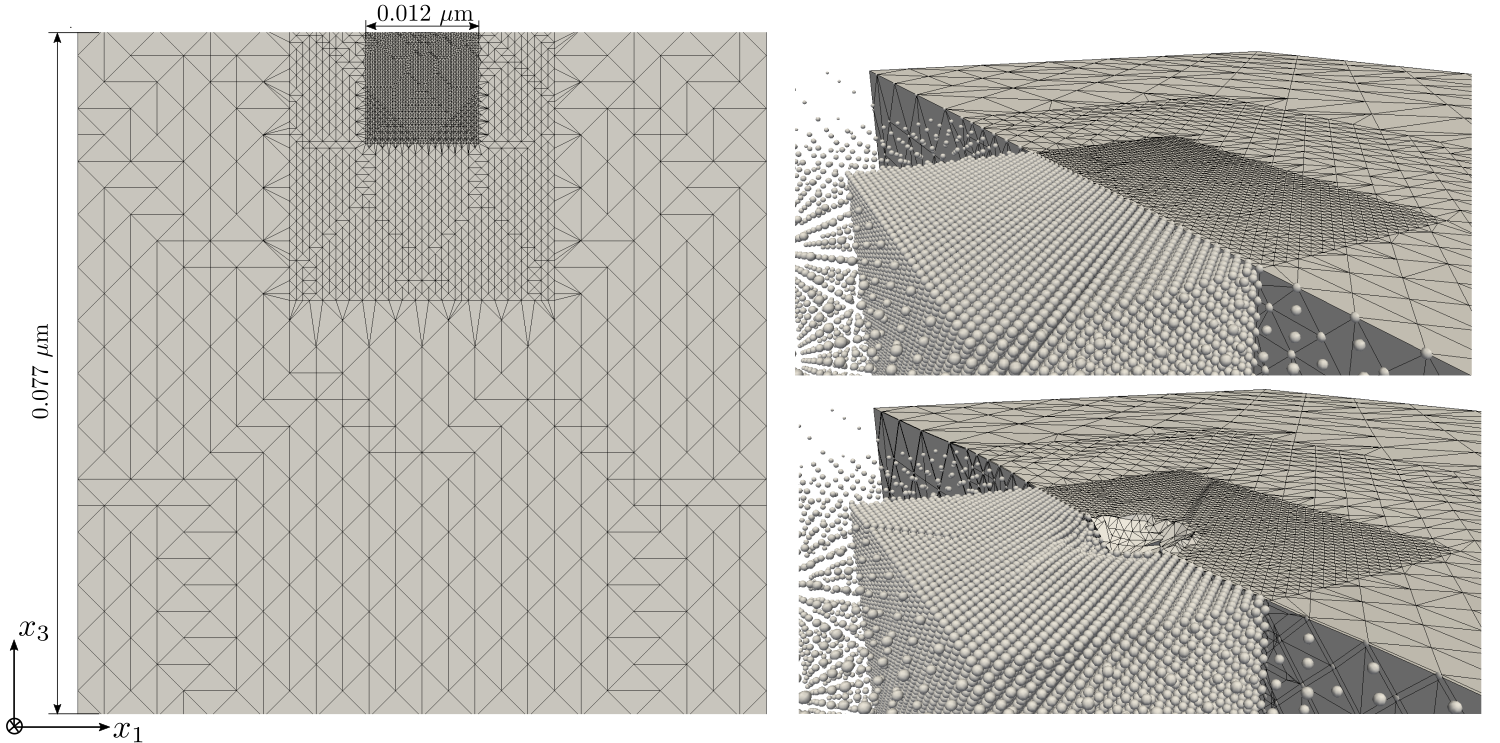}}
 \put(-485,235){$(a)$}
  \put(-235,235){$(b)$}
 \put(-235,105){$(c)$}
 \caption{Illustration of the undeformed and deformed QC meshes of a $0.077$~$\mu$m FCC single-crystal of pure Cu. $(a)$ Cross-section of the initial, undeformed mesh, $(b, c)$ zoomed-in perspective views of the atomistic region (33$\times$33$\times$33 unit cells) and the surrounding coarsened regions in the ($b$) undeformed and $(c)$ deformed configurations underneath a 5~nm spherical indenter at an indentation depth of $0.75$~nm.}
\label{fig: QC_nanoindentation_mesh}
\end{figure}

\begin{figure}[!b]
 \centering
 {\includegraphics[width = 0.7\textwidth]{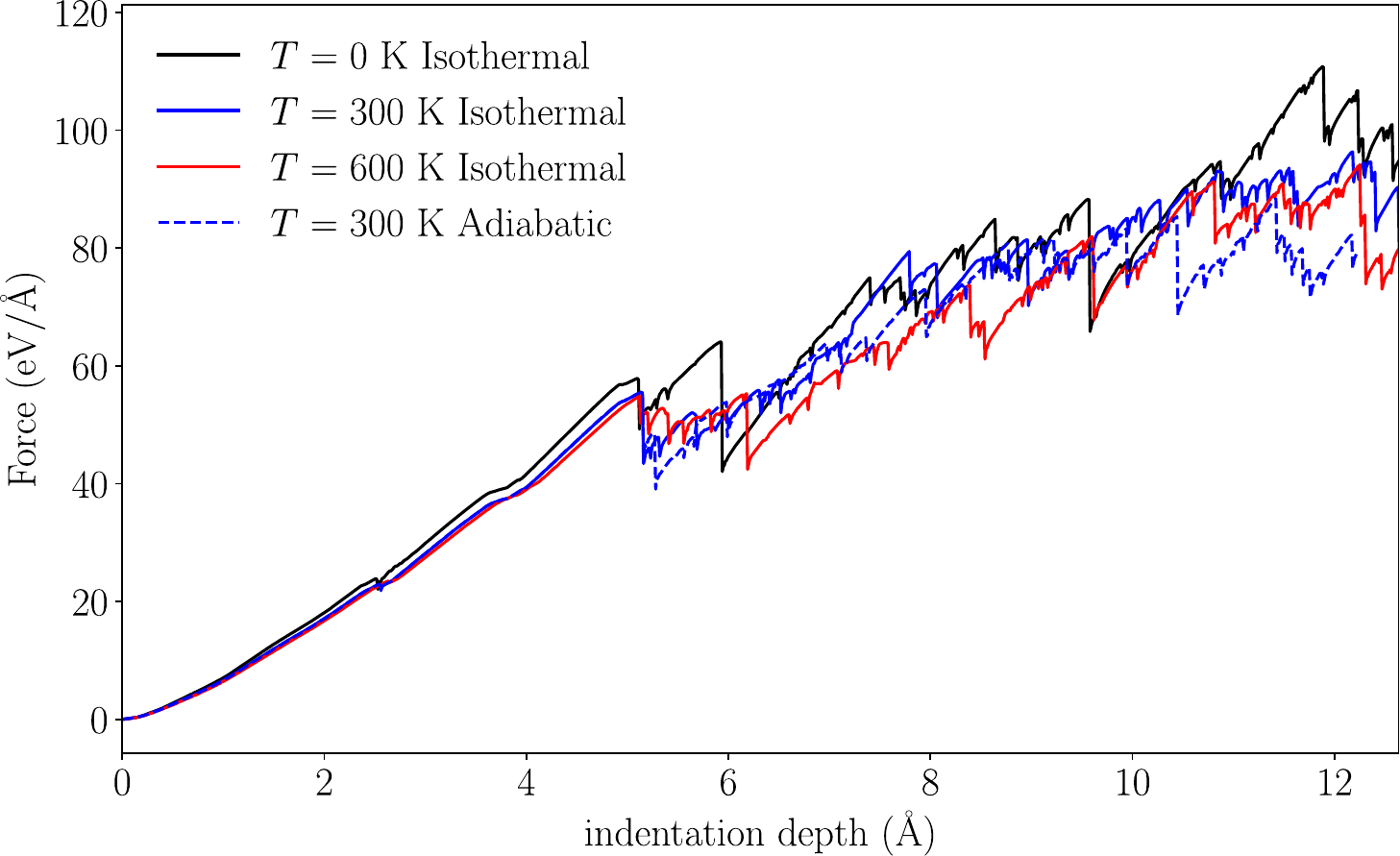}}
 \caption{Variation of the indenter force with the indenter depth for isothermal ($T=0~\mathrm{K},~300~\mathrm{K},~600~\mathrm{K}$) and adiabatic (initially at $T=300~\mathrm{K}$) conditions. After an initial elastic \Revision{regime}, the curve shows the typical serrated behavior due to dislocation activity underneath the indenter.}
 \label{fig: FvsIndenter}
\end{figure}

\subsection{Thermal effects on nanoindentation of copper}

\begin{figure}[!t]
 \centering
 {\includegraphics[width = \textwidth]{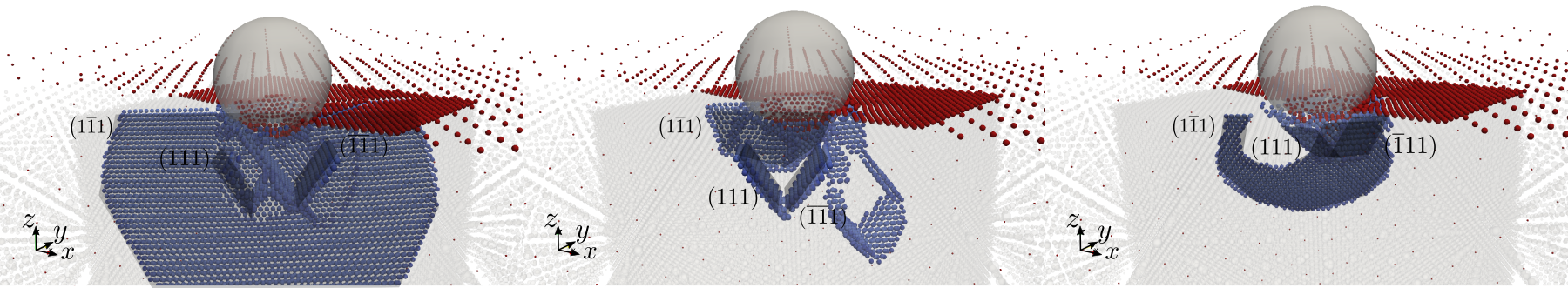}}
 \caption{Microstructure generated below the $5$~nm spherical indenter at an indentation depth of $1$~nm for \textit{isothermal} deformation of a Cu single-crystal. The generated dislocations move towards the boundaries of the atomistic region, creating stacking faults in the crystallographic planes of the $\{111\}$-family (shaded in gray). Red repatoms are top-surface repatoms and blue atoms denote those within the microstructure, identified using the centrosymmetry parameter (values $>5$ identify \Revision{surface} and microstructure atoms only). All repatoms are shown with reduced opacity for comparison of the size of the microstructure with the atomistic domain. 
 }
 \label{fig: microstructureIsothermal}
\end{figure}

\begin{figure}[!b]
 \centering
 {\includegraphics[width = \textwidth]{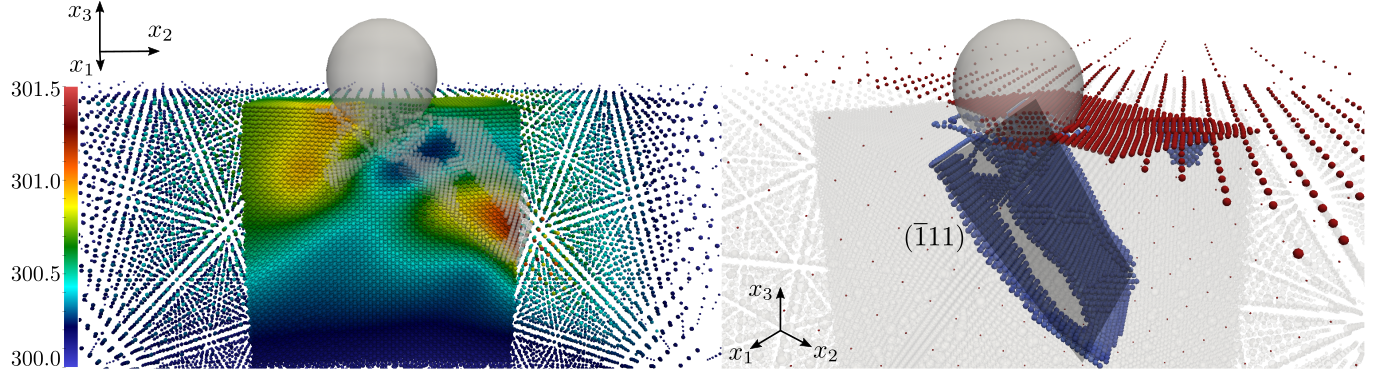}}
 \put(-490, 135){$(a)$}
 \put(-240, 135){$(b)$}
 \caption{Adiabatic deformation of the Cu single-crystal under a spherical indenter. $(a)$ Spatial variation of temperature in the cross-section of the atomistic region. As dislocations and stacking faults are created, local thermal gradients are generated which are diffused via the thermal transport model.  $(b)$ Microstructure generated below the spherical indenter at an indentation depth of $1$~nm. The generated dislocations move towards the boundaries of the atomistic region, creating stacking faults in the crystallographic planes of the $\{111\}$-family (shaded in gray). Red repatoms are top-surface repatoms and blue atoms denote those within the microstructure, identified using the centrosymmetry parameter (values $>5$ identify \Revision{surface} and microstructure atoms only). Shaded repatoms denote all repatoms with reduced opacity, shown here for comparison of the size of the microstructure with the atomistic domain.}
 \label{fig: microstructureAdiabatic}
\end{figure}

Finally, we apply the thermomechanical transport model to the case of nanoindentation into a Cu single-crystal. While the problem is well studied in 2D using finite-temperature QC implementations~\citep{tadmor2013finite}, only few QC studies exist that study finite-temperature effects in 3D. \citet{kulkarni2007coarse} studied nanoindentation of a $32\times 32\times 32$ unit cell FCC nearest-neighbor Lennard-Jones crystal, using the Wentzel-Kramers-Brillouin (WKB) approximation, which captures the thermomechanical coupling at comparably low temperature only. We here perform nanoindentation simulations of a Cu cube of $0.077$~$\mu$m side length (approximately $215\times215\times215$ unit cells, see Figure~\ref{fig: QC_nanoindentation_mesh}), modeled by the EAM potential of~\citet{dai2006extended}, underneath a $5$~nm spherical indenter modeled using the indenter potential of~\citet{kelchner1998dislocation} with a force constant of $1000$~eV/\AA$^3$ and a maximum displacement of $1.26$~nm or approximately $3.5$~times the lattice parameter at $0~$K. The crystal consists of approximately 50 million atoms, represented in the QC framework by approximately 0.2~million repatoms. The top surface is modeled as a free \Revision{surface}, while all other boundaries suppress wall-normal displacements, allowing only in-plane motion. 
Below, we discuss the results for isothermal deformation at $T=0~\mathrm{K},~300~\mathrm{K},~600~\mathrm{K}$ and for quasistatic adiabatic deformation, the latter being initially at $T=300~\mathrm{K}$. 
Figure~\ref{fig: FvsIndenter} shows the variation of the total force on the spherical indenter vs.\ indentation depth for both isothermal and quasistatic adiabatic conditions. The force increases nonlinearly with indentation depth, showing the typical Hertzian-type initial elastic section. With increasing indentation depth, atoms underneath the indenter generate dislocations and stacking faults, overall creating a complex microstructure \Revision{consisting of prismatic dislocation loops (PDLs), as also observed by~\citet{ponga2015finite} in nano-void growth in Cu. These PDLs are created in the $\{111\}$ slip planes of the crystal}, as shown in Figure~\ref{fig: microstructureIsothermal}. At the first dislocation activation, the indenter force drops after reaching a critical force. This critical force decreases with increasing temperature, indicating plastic softening. The dislocations move towards the boundaries of the atomistic region, gliding in the crystallographic planes of the $(111)$-family, giving way to stacking faults in those planes. As shown in Figure~\ref{fig: microstructureIsothermal}, while the initial dislocations maintain their structure, the stacking fault structure changes significantly at the same indentation depth as temperature increases.

Figure~\ref{fig: microstructureAdiabatic} shows the spatial temperature distribution and the emergent microstructure during the quasistatic adiabatic deformation of the Cu single-crystal. With the dislocations, local temperature gradients are generated \Revision{along the PDLs generated in the $\{111\}$ slip planes} due to large gradients in $S_{\Sigma}$, which are triggered due to large deviations from a centrosymmetric neighborhood (\Revision{as identified in Figure~\ref{fig: DislocationActivation_Atomistic_Adiabatic}}). These temperature gradients are diffused as a result of the thermal transport. 
We note that, for a thorough quantitative analysis, one may want to obtain results averaged over multiple simulations with initial conditions and/or indenter position slightly perturbed, since the emergence of microstructure below the indenter within the highly-symmetric single-crystal is associated with instability and strongly depends on local fluctuations and initial conditions. Such a statistical analysis is deferred to future work.

\section{Conclusion and discussion}
\label{sec: Conclusions}

We have presented a Gaussian phase packets-based (GPP-based) formulation of finite-temperature equilibrium and nonequilibrium thermomechanics applied to atomistic systems. \Revision{Application of the Liouville equation to an ansatz of the probability distribution yields equations of motion for all phase-space parameters, thus aiming for temporal upscaling by separating the slow mean motion of atoms from their statistical position and momentum variances at finite temperature. For a general parametrization (or a sequence of ans\"atze) of the Hamiltonian systems, we have presented the variational structure and weak formulation of the Liouville equation, which alternatively yields the phase-space parameter evolution equations. As a special choice, we} have shown that approximating the global statistical distribution function by a multivariate Gaussian \Revision{ansatz can capture thermal transport at the atomic scale} only via interatomic correlations. Due to high computational costs, we have neglected \Revision{such} interatomic correlations, which results in a local GPP approximation of the system. Such a system exhibits reversible dynamics with thermomechanical coupling, causing local heating and cooling upon movement of atoms relative to local neighborhoods. Moreover, in the quasistatic limit, we have shown that the equations yield local mechanical and thermal equilibrium conditions, the latter yielding the local equation of state of the atoms based on the interatomic force field. \Revision{We have shown that the thus-obtained equilibrium relations are identical to those of the maximum-entropy (\textit{max-ent}) ansatz (even though the latter postulates a probability distribution in the quasistatic limit). However, unlike max-ent, our approach yields the full dynamic evolution equations, thereby allowing us to gain insight into the thermomechanical evolution in phase space. We furthermore derived different constraint conditions to be applied to the equilibrium relations for modeling isothermal, isobaric, and isentropic processes. The setup has been validated by computing thermal expansion and elastic constants of a copper single-crystal.}

To capture the irreversibility due to local thermal transport triggered by the adiabatic heating/cooling of atoms, we have coupled the quasistatic framework with linear Onsager kinetics. Such a model involves an empirical coefficient fitted to \Revision{match} approximate bulk conductivity measurements and captures the experimentally observed size-effects of the thermal conductivity, as shown by \cite{venturini2014atomistic}. Moreover, we have shown that the time scale imposed by the atomic-scale transport is \Revision{approximately 100 times} that of atomic vibrations. While the atomic-scale thermal transport imposes a small time scale as the system reaches a non-uniform steady state, the local heat flux imbalance decreases. Below a tolerance value, the heat transport can be terminated, yielding a steady-state solution. Based on the global multivariate Gaussian \Revision{ansatz}, interatomic correlations may be fitted to obtain correlation functions (akin to interatomic potentials), which can help develop the transport constitutive properties of atomistic systems and also advance current understanding of the long-standing nanoscale thermal transport problem. 

Finally, we have combined the quasistatic thermomechanical equations based on the local GPP approximation with thermal transport in a high-performance, distributed-memory, updated-Lagrangian 3D QC solver, which is capable of modeling thermomechanical deformation of large-scale systems by coarse-graining the atomistic ensemble in space. Benchmark simulations of dislocation nucleation and nanoindentation under isothermal and adiabatic \Revision{constraints} showed \Revision{reasonable} agreement between coarse-grained and \Revision{fully resolved} simulations. Since the time integration of the atomic transport can be terminated for a small heat flux imbalance (discussed in Algorithm~\ref{alg: thermomechanical_algo}), the quasistatic simulations offer significant advantages over traditional MD studies, which can tackle only high strain rates. The presented methodology of coupling local thermal equilibrium with a surrogate empirical model of thermal transport and spatial coarse-graining (by the QC method) can model deformation of large crystalline systems at mesoscales and at quasistatic loading rates. Due to the time-scale limitations of MD, a one-to-one comparison of the presented simulations with finite-temperature MD simulations is prohibitively costly \Revision{(except for equilibrium properties like the presented thermal expansion coefficient)}. A detailed analysis of the accuracy of the spatial coarse-graining of the thermomechanical model presented here, and a comparison with suitable MD simulations qualifies as a possible extension of this work. For such comparisons, however, large-scale nonequilibrium molecular dynamics (NEMD) simulations are required at sufficiently slow strain rates\Revision{, which go beyond the scope of this study}.

\section*{Acknowledgments}
The support from the European Research Council (ERC) under the European Union’s Horizon 2020 research and innovation program (grant agreement no.~770754) is gratefully acknowledged. The authors thank \Revision{Miguel Angel Spinola Fandila for help with the LAMMPS simulations used for numerical validation}.

\bibliographystyle{elsarticle-harv}
\bibliography{./references}

\appendix
\section{Time evolution of phase averaged quantities}
\label{sec: timeEvoPhaseAVG}
The time evolution of the phase average of a phase space quantity $A(\bfz)$ can be derived using the representative solution of the Liouville equation,
\begin{equation}
 \frac{\partial f}{\partial t} + i\mathcal{L}f = 0 \ \implies \ 
 f(\bfz, t) = e^{-i\mathcal{L}t}f(\bfz_0, 0) = e^{-i\mathcal{L}t}f(\bfz),
 \label{eq: Liouville_App}
\end{equation}
where $f(\bfz_0, 0)$ is the initial condition and $e^{-i\mathcal{L}t}$ is the propagating operator, which transforms the probability distribution initially defined at phase space coordinate $\bfz_0$ to the probability distribution at $\bfz(t)$~\citep{j2007statisticalEvans, zubarev1996statistical}. Furthermore, the time evolution of the phase space quantity $A(\bfz)$ is given by
\begin{equation}
 \frac{\dd A}{\dd t} = i\mathcal{L}A
 \ \implies \
 A(\bfz, t) =  e^{i\mathcal{L}t}A(\bfz_0, 0) = e^{i\mathcal{L}t}A(\bfz).
  \label{eq: PhaseQauntity_App}
\end{equation}
Equations~\eqref{eq: Liouville_App} and \eqref{eq: PhaseQauntity_App} reveal that the operators $e^{\pm i\mathcal{L}t}$ transport the probability distribution $f(\bfz)$ and phase space quantities $A(\bfz)$ defined in terms of $\bfz$ of a system of particles, given that $\bfz$ also changes in time as the system of particles evolves.
\Revision{
Operator $i\mathcal{L}$ satisfies the property
\begin{equation}
 \int_{\Gamma}A(\bfz)i\mathcal{L}f(\bfz)d\bfz = \int_{\Gamma}(-i\mathcal{L})A(\bfz)f(\bfz)d\bfz
 \label{eq: SelfAdjoint}
\end{equation}
for real-valued $A$ and $f\to 0$ as $\bfz \to \partial \Gamma$ where $\partial \Gamma $ is the boundary of $\Gamma\subseteq\mathbb{R}^{6N}$. For $\Gamma$ almost covering $\mathbb{R}^{6N}$, $f(\bfz)$ approaches 0 as any component of momentum approaches infinity or any spatial dimension approaches infinity (the probability of finding classical atoms far away from their mean positions must decay to 0) .} Using this property, the time evolution of the phase average of a phase space quantity $A(\bfz)$, defined in terms of $\bfz$, is obtained from
\begin{equation}
N!\,h^{3N}\frac{\dd \avg{A}}{dt} = \frac{\dd}{\dd t}\int_{\Gamma} f(\bfz, t)A(\bfz)\dd\bfz = \int_{\Gamma}\frac{\dd}{\dd t}\left[ e^{-i\mathcal{L}t} f(\bfz)\right] A(\bfz) \dd\bfz =  \int_{\Gamma} e^{-i\mathcal{L}t} f(\bfz) i\mathcal{L}t A(\bfz) \dd\bfz.
\end{equation}
Using equation~\eqref{eq: PhaseQauntity_App}, we arrive at
\begin{equation}
 \frac{\dd\avg{A}}{\dd t} = \frac{1}{N!\,h^{3N}}\int_{\Gamma} f(\bfz, t) \frac{\dd A}{\dd t} \dd\bfz = \avg{\frac{\dd A}{\dd t}}.
 \label{eq: PhaseSpaceDiff_appA}
\end{equation}
We note that equation~\eqref{eq: PhaseSpaceDiff_appA} is obtained using only the evolution equations~\eqref{eq: Liouville_App}, \eqref{eq: PhaseQauntity_App} and property~\eqref{eq: SelfAdjoint}, which hold for any Hamiltonian system and thus contain no time-coarsening approximations. Accordingly, time-variational formulations such as the Frenkel-Dirac-McLachlan variational principle~\citep{mclachlan1964variational} lead to identical equations.

\Revision{
\section{Time evolution of GPP parameters using the variational formulation}
\label{sec: variationalFormulationGaussian}

For the distribution function $f(\bfz, t)$ parametrized as a multivariate Gaussian, the parameter array $\xi(t)$ can be considered separately as $\xi(t) = \left(\lavg{\bfz}, \bfSigma\right)$. Correspondingly, the test functions $u_i$ in \eqref{eq: timeDerivParams} become
\begin{equation}
    \bfu = \frac{\partial \ln f}{\partial \lavg{\bfz}}, 
    \qquad
    \bfU = \frac{\partial \ln f}{\partial \bf\Sigma}.
\end{equation}
The equations of motion for $\lavg{\bfz}$ and $\bfSigma$ are obtained from \eqref{eq: EOM_Parameters} as, respectively,
\begin{subequations}
\begin{equation}
\avg{\bfu\otimes\bfu}\dot{{\lavg{\bfz}}} = \avg{\{\bfu, H\}},
\label{eq: EOM_Mean}
\end{equation}
and
\begin{equation}
 \avg{\bfU\otimes\bfU}:\dot{\bfSigma} = \avg{\{\bfU,H\}}.
 \label{eq: EOM_Sigma}
\end{equation}
\end{subequations}
Substituting the multivariate Gaussian distribution $f(\bfz, t)$ from \eqref{eq: GPP_FULL}, we obtain vector $\bfu$ as, respectively,
\begin{subequations}
\begin{align}
    \bfu = \frac{1}{2}\left(\bfSigma^{-1} + \bfSigma^{-\mathrm{T}}\right)(\bfz - \lavg{\bfz}) = \bfSigma^{-1}(\bfz - \lavg{\bfz}),
\end{align}
and
\begin{align}
\bfU &= -\frac{1}{2\text{det}\bfSigma}\frac{\partial \left(\text{det}\bfSigma\right)}{\partial \bfSigma} - \frac{1}{2}\frac{\partial }{\partial \bfSigma}\left(\left(\bfz - \lavg{\bfz}\right)\T\bfSigma^{-1}\left(\bfz - \lavg{\bfz}\right)\right)
=-\frac{1}{2}\bfSigma^{-1} + \frac{1}{2}\bfu\otimes\bfu.
\end{align}
\end{subequations}
Equations \eqref{eq: EOM_Mean} and \eqref{eq: EOM_Sigma} can be simplified, using
\begin{align}
\avg{\bfu\otimes\bfu} &= \bfSigma^{-1}\avg{(\bfz - \lavg{\bfz})\otimes(\bfz - \lavg{\bfz})}\bfSigma^{-\mathrm{T}} = \bfSigma^{-\mathrm{T}} = \bfSigma^{-1},\\
\avg{\bfU\otimes\bfU} &= \frac{1}{4}\avg{\bfSigma^{-1}\otimes\bfSigma^{-1} - \bfSigma^{-1}\otimes \bfu\otimes\bfu - \bfu\otimes\bfu\otimes\bfSigma^{-1} + \bfu\otimes\bfu\otimes\bfu\otimes\bfu}=\frac{1}{2}\left(\bfSigma^{-1}\otimes\bfSigma^{-1}\right),\\
\avg{\{\bfu, H\}} &= \avg{\dot{\bfz}\cdot\frac{\partial \bfu}{\partial \bfz}} = \bfSigma^{-1}\avg{\dot{\bfz}},\\
\avg{\{\bfU, H\}} &= \avg{\dot{\bfz}\cdot\frac{\partial \bfU}{\partial \bfz}} = \frac{1}{2}\avg{\bfSigma^{-1}\dot{\bfz}\otimes\bfu + \bfu\otimes\bfSigma^{-1}\dot{\bfz}}.
\end{align}
Substituting the above relations and using the identity $(\bfA\otimes\bfB):\bfC = \bfA\bfC\bfB$, we obtain, 
\begin{align}
    \bfSigma^{-\mathrm{T}}\dot{\lavg{\bfz}} &= \bfSigma^{-1}\avg{\dot{\bfz}},\\
    \bfSigma^{-1}\dot{\bfSigma}\bfSigma^{-1} &= \bfSigma^{-1}\avg{\dot{\bfz}\otimes(\bfz - \lavg{\bfz})}\bfSigma^{-1} + \bfSigma^{-1}\avg{(\bfz - \lavg{\bfz})\otimes\dot{\bfz}}\bfSigma^{-1},
\end{align}
which yield the equations of motion in \eqref{eq: phaseSpaceEq}. 

}
\section{Quasistatic GPP as Helmholtz free energy minimization}
\label{sec: GPP_QC_FreeEnergy}

The Helmholtz free energy $\mathcal{F}$ as a function of parameter set $(\lavg{\bfq}, S_\Sigma, S_\Omega)$ is defined as
\begin{equation}
 \mathcal{F}(\lavg{\bfq}, S_\Sigma, S_\Omega) = E(\lavg{\bfq}, S_\Sigma, S) - \sum_i\frac{\Omega_iS_i}{k_B m_i}
 \label{eq: FreeEnergy_App}
\end{equation}
with the relation
\begin{equation}
 \frac{\Omega_i}{k_B m_i} = \frac{\partial E}{\partial S_i}.
\end{equation}
Minimization of $\mathcal{F}$ with respect to the set $\lavg{\bfq}$ yields
\begin{equation}
 -\frac{\partial \mathcal{F}}{\partial \lavg{\bfq}_i} = 0
 \ \implies \ 
 -\avg{\frac{\partial V(\bfq)}{\partial \bfq_i}} =  \avg{F_i(\bfq)} = 0.
\label{eq: Free_MinimizationQ_App}
\end{equation}
To minimize $\mathcal{F}$ with respect to the set $S_{\Sigma}$, we consider the relation
\begin{equation}
 \bfq_i - \lavg{\bfq}_i = \sqrt{\Sigma_i}\bfx_i,
 \label{eq: standardDeviation}
\end{equation}
for some normalized vector $\bfx_i$. From equation~\eqref{eq: standardDeviation} it follows that
\begin{equation}
    \frac{\partial \bfq_i}{\partial S_{\Sigma,i}} =  \frac{\partial \bfq_i}{\partial \sqrt{\Sigma_i}}\frac{\partial \sqrt{\Sigma_i}}{\partial S_{\Sigma,i}}= \sqrt{\Sigma_i}\bfx_i = \bfq_i - \overline{\bfq}_i. 
\label{eq: std_diff_app}
\end{equation}
Finally, minimization of $\mathcal{F}$ with respect to the set $S_{\Sigma}$ yields the following set of equations:
\begin{equation}
-\frac{\partial \mathcal{F}}{\partial S_{\Sigma,i}} = 0
\ \implies\ 
\frac{3\Omega_i}{m_i} - \avg{\frac{\partial V(\bfq)}{\partial \bfq_i}\cdot\frac{\partial \bfq_i}{\partial S_{\Sigma,i}}} = \frac{3\Omega_i}{m_i} + \avg{\bfF_i(\bfq)\cdot\left(\bfq_i-\lavg{\bfq}_i\right)}=0.
\label{eq: Free_Minimization_App}
\end{equation}
Equations~\eqref{eq: Free_MinimizationQ_App} and~\eqref{eq: Free_Minimization_App} are identical to the quasistatic GPP equations~\eqref{eq: GPP_QS}.

\section{Time step stability bounds for entropy transport}
\label{sec: Entropy_transport_stability}

Applying a forward-Euler explicit time discretization to equation~\eqref{eq: temperature_transport}, we obtain
\begin{equation}
\frac{1}{\Delta t^{(k)}}\left(\left(
\begin{matrix}
T_i \\
T_j
\end{matrix}
\right)^{(k+1)} - 
\left(
\begin{matrix}
T_i \\
T_j
\end{matrix}
\right)^{(k)}\right)=\frac{2A_0}{3k_B}\frac{T^{2,(k)}_{ij}}{T^{(k)}_iT^{(k)}_j}\left(
\begin{matrix}
T_i - T_j \\
T_j - T_i
\end{matrix}
\right)^{(k)},
\end{equation}
where superscript $(k)$ implies a quantity evaluated at the $k^\text{th}$ time step. Rearranging the above equation yields
\begin{equation}
\left(
\begin{matrix}
T_i \\
T_j
\end{matrix}
\right)^{(k+1)} =\bfT^{(k)}\left(
\begin{matrix}
T_i \\
T_j
\end{matrix}
\right)^{(k)},
\end{equation}
where $\bfT^{(k)}$ is the transition matrix at the $k^{\mathrm{th}}$ time step, defined by
\begin{equation}
\bfT^{(k)} = \bfI + \frac{2A_0\Delta t^{(k)}}{3k_B}\frac{T^{2,(k)}_{ij}}{T^{(k)}_iT^{(k)}_j}\left(
\begin{matrix}
1 & -1 \\
-1 & 1
\end{matrix}
\right).
\end{equation}
For numerical stability, the transition matrix must have eigenvalues with magnitude $\leq 1$, which yields the bound
\begin{equation}
    \frac{2A_0 \Delta t^{(k)} }{3k_B}\left(\frac{T^{2,(k)}_{ij}}{T^{(k)}_iT^{(k)}_j}\right)\leq 1.
    \label{eq: time_step_app}
\end{equation}
Applying the above limit to a system in which the $i^{\mathrm{th}}$ atom has multiple neighbors, we obtain the constraint in equation~\eqref{eq: time_step_constraint_nonlinear}.
\end{document}